\documentclass[a4paper,fleqn]{cas-sc}

\usepackage[authoryear]{natbib}
\usepackage[utf8]{inputenc}
\usepackage{tabularx}
\usepackage{xltabular}
\usepackage{longtable}
\usepackage{subcaption}
\usepackage{placeins}
\usepackage[expansion=false]{microtype}

\usepackage{xcolor}

\AtBeginEnvironment{longtable}{\sffamily}
\newcommand{\tabhead}[1]{\begin{tabular}[c]{@{}c@{}}#1\end{tabular}}
\newcommand{\reviewfigure}[2]{\includegraphics[width=#1]{#2}}
\NewDocumentCommand\Ouml{}{\"O}

\newtheorem{remark}{Remark}

\begin{document}
\let\WriteBookmarks\relax
\def\floatpagepagefraction{1}
\def\textpagefraction{.001}

\shorttitle{From asymptotic to designer-assigned-time control}
\shortauthors{\"{O}. Bing\"{o}l}

\title[mode=title]{From Asymptotic to Designer-Assigned-Time Control: A Review of Stability Notions, Design Mechanisms, and Controller Architectures}

\author[1]{\Ouml zhan Bing{\"o}l}
\cormark[1]
\ead{ozhan.bingol@gumushane.edu.tr}

\affiliation[1]{organization={Department of Electrical and Electronic Engineering, Faculty of Engineering and Natural Sciences, G\"{u}m\"{u}\c{s}hane University},
            city={G\"{u}m\"{u}\c{s}hane},
            country={Turkey}}

\cortext[1]{Corresponding author}

\begin{abstract}
Many control tasks require a target to be reached not only eventually but on time. Finite-, fixed-, predefined-, and prescribed-time control address this need, yet the labels are used loosely: a settling time that grows with the initial condition, a bound that holds for all initial conditions, a deadline chosen by the designer, and a limit attained only at the terminal instant often share one name. This review aims to make such claims comparable. It separates three questions that are often conflated: what temporal property is promised, which feature of the Lyapunov analysis produces it, and which controller architecture carries it into the closed loop. A recurring question is whether a guarantee proven for an idealized loop survives once observers, adaptation, disturbances, actuator limits, and digital implementation are included. The examined studies are therefore audited one by one, recording what each claims, which variable is actually certified, and how the result is validated. The audit shows that estimation and approximation errors often reduce an exact guarantee to a practical one, and that experimental evidence comes mostly from fast electromechanical systems. A scalar benchmark with two complementary tunings shows how much of the apparent difference between methods stems from conservative bounds, initial-condition dependence, and numerical tolerance, and how saturation and sampling can delay or remove a deadline. The review closes with open problems, among them deciding which deadlines a given plant can meet, combining certificates across interconnected subsystems, and preserving a guarantee through implementation.
\end{abstract}

\begin{keywords}
finite-time stability \sep fixed-time stability \sep predefined-time stability \sep prescribed-time control \sep convergence-time certificates \sep Lyapunov methods
\end{keywords}

\NewCommandCopy\origprintorcid\printorcid
\ExplSyntaxOn
\bool_gset_true:N \g_stm_nologo_bool
\RenewDocumentCommand \printorcid { } { \seq_if_empty:NF \g_stm_orcid_seq { \origprintorcid } }
\ExplSyntaxOff
\maketitle

\section{Introduction and Review Scope}
\label{sec:introduction}

Lyapunov stability provides a framework for analyzing whether trajectories remain near a target and converge to it \citep{slotine1991applied,lyapunov1992general}. Asymptotic stability specifies eventual convergence; exponential stability adds a decay-rate estimate. These notions support robust and adaptive designs in applications ranging from servomechanisms \citep{huang1994robust} to satellite control \citep{burroni2024distributed}, but their standard estimates alone do not certify exact settling within a finite deadline.

Time-critical control requires the transient to be evaluated alongside steady-state accuracy. Prescribed-performance and safety-critical designs illustrate the importance of explicit error and state constraints \citep{xu2023output,cohen2024safety}, while resource allocation for time-sensitive services imposes a related implementation constraint \citep{zhao2021energy}. Lane changes \citep{nilsson2016lane}, assembly-line manipulation \citep{ozkan2018guidance}, and recovery from power-grid disturbances \citep{frasca2021control} motivate explicit temporal requirements, although these application needs do not themselves imply an exact settling-time theorem. The question addressed here is what each convergence-time guarantee actually states and under which closed-loop conditions it can be realized.

The first major step toward explicit convergence-time guarantees was the formalization of finite-time stability (FTS) \citep{bhat2000finite}. In contrast to asymptotic and exponential stability, a finite-time stable system reaches the equilibrium in a finite amount of time. This property is typically obtained by incorporating nonlinear feedback terms, such as fractional powers of the state or error variables, which generate non-Lipschitz closed-loop dynamics near the equilibrium. Finite-time stability offers finite settling through stronger dissipation near the equilibrium and can provide favorable disturbance-rejection properties. However, its main limitation is that the settling time generally depends on the initial system states \citep{lin2020enhancing,zuo2018overview,sui2022novel,ren2023adaptive}. Consequently, when the initial condition is unknown, uncertain, or far from the equilibrium, the convergence time may be difficult to predict and may become unacceptably large for time-critical applications.

To overcome the dependence of the settling time on initial conditions, fixed-time stability (FxTS) was introduced \citep{polyakov2011nonlinear}. A fixed-time stable system is finite-time stable and, additionally, admits a uniform upper bound on the settling time that is independent of the initial condition. This property is commonly achieved by combining nonlinear terms that accelerate convergence both far from and near the equilibrium. Fixed-time stability therefore offers a different guarantee from finite-time stability: a predictable worst-case convergence-time bound. This feature is particularly valuable in applications where the initial condition cannot be accurately measured or bounded in advance. Nevertheless, an important limitation remains: the fixed-time upper bound is generally determined as a consequence of the selected controller gains and Lyapunov parameters and may be conservative or difficult to adjust directly to a desired value \citep{aldana2019enhancing,liu2022overview}. Hence, although fixed-time stability removes initial-condition dependence, it does not necessarily make the desired convergence time an explicit and independently assignable design specification.

Predefined-time stability (PdTS) was developed as a strengthening of fixed-time stability in which an upper bound of the settling time can be assigned a priori as an explicit design parameter \citep{sanchez2015predefined,jimenez2017optimal,jimenez2018note,sanchez2018class,munoz2019predefined,jimenez2020lyapunov,sanchez2020predefined}. In a predefined-time stable system, the designer selects a positive constant $T_c$ before operation, and the controller is constructed so that the settling-time function satisfies an appropriately normalized upper-bound condition associated with $T_c$, independently of the admissible initial condition. This transforms the settling-time bound from an implicit consequence of controller gains into an explicit design specification. Predefined-time approaches commonly exploit normalized nonlinear Lyapunov dissipation laws, fixed-time-type nonlinearities, or related Lyapunov-like constructions in which the desired time parameter appears directly in the stability condition. Such formulations are particularly attractive when a guaranteed upper bound on convergence must be available before the system is initialized.

In parallel, prescribed-time (PT) control has emerged as another major framework for designer-assigned convergence \citep{song2017time}. Its central objective is to drive the system state or tracking error to the desired equilibrium within a finite time horizon selected arbitrarily in advance by the designer and independently of the initial condition \citep{song2019time,li2021stochastic}. A substantial branch of prescribed-time control employs explicitly time-varying mechanisms \citep{song2019time}, including time scaling \citep{yuan2025time}, state scaling \citep{tian2025prescribed}, nonlinear time transformations \citep{hao2026prescribed}, and time-varying feedback gains \citep{zhou2021prescribed}, to reshape an infinite-time stabilization problem over a finite prescribed interval. Other formulations seek bounded \citep{su2025prescribed} or nonsingular prescribed-time controllers \citep{shi2024prescribed} in order to alleviate the gain-growth and implementation difficulties that may arise as the prescribed terminal time is approached. Prescribed-time techniques have subsequently been extended from state-feedback stabilization \citep{song2017time} to output-feedback \citep{li2022prescribed}, adaptive \citep{hua2021adaptive}, robust \citep{obuz2024robust}, stochastic \citep{li2021stochastic}, multi-agent \citep{gong2020distributed}, constrained \citep{yao2024novel}, and optimal-control settings \citep{song2023prescribed}.

\begin{figure*}[pos=!t]
\centering
\reviewfigure{\textwidth}{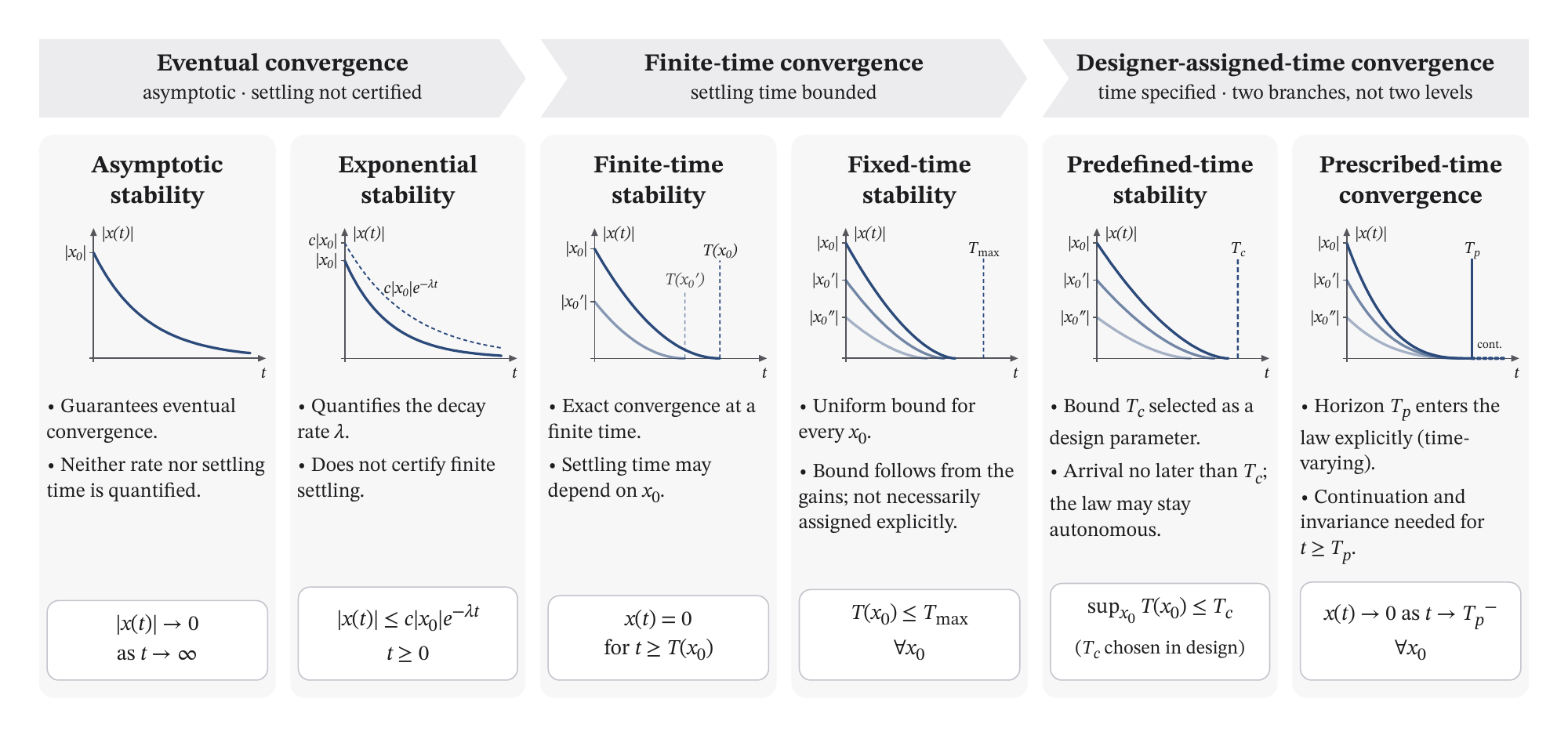}
\caption{Conceptual overview of convergence-time specifications (schematic curves).}
\label{fig:temporal_overview}
\end{figure*}

Although predefined-time and prescribed-time approaches share the fundamental objective of allowing the convergence horizon to be selected in advance, their dominant formulations emphasize different mechanisms. Predefined-time stability is commonly expressed through settling-time functions and Lyapunov inequalities normalized with respect to a designer-selected upper bound, whereas prescribed-time control frequently employs explicitly time-dependent control structures or transformations constructed around a prescribed terminal horizon. These distinctions are not absolute, since contemporary developments increasingly combine bounded time-varying gains, Lyapunov normalization, scaling transformations, and fixed-time mechanisms. Accordingly, rather than arranging predefined-time and prescribed-time control as two successive levels in a strictly ordered stability hierarchy, this review treats them as two important branches of a broader class of \emph{designer-assigned-time control} (Remark~\ref{rem:no_hierarchy} makes the point precise). This is also the only sense in which the progression in the title is meant: it orders the temporal information supplied by a guarantee, not the speed, robustness, or control effort of the controllers that realize it. This perspective enables their common objectives, mathematical differences, controller-design mechanisms, and implementation trade-offs to be examined within a unified framework. Figure~\ref{fig:temporal_overview} provides a schematic overview of the temporal specifications. In its fixed-time panel, $T_{\max}$ is a common upper bound, not a common arrival time; in its predefined-time panel, the definition requires only $\sup_{x_0}T(x_0)\le T_c$, and for a tight law the settling time approaches $T_c$ only as $|x_0|\to\infty$; and zero-state behavior after the horizon presupposes the continuation and invariance conditions of Section~\ref{sec:notions}.

\begin{figure*}[pos=!t]
\centering
\reviewfigure{0.90\linewidth}{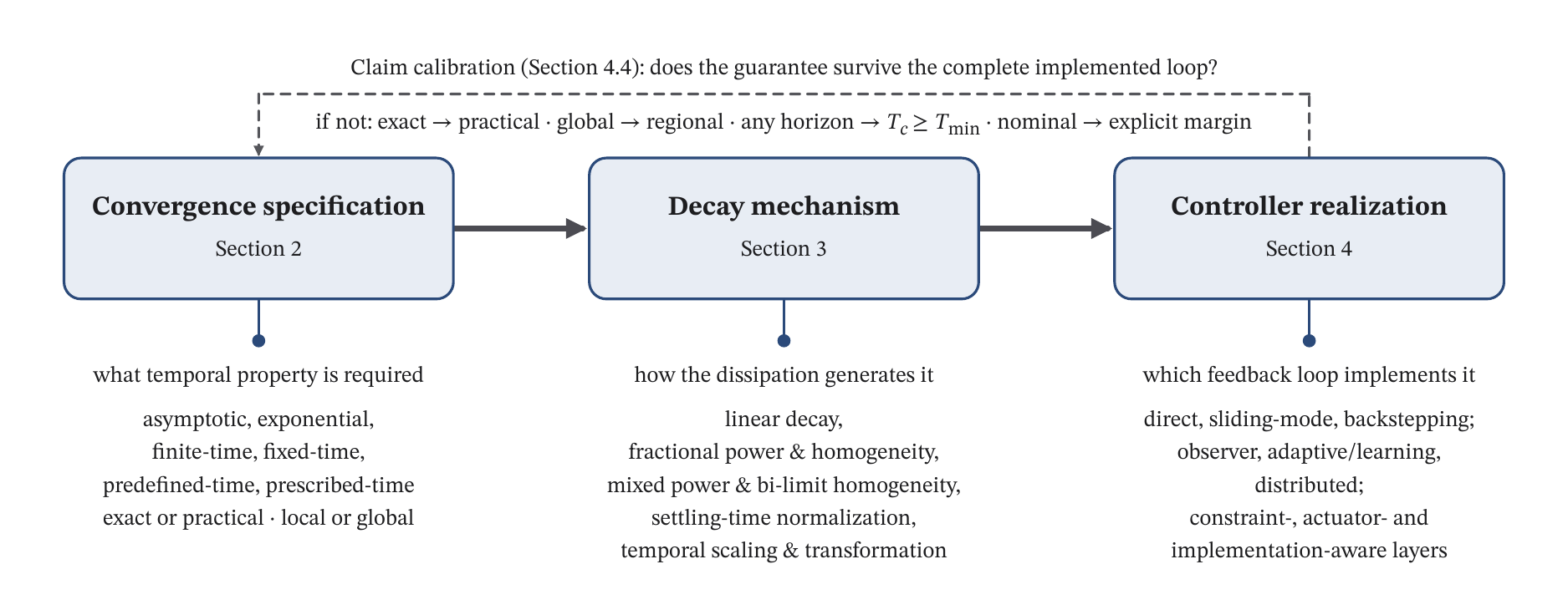}
\caption{Three-layer organization: specifications, mechanisms, and controller realizations.}
\label{fig:three_layers}
\end{figure*}

The organizing principle is a three-layer distinction, summarized in Fig.~\ref{fig:three_layers}. A \emph{convergence specification} states the required temporal property. A \emph{design mechanism} explains how that property is generated, for example through state-dependent dissipation, homogeneity, normalization, or temporal transformation. A \emph{controller architecture} realizes the mechanism through physical feedback, potentially together with estimation, adaptation, filtering, or communication dynamics. The same mechanism can appear in different architectures, and a single architecture can accommodate several temporal specifications. The central question is whether the guarantee survives the passage from the nominal mechanism to the complete implemented closed loop. A stability notion alone determines neither the mechanism through which the temporal property is generated nor the architecture used to realize it; Sections~\ref{sec:notions}, \ref{sec:mechanisms}, and~\ref{sec:architectures} correspond to the three layers. The dashed return path in Fig.~\ref{fig:three_layers} is the claim calibration of Section~\ref{subsec:design_guidance}: when the guarantee is not preserved in the implemented loop, the specification is weakened explicitly, as in \eqref{eq:claim_calibration}.

\subsection{Scope and Contributions}
\label{subsec:scope_contributions}

Existing surveys provide complementary perspectives on finite- and fixed-time control and its engineering applications \citep{liu2022overview}, finite- and fixed-time synchronization of complex networks \citep{anand2020brief}, cooperative and multiagent systems \citep{zuo2018overview,zuo2026fixed}, including finite-time coordination of underwater vehicles \citep{chen2024finite}, finite-time stability analysis in both lineages of the term \citep{xu2020finite} (Remark~\ref{rem:two_fts}), prescribed-time control \citep{song2023prescribed}, and distributed fixed- and prescribed-time optimization \citep{ning2026overview}. Neighboring topics are covered by surveys of safe learning with control Lyapunov and control barrier functions \citep{anand2021safe} and of Lyapunov-Krasovskii functionals for time-delay systems \citep{zhang2022construction}. The present review connects these perspectives through the specification-mechanism-architecture distinction. Its scope is a comparative synthesis of representative analytical and design structures, rather than an exhaustive enumeration of every application or a claim that all methods reduce to one Lyapunov formula. Table~\ref{tab1} places the review alongside representative surveys; it compares scope without treating specialization as a limitation.

\begin{table}[pos=!t]
\centering
\caption{Positioning of the present review with respect to representative surveys. The last column states the complementary emphasis of this review, not a limitation of the cited survey, each of which serves its own purpose.}\label{tab1}
{\footnotesize
\setlength{\tabcolsep}{3pt}
\renewcommand{\arraystretch}{1.15}
\begin{tabularx}{\textwidth}{@{}>{\raggedright\arraybackslash}p{0.15\textwidth} >{\raggedright\arraybackslash}p{0.24\textwidth} >{\raggedright\arraybackslash}p{0.20\textwidth} >{\raggedright\arraybackslash}X@{}}
\toprule
\textbf{Survey} & \textbf{Main scope} & \textbf{Emphasis} & \textbf{Complementary focus of this review}\\
\midrule
\citealp{liu2022overview} & Finite- and fixed-time control and engineering applications & Theory, methods, and applications & The temporal guarantee assessed together with its mechanism and architecture\\
\addlinespace
\citealp{zuo2018overview} & Fixed-time cooperative control of multiagent systems & Cooperative and coordination tasks & The certified variable and the conditions of validity compared across architectures\\
\addlinespace
\citealp{song2023prescribed} & Prescribed-time control & Design mechanisms and problem classes & Predefined- and prescribed-time results read with common temporal criteria\\
\addlinespace
\citealp{ning2026overview} & Distributed fixed- and prescribed-time optimization & Optimization problems and algorithms & Convergence of the optimization variable separated from convergence of the plant or regulation error\\
\addlinespace
Present review & Convergence guarantees from asymptotic to designer-assigned time & Specification-mechanism-architecture framework & Preservation of the nominal guarantee under auxiliary dynamics and implementation conditions\\
\bottomrule
\end{tabularx}
}
\end{table}

The review develops three connected contributions:
\begin{enumerate}
\item A consistent account of convergence-time specifications that distinguishes actual settling times, uniform bounds, designer-selected horizons, terminal-time limits, and practical convergence, together with a map from the terminology used in the literature to these classes (Table~\ref{tab:terminology}).
\item A mechanism-oriented organization of representative Lyapunov inequalities and transformations, retaining their mathematical conditions alongside the corresponding temporal interpretation.
\item An architectural analysis of how reaching, recursive, estimation, adaptation, and information structures affect the scope and preservation of a temporal guarantee, supported by an audit of the examined studies (Appendix~\ref{app:audit}, with boundary cases in Table~\ref{tab:audit}).
\end{enumerate}

Implementation constraints belong to the architectural analysis: a desired decay law must be compatible with available control authority, measured information, and auxiliary dynamics. This perspective supports the subsequent numerical and research-direction sections without presupposing that a stronger nominal guarantee always requires greater control effort.

Studies are compared in this review by the temporal property they prove, not by the name of the class they use. The comparison considers the certified variable and its accuracy, the dependence of the settling time on the initial condition, and the assumptions under which the guarantee holds; auxiliary dynamics and implementation constraints are examined to decide whether the nominal result is preserved in the closed loop. The evaluation criteria are stated in Section~\ref{subsec:comparison_criteria}, where they are applied to the study-level audit.

The remainder of the review is organized as follows. Section~\ref{sec:notions} defines the convergence specifications and maps the terminology of the literature onto them. Section~\ref{sec:mechanisms} examines their generating mechanisms and the associated inequality catalogues. Section~\ref{sec:architectures} considers architectural realizations, implementation constraints, and design guidance. Section~\ref{sec:numerical} presents illustrative comparisons. Section~\ref{sec:open-problems} collects open research directions, and Section~\ref{sec:conclusions} concludes the review.

\section{Stability and Convergence-Time Notions}
\label{sec:notions}

This section defines the temporal properties used throughout the review. Each notion is stated with its definition and one representative Lyapunov certificate; the derivations, the extended inequality catalogues, and their mechanism-level interpretation are deferred to Section~\ref{sec:mechanisms}. Unless stated otherwise, the guarantee concerns the equilibrium of the complete system under the stated solution concept and initial-condition domain. The separation of predefined-time stability from prescribed-time convergence in Sections~\ref{subsec:predefined_time_stability} and~\ref{subsec:prescribed_time_control} is the analytical convention of this review, based on the mechanism that produces the assigned time; it is not a literature-wide definition, and studies published under either name are compared through their certificates rather than their labels (Remark~\ref{rem:no_hierarchy}).

\subsection{System Description and Notation}
\label{subsec:system_description}

Consider the autonomous nonlinear system
\begin{equation}
\dot{x}=f(x), \qquad f(0)=0,
\label{eq1}
\end{equation}
where $x\in D\subseteq\mathbb{R}^{n}$ is the state vector, $D$ is an open neighborhood of the origin, and $f:D\rightarrow\mathbb{R}^{n}$ is continuous in the continuous-feedback setting. Local Lipschitz continuity away from the equilibrium, together with appropriate existence, forward-uniqueness, and continuation conditions, is assumed where required by the cited result. Local Lipschitz continuity at the equilibrium is not imposed as a common hypothesis for autonomous finite-time convergence. The condition $f(0)=0$ makes $x=0$ an equilibrium point. The Euclidean norm is denoted by $|\cdot|$, and $x(t,x_0)$ denotes the solution of \eqref{eq1} from $x(0)=x_0$.

A continuously differentiable function $V:D\rightarrow\mathbb{R}_{\geq 0}$ is positive definite if $V(0)=0$ and $V(x)>0$ for all $x\in D\setminus\{0\}$, and radially unbounded if $V(x)\rightarrow\infty$ as $|x|\rightarrow\infty$. Its derivative along the trajectories of \eqref{eq1} is
\begin{equation}
\dot{V}(x)=\nabla V(x)f(x).
\label{eq2}
\end{equation}

\begin{remark}[Solution concepts and invariance]
Classical Lyapunov arguments are understood along classical or Carath\'eodory solutions as appropriate. For discontinuous feedback, a specified differential-inclusion interpretation, such as the Filippov interpretation, and an appropriate generalized derivative are required. Existence and invariance must be established for that interpretation. When $\dot V\leq0$, LaSalle-type invariance arguments and the related Barbashin-Krasovskii criterion require the corresponding invariant-set hypotheses; nonincrease alone does not establish attraction \citep{khalil2002nonlinear}. For the finite-time notions below, reaching the target must be accompanied by its subsequent invariance.
\end{remark}

\subsection{Asymptotic Stability}
\label{subsec:asymptotic_stability}

The equilibrium point $x=0$ of \eqref{eq1} is stable in the sense of Lyapunov if, for every $\epsilon>0$, there exists $\delta(\epsilon)>0$ such that
\begin{equation}
|x(0)|<\delta
\quad \Rightarrow \quad
|x(t)|<\epsilon,
\qquad \forall t\geq 0.
\label{eq3}
\end{equation}
It is asymptotically stable if it is Lyapunov stable and there exists $\delta_a>0$ such that
\begin{equation}
|x(0)|<\delta_a
\quad \Rightarrow \quad
\lim_{t\rightarrow\infty}x(t)=0,
\label{eq4}
\end{equation}
and globally asymptotically stable if these properties hold for all $x(0)\in\mathbb{R}^{n}$ \citep{slotine1991applied,khalil2002nonlinear}. A sufficient certificate is a continuously differentiable $V$ with \citep{slotine1991applied,khalil2002nonlinear}
\begin{equation}
V(0)=0,\qquad V(x)>0,\qquad \dot V(x)<0,\qquad \forall x\in D\setminus\{0\},
\label{eq:as_certificate}
\end{equation}
the guarantee being global when $D=\mathbb{R}^n$ and $V$ is radially unbounded. Asymptotic stability is a qualitative guarantee: it ensures eventual approach to the equilibrium but quantifies neither the convergence rate nor a settling time.

\subsection{Exponential Stability}
\label{subsec:exponential_stability}

The equilibrium $x=0$ of \eqref{eq1} is exponentially stable if there exist positive constants $c$, $\lambda$, and $r$ such that, for all $|x(0)|<r$,
\begin{equation}
|x(t)|\leq c|x(0)|e^{-\lambda t},
\qquad \forall t\geq 0,
\label{eq5}
\end{equation}
where $\lambda$ is the convergence rate and $c$ bounds the transient overshoot; the property is global if \eqref{eq5} holds for all $x(0)\in\mathbb{R}^n$ \citep{slotine1991applied}. A common sufficient certificate is a continuously differentiable $V$ and constants $c_1,c_2,c_3>0$ with
\begin{equation}
c_1|x|^2\leq V(x)\leq c_2|x|^2,
\qquad
\dot{V}(x)\leq -c_3|x|^2,
\label{eq6}
\end{equation}
under which the rate can be related explicitly to $c_1$, $c_2$, and $c_3$. Exponential stability supplies a decay envelope but does not certify finite settling: the estimate \eqref{eq5} is compatible with asymptotic approach, and finite settling, when it occurs, must be established separately.

\subsection{Finite-Time Stability}
\label{subsec:finite_time_stability}

FTS requires that trajectories reach the equilibrium in a finite amount of time \citep{bhat2000finite}, rather than only as $t\rightarrow\infty$. The equilibrium $x=0$ of \eqref{eq1} is finite-time stable if it is Lyapunov stable and, for every initial condition in some neighborhood $U\subseteq D$ of the origin, there exists a finite settling time $T(x_0)<\infty$ such that
\begin{equation}
\lim_{t\rightarrow T(x_0)}x(t)=0,
\qquad
x(t)=0\quad \forall t\geq T(x_0).
\label{eq14}
\end{equation}
If this holds for every $x_0\in\mathbb{R}^{n}$, the equilibrium is globally finite-time stable. A representative certificate is a continuously differentiable positive-definite $V$ with
\begin{equation}
\dot{V}(x)\leq -cV^\alpha(x),
\qquad c>0,\quad \alpha\in(0,1),\quad \forall x\neq 0,
\label{eq16}
\end{equation}
which gives
\begin{equation}
T(x_0)\leq \frac{V^{1-\alpha}(x_0)}{c(1-\alpha)}.
\label{eq17}
\end{equation}
The bound depends on $V(x_0)$ and does not by itself establish a uniform global bound; Section~\ref{subsec:homogeneous_mechanisms} develops the mechanism and Table~\ref{tab:finite_catalogue} collects extensions and practical variants. Finite-time stability is a system-level property; it should not be identified with terminal sliding-mode control \citep{venkataraman1993control}, a design methodology that often yields finite-time convergence on the sliding surface.

\begin{remark}[Two notions sharing one name]
\label{rem:two_fts}
Two structurally different properties are called \emph{finite-time stability} in the literature. The first, due to \citet{kamenkov1953stability} and \citet{dorato1961short}, is a boundedness property over a prescribed window: for given $(c_1,c_2,T,R)$ with $0<c_1<c_2$ and $R>0$, $x^{\top}(0)Rx(0)\le c_1$ must imply $x^{\top}(t)Rx(t)<c_2$ for all $t\in[0,T]$. It presupposes no equilibrium, requires no convergence, is independent of asymptotic stability, and its Lyapunov certificate \emph{relaxes} the classical condition to $\dot V\le\beta V$ with $\beta\ge0$, combined with a level-set condition linking $c_1$, $c_2$, $T$, and $\beta$. The second, formalized by \citet{bhat2000finite}, is the equilibrium property defined above: Lyapunov stability with a settling-time function $T(\cdot)$; it is strictly stronger than asymptotic stability, and its certificate \emph{strengthens} the classical condition to \eqref{eq16}. The distinction matters here because the window $T$ of the first notion resembles the assigned horizon $T_c$ of the second lineage only superficially: in the former, $T$ is an interval over which a bound must hold and carries no convergence guarantee; in the latter, the terminal time is an outcome of the dynamics or a design choice. All designer-assigned-time notions in this review belong to the second lineage. Throughout, ``finite-time stability'' means the Bhat-Bernstein notion; the first is called \emph{short-time stability} whenever it is invoked.
\end{remark}

\begin{remark}
Finite-time stability is also conceptually distinct from terminal sliding-mode control (TSMC) \citep{venkataraman1993control}. While TSMC provides a design methodology that often yields finite-time convergence on the sliding surface, FTS is a system-level property defined independently of any specific control structure. Not all terminal sliding-mode controllers guarantee global finite-time stability, and not all finite-time stable systems rely on terminal sliding surfaces.
\end{remark}

\subsection{Fixed-Time Stability}
\label{subsec:fixed_time_stability}

FxTS strengthens FTS by requiring a uniform, state-independent settling-time bound \citep{polyakov2011nonlinear}. The equilibrium $x=0$ of \eqref{eq1} is fixed-time stable if it is finite-time stable and there exists a constant $T_{\max}<\infty$ such that
\begin{equation}
T(x_0)\leq T_{\max},
\qquad \forall x_0\in D,
\label{eq18}
\end{equation}
and globally fixed-time stable if this holds for every initial state in $\mathbb{R}^{n}$. The uniformity concerns the upper bound, not the actual settling-time function: $T(x_0)$ may still vary with $x_0$. A common certificate is a continuously differentiable positive-definite $V$ with
\begin{equation}
\dot{V}(x)\leq -c_1V^\alpha(x)-c_2V^\beta(x),
\qquad c_1,c_2>0,\quad \alpha\in(0,1),\quad \beta>1,\quad \forall x\neq 0,
\label{eq19}
\end{equation}
for which one common upper bound is
\begin{equation}
T_{\max}\leq \frac{1}{c_1(1-\alpha)}
+
\frac{1}{c_2(\beta-1)}.
\label{eq20}
\end{equation}
The term $V^\alpha$ governs the near-origin behavior and $V^\beta$ the large-state behavior; Section~\ref{subsec:mixed_power_mechanisms} develops the mechanism and Table~\ref{tab:fixed_catalogue} collects additional structures.

\subsection{Predefined-Time Stability}
\label{subsec:predefined_time_stability}

PdTS is treated here as an assignability property of a parameterized fixed-time family. Consider
\begin{equation}
\dot x=f(x;\vartheta),\qquad f(0;\vartheta)=0,\qquad \vartheta\in\Theta,
\label{eq:pdt_family}
\end{equation}
with settling-time function $T(x_0;\vartheta)$ on a stated initial-condition domain $D$. For each admissible designer-selected $T_c>0$, the family has the predefined-time property if a parameter choice $\vartheta=\vartheta(T_c)$ gives Lyapunov stability and finite settling with
\begin{equation}
\sup_{x_0\in D}T(x_0;\vartheta(T_c))\leq T_c.
\label{eq21}
\end{equation}
The definition is global when $D=\mathbb R^n$ and the corresponding global solution conditions hold \citep{sanchez2015predefined,jimenez2020lyapunov}. Restrictions on admissible horizons, initial conditions, or controller parameters must be stated rather than hidden in the tuning map.

Property \eqref{eq21} is the \emph{weak} predefined-time property. The \emph{strong} version in \citet{jimenez2020lyapunov} additionally requires the bound to be tight, $\sup_{x_0\in D}T(x_0;\vartheta(T_c))=T_c$. Tightness is therefore a separate property: \eqref{eq21} neither imposes equality nor requires every nonzero trajectory to arrive exactly at $T_c$ \citep{jimenez2020lyapunov,jimenez2017optimal}. The relevant distinction is between possessing a uniform bound for a fixed tuning and providing a parameterization that certifies a selected bound; some fixed-time families acquire the latter through gain scaling, so predefined-time assignment need not introduce a different state-dependent convergence mechanism.

A representative sufficient certificate uses a positive-definite continuously differentiable $V$ and a normalized profile $\widehat\Phi$ with $\widehat\Phi(0)=0$, $\widehat\Phi(v)>0$ for $v>0$, and
\begin{equation}
\dot V\leq-\frac{1}{T_c}\widehat\Phi(V),
\qquad
\int_0^\infty\frac{dv}{\widehat\Phi(v)}\leq1,
\label{eq22}
\end{equation}
the integral being improper at zero and at infinity. Under the required existence and invariance conditions,
\begin{equation}
T(x_0;\vartheta(T_c))\leq
T_c\int_0^{V(x_0)}\frac{dv}{\widehat\Phi(v)}\leq T_c.
\label{eq:pdt_normalized_bound}
\end{equation}
For a global result, properness and forward-continuation conditions are also required. Section~\ref{subsubsec:time_normalized_design} relates this convention to the explicit factor $G_c$, and Table~\ref{tab:assigned_catalogue} collects representative structures.

\subsection{Prescribed-Time Convergence}
\label{subsec:prescribed_time_control}

PT control is the other major designer-assigned-time approach. Unlike FTS, where the settling time may depend on the initial condition, and FxTS, where only a uniform upper bound is guaranteed, PT control constructs the closed-loop dynamics around a finite terminal horizon $T_p>0$ selected in advance \citep{song2017time,song2023prescribed}; the horizon is independent of the initial condition and enters the control or transformation mechanism explicitly. Since PT controllers are typically time-varying, consider the nonautonomous system
\begin{equation}
    \dot{x}=f(x,t), \qquad f(0,t)=0,
    \label{eq:pt_system}
\end{equation}
with $x\in\mathbb{R}^{n}$ and $t\in[0,T_p)$. For a fixed initialization time $t_0=0$, a representative prescribed-horizon convergence estimate uses a monotonically increasing function
\begin{equation}
    \tau_p:[0,T_p)\rightarrow\mathbb{R}_{+},
    \qquad
    \lim_{t\rightarrow T_p^-}\tau_p(t)=\infty,
    \label{eq:pt_mu}
\end{equation}
and a class-$\mathcal{KL}$ function $\beta$ such that
\begin{equation}
    \|x(t)\|
    \leq
    \beta\!\left(\|x(0)\|,\tau_p(t)\right),
    \qquad
    t\in[0,T_p),
    \label{eq:pt_definition}
\end{equation}
so that
\begin{equation}
    \lim_{t\rightarrow T_p^-}x(t)=0.
    \label{eq:pt_terminal}
\end{equation}
Uniformity with respect to a variable initialization time requires the corresponding estimate for the shifted horizon and is not inferred from the fixed-$t_0$ notation alone. When the closed-loop solution is continuously extended beyond $T_p$ and the origin remains invariant, the state can be maintained at
\begin{equation}
    x(t)=0,\qquad t\geq T_p.
    \label{eq:pt_after}
\end{equation}

A representative Lyapunov characterization introduces a positive time-varying function $\mu(t)$ with
\begin{equation}
    \int_{0}^{T_p}\mu(s)\,ds=\infty,
    \label{eq:pt_integral}
\end{equation}
and a continuously differentiable positive-definite $V(x)$ with
\begin{equation}
    \dot V(x,t)
    \leq
    -c\mu(t)V(x),
    \qquad
    c>0,
    \label{eq:pt_lyapunov}
\end{equation}
so that $V(t)\leq V(0)\exp\!\left(-c\int_0^t\mu(s)\,ds\right)\rightarrow0$ as $t\rightarrow T_p^-$ under the stated positive-definiteness and solution conditions \citep{song2023prescribed,hua2021adaptive}. This inequality establishes a terminal limit on $[0,T_p)$; maintaining the state at zero after $T_p$ additionally requires a well-defined continuation and equilibrium invariance. Divergence of an auxiliary scaling function such as $\mu(t)=T_p/(T_p-t)$ is a property of the certificate, not of the physical input: admissibility of the input and of the internal signals, and the continuation of the solution at and after $T_p$, are separate conditions examined in Section~\ref{subsubsec:time_varying_pt}. Three cases should therefore be distinguished. Time-scaling certificates of the form \eqref{eq:pt_lyapunov} typically give a terminal limit: $x(t)\neq0$ for $t<T_p$ and $x(t)\to0$ only as $t\to T_p^-$; for example, $\dot x=-cT_p x/(T_p-t)$ yields $x(t)=x_0\bigl((T_p-t)/T_p\bigr)^{cT_p}$. Some designs settle strictly before $T_p$ with an explicit margin \citep{Cui2026PrescribedTimeMorphingHMV}. Exact settling at $T_p$ with bounded signals and a solution defined at and after $T_p$, as in terminal-time-regulator (nominally) and periodic-delayed-feedback designs \citep{Shao2026TTRPrescribedPerformanceSMC,Ding2026PDFSuperTwistingPsT}, is a stronger and distinct specification.

\begin{remark}[Predefined- and prescribed-time are not consecutive levels]
\label{rem:no_hierarchy}
PdTS and PT control should not be treated as consecutive levels of a strictly ordered stability hierarchy. In PdTS, the principal specification is a designer-selected upper bound, $T(x_0)\leq T_c$, established through a normalized settling-time or Lyapunov condition, and the construction may remain autonomous. In PT control, the controller is constructed explicitly around a terminal horizon $T_p$ through time-varying scaling or transformation. The two frameworks share the designer-assigned-time objective but emphasize different analytical and controller-design mechanisms, and a given design may draw on both. The division is therefore a classification convention for comparing analyses: a result in which the assigned time does not enter the law explicitly is not thereby excluded from prescribed-time control as its authors name it, and it is read here through the mechanism its certificate uses. The same holds for the practical and input-to-state variants of Section~\ref{subsec:iss_practical}, which modify a nominal guarantee rather than adding a level.
\end{remark}

\begin{remark}[Arbitrary-time stabilization]
\label{rem:arbitrary_time}
\citet{pal2020design} introduced \emph{arbitrary-time stability} as a fixed-time property whose convergence-time quantity $T_a$ can be evaluated in advance and adjusted through admissible variations of the design parameters, and \emph{free-will arbitrary-time stability}, where $T_a$ is to be chosen independently of both the initial condition and the system parameters. \citet{orlov2022comments} showed that the representative arbitrary-time dynamics are a special case of the earlier prescribed-time construction \citep{song2017time} under a logarithmic change of variables, and raised separate technical objections to the free-will definition. This review therefore does not treat arbitrary-time stabilization as a distinct notion: studies using the term are classified with prescribed-time approaches when their mechanism and temporal specification coincide, as recorded in Table~\ref{tab:terminology}.
\end{remark}

\subsection{Input-to-State and Practical Stability Perspectives}
\label{subsec:iss_practical}

The notions above describe exact convergence of the nominal closed loop. In practical implementations, disturbances, approximation errors, observer residuals, and sampling effects may remain nonvanishing \citep{li2023adaptive}, and exact convergence is replaced by an input-dependent or practical property. For the disturbed system $\dot{x}=f(x,d)$, input-to-state stability (ISS) provides the standard robustness characterization \citep{sontag1995characterizations, sontag2008input},
\begin{equation}
    \|x(t)\|
    \leq
    \beta(\|x_0\|,t)
    +
    \gamma(\|d\|_{\infty}),
    \label{eq:iss_estimate}
\end{equation}
with $\beta$ of class $\mathcal{KL}$ and $\gamma$ of class $\mathcal{K}$; the associated ISS-Lyapunov inequality $\dot V\leq-\alpha(V)+\sigma(\|d\|)$ separates the stabilizing dissipation from the effect of the disturbance. Trajectories then converge to and remain within a residual neighborhood, which leads to the practical-stability \citep{lakshmikantham1990practical} and ultimate-boundedness \citep{barmish1982ultimate} concepts.

The same distinction applies to finite- and fixed-time convergence. Such results are frequently described as semi-global practical finite-time stability, for which the entrance time may depend on the initial condition \citep{xi2023practical}; the fixed-time counterpart additionally requires a uniform bound on the entrance time and is commonly called semi-global practical FxTS \citep{poveda2021fixed}, although the terminology is not uniform. Designer-assigned-time frameworks admit analogous practical interpretations: predefined-time ultimate boundedness requires entry into a prescribed neighborhood within a designer-selected time $T_c$ \citep{jimenez2020lyapunov}, and practical PT formulations replace exact convergence at $T_p$ by convergence to a specified accuracy region \citep{cao2022practical}. Such residual-set results dominate much of the recent learning-based, event-triggered, and networked literature: uniform ultimate boundedness of implemented adaptive-dynamic-programming loops \citep{Xu2026ZeroSumDSTC,Yue2026SMSEventTriggeredFTC,Zhang2026HumanMachineDynamicAuthority}, practical stability certified from data \citep{Mei2026DataDrivenIntermittent}, and input-to-state practical stability under disturbances and denial-of-service attacks \citep{Shangguan2026DISMLFC}. These concepts provide a complementary robustness perspective on how a nominal guarantee is modified; Section~\ref{subsec:specification_to_design} states the corresponding qualification once, at the mechanism level.

\subsection{Terminology Used in the Literature}
\label{subsec:terminology}

Table~\ref{tab:terminology} maps the terms encountered in the literature to the temporal property they certify in the terms of this section; the mechanisms that generate these properties are left to Section~\ref{sec:mechanisms}. Each row answers one question: a source uses this term, so what temporal property is actually established? The rows are grouped into core temporal notions, alternative names for them, and qualifiers that modify a guarantee rather than add a class. The map is used when a study's own terminology differs from the class recorded for it in the audit of Section~\ref{subsec:design_guidance} and Appendix~\ref{app:audit}.

\noindent Terminology map. Original terminology is retained when a study is cited; the interpretation column states how the term is read in this review, and the third column the condition that distinguishes it or must be checked.

{
\footnotesize
\setlength{\tabcolsep}{3pt}
\renewcommand{\arraystretch}{1.12}
\begin{xltabular}{\textwidth}{@{}>{\raggedright\arraybackslash}p{0.20\textwidth} >{\raggedright\arraybackslash}p{0.20\textwidth} >{\raggedright\arraybackslash}X >{\raggedright\arraybackslash}p{0.19\textwidth}@{}}
\caption{Convergence-time terminology.}\label{tab:terminology}\\
\toprule
\textbf{Term in the literature} & \textbf{Interpretation in this review} & \textbf{Distinguishing condition} & \textbf{Representative references}\\
\midrule
\endfirsthead
\multicolumn{4}{@{}l}{\footnotesize\textit{Table~\ref{tab:terminology} (continued)}}\\
\toprule
\textbf{Term in the literature} & \textbf{Interpretation in this review} & \textbf{Distinguishing condition} & \textbf{Representative references}\\
\midrule
\endhead
\midrule\multicolumn{4}{r@{}}{\footnotesize\textit{continued on next page}}\\
\endfoot
\bottomrule\endlastfoot

\multicolumn{4}{@{}l}{\textit{Core temporal notions}}\\*
Finite-time stability (Bhat-Bernstein); fast finite-time & Finite settling time, possibly dependent on $x_0$ & Arrival in finite time and retention afterwards are both required; a uniform bound is a separate claim. ``Fast'' adds a linear term and does not change the class. & \citealp{bhat2000finite,huang2005global,sun2017fast,sun2019fast}\\
Fixed-time stability; fast fixed-time; nonsingular fast terminal & Uniform settling-time bound, independent of $x_0$ & Trajectories need not arrive together; tightness or direct assignability of the bound is not part of the definition & \citealp{polyakov2011nonlinear,polyakov2015finite,ni2016fast,zhang2022novel,barman2025novel}\\
Predefined-time stability (weak / strong) & Upper bound $T_c$ selected by the designer & $T(x_0)\le T_c$ (weak) and $\sup_{x_0}T(x_0)=T_c$ (strong) are different claims; the second adds tightness & \citealp{sanchez2015predefined,jimenez2017optimal,jimenez2020lyapunov}\\
Prescribed-time control; prescribed finite time & Convergence tied to a selected deadline $T_p$ & The terminal limit $t\to T_p^-$ (typical of time scaling), settling before $T_p$ with a margin, exact settling at $T_p$ with bounded signals, and retention after $T_p$ are assessed separately & \citealp{song2017time,song2019time,song2023prescribed,zuo2022adaptive}; exact settling at $T_p$: \citealp{Shao2026TTRPrescribedPerformanceSMC,Ding2026PDFSuperTwistingPsT}\\
\multicolumn{4}{@{}l}{\textit{Alternative names}}\\*
Preassigned-time; appointed(-fixed)-time & Read as predefined-time by the role of the horizon & Whether the bound is selectable for every tuning or only computable after the gains are fixed & \citealp{liu2019direct,hu2020fixed,wei2022fixed,jiang2025composite,sun2023appointed,li2021appointed}\\
Predefined-time-synchronized & Predefined-time with simultaneous arrival of all components & Ratio persistence is an additional property, not implied by a predefined bound & \citealp{Yu2026ContinuumPTSS,Ma2026PredefinedSynchronizedBipartite}\\
Time-base generator; predefined-time quasi-synchronization & Prescribed-time mechanism with a regularized clock or bounded gain & Whether the terminal claim is exact or a residual set & \citealp{Sun2026PredefinedUAVSpatiotemporal,Zhou2026PTQuasiSynchronization,Li2026PsTClusterConsensusHMAS}\\
Arbitrary-time; free-will arbitrary-time & Classified with prescribed-time approaches when mechanism and specification coincide (Remark~\ref{rem:arbitrary_time}) & A new name alone does not establish a new stability property; definition and certificate are compared case by case & \citealp{pal2020design,orlov2022comments,asadollahi2024fixed}\\
Finite-time / short-time stability (Kamenkov-Dorato); finite-time bounded consensus & Not a convergence notion (Remark~\ref{rem:two_fts}) & Bounded state over a finite window; no settling is implied & \citealp{kamenkov1953stability,dorato1961short,you2024finite,Liu2026MultiLoopNCSFiniteTime,Chen2026FTBCMultiAgentAttacks}\\
\multicolumn{4}{@{}l}{\textit{Qualifiers of the guarantee}}\\*
Practical predefined-time; predefined time and accuracy; predefined-time ultimate boundedness & Entry into a residual set within the selected bound & Size of the set and retention after entry must be stated & \citealp{wang2022adaptive,ni2021global,zhang2022practically,li2023adaptive,zhang2026predefined,jimenez2020lyapunov}\\
Practical prescribed-time & Entry into an accuracy region by $T_p$ & Exact convergence is replaced by a stated region and its retention & \citealp{cao2022practical,Golestani2026PracticalPrescribedTimePPC,Kim2026EventImpulsivePrescribedTimeUUV,Cai2026MSVPracticalPsT,Zhou2026PracticalPsTRigidFormation}\\
Finite-, fixed-, specified-time prescribed performance & Envelope reaching its terminal width at a selected time & Exact only for zero terminal width; may coexist with only asymptotic state convergence \citep{Yao2026FlexibleManipulatorsPrescribedPerformance} & \citealp{Stihi2026FTPPFTVSMC,Xu2026NeuroadaptiveFixedTimePrescribedPerformance,Lin2026FixedTimePPCMMCPS,Lv2026PredefinedLFCFOSMC,Luo2026HiTLUAVUSV,Xu2026AdaptivePrescribedPerformanceMAS,Zhu2026OptimizedSlidingModePsTFormation,An2026PsTPerformancePlatoon}\\
Finite- / fixed-time ISS; semi-global practical finite- / fixed-time stability & Temporal guarantee for a residual set scaled by the input & Settling under zero input is distinct from the accuracy obtained under a disturbance & \citealp{Cheah2026FTISSFXISSFormation,xi2023practical,poveda2021fixed}\\
Stochastic finite-, fixed-, predefined-time & Guarantee in probability or in expectation & $\mathbb E[\tau]\le T_c$ does not mean that every sample path settles by $T_c$ & \citealp{yin2011finite,Peng2026NoiseControlFiniteTime,Liu2026DynamicMemoryPT}\\
\end{xltabular}
}

The notions of this section form the first layer of Fig.~\ref{fig:three_layers}: they say what temporal property is required, but not how it is generated or realized. Section~\ref{sec:mechanisms} takes up the mechanisms, and Section~\ref{sec:architectures} their realization within controller architectures.

\FloatBarrier
\section{Convergence-Shaping Design Mechanisms}
\label{sec:mechanisms}

\begin{figure*}[pos=!t]
\centering
\reviewfigure{\textwidth}{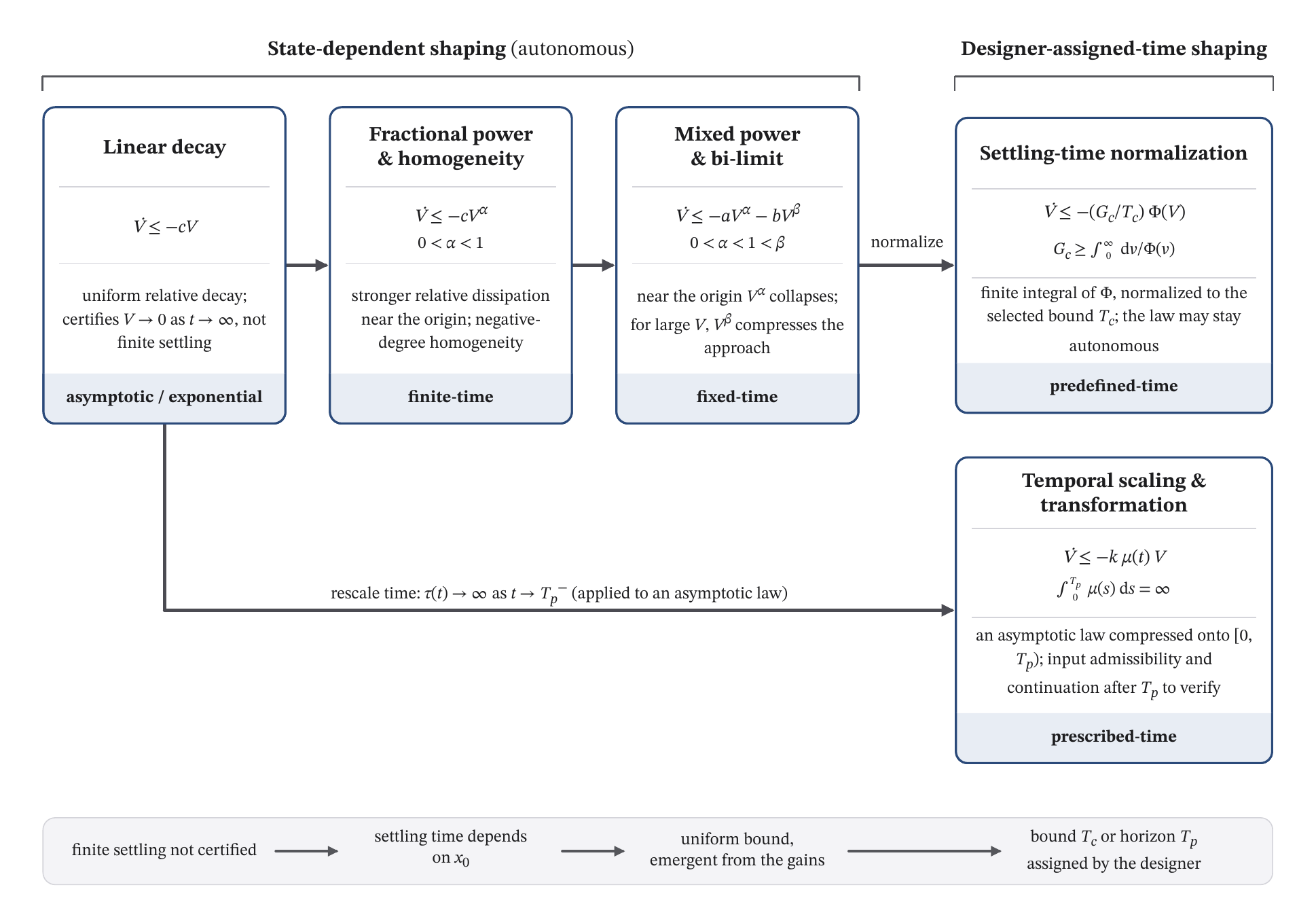}
\caption{Representative convergence-shaping mechanisms.}
\label{fig:mechanism_map}
\end{figure*}

Having established the convergence specifications in Section~\ref{sec:notions}, this section turns to the second layer of Fig.~\ref{fig:three_layers}: the mathematical mechanisms through which the corresponding temporal properties are generated. The objective is not to associate a stability notion with a specific controller architecture, but to identify the modifications of the closed-loop dissipation that distinguish the notions. From a Lyapunov perspective, these modifications concern the structure of the decay imposed on a positive-definite function $V$. The discussion proceeds from state-dependent and multi-region dissipation to settling-time normalization and explicitly time-dependent constructions; the mechanisms are analytical routes rather than mutually exclusive control classes. Figure~\ref{fig:mechanism_map} summarizes them. Settling-time normalization requires a dissipation profile with a finite integral and may keep the law autonomous; temporal scaling and transformation compress an asymptotic law onto $[0,T_p)$. Both are prominent, nonexclusive routes to designer-assigned-time guarantees. The linear-decay panel of the figure is an exponential Lyapunov baseline, and its annotations concern relative decay, not a universal ranking of actual convergence speed.

\subsection{From Convergence Specifications to Decay-Law Design}
\label{subsec:specification_to_design}

A convenient way to connect a temporal specification with feedback synthesis is to express the desired closed-loop behavior through a target Lyapunov decay law. Consider the control-affine nonlinear system
\begin{equation}
    \dot{x}
    =
    f(x,t)+g(x,t)u+d(x,t),
    \label{eq:general_control_affine_sec3}
\end{equation}
where $x\in\mathbb{R}^{n}$ is the state, $u\in\mathbb{R}^{m}$ the control input, and $d(x,t)$ the uncertainty or external disturbance when present. For a positive-definite Lyapunov or Lyapunov-like function $V(x,t)$, the decay-law design problem reads
\begin{equation}
    \dot V(x,t)
    \leq
    -\Psi(V,t)+\Delta(x,t),
    \label{eq:general_decay_sec3}
\end{equation}
where $\Psi(V,t)$ is the intended dissipative action and $\Delta(x,t)$ collects uncompensated uncertainty, approximation errors, sampling effects, and other residual terms.

The temporal behavior of the closed loop is determined primarily by the structure imposed on $\Psi$. For classical exponential stabilization the reference case is \citep{slotine1991applied}
\begin{equation}
    \Psi(V,t)=cV,
    \qquad
    c>0,
    \label{eq:sec3_exp_decay}
\end{equation}
which gives $\dot V\leq-cV$ and therefore
\begin{equation}
    V(t)
    \leq
    V(0)e^{-ct}.
    \label{eq:sec3_exp_solution}
\end{equation}
This linear-in-$V$ dissipation is the baseline: $V$ approaches zero exponentially but does not generally reach zero at a finite time. Stronger temporal requirements are obtained not by increasing $c$ but by changing the functional structure of the dissipation. In general,
\begin{equation}
    \Psi(V,t)
    =
    \Psi_{\mathrm{s}}(V)
    +
    \Psi_{\mathrm{t}}(V,t),
    \label{eq:sec3_general_shaping}
\end{equation}
where $\Psi_{\mathrm{s}}$ is state-dependent shaping, which modifies how the convergence rate varies across the state space, and $\Psi_{\mathrm{t}}$ is any explicitly time-dependent contribution, which reshapes the temporal evolution directly. The mechanisms below are different ways of constructing these two components.

Equation~\eqref{eq:general_decay_sec3} also separates nominal convergence shaping from robustness, and this separation is stated here once for all mechanisms of this section. When $\Delta=0$, the temporal property follows from the decay law itself. When $\Delta\neq0$ and the stabilizing term vanishes at the target while a nonvanishing residual remains, the inequality alone generally supports a residual-set conclusion rather than exact convergence, unless the residual is cancelled, dominated, or incorporated explicitly into the analysis. For an autonomous increasing decay profile and a constant residual bound, the corresponding set is $\{\Psi(V)\le\Delta\}$. Two sets must then be kept apart. The trajectory \emph{approaches} the boundary set $\{\Psi=\Delta\}$, but in general reaches it only asymptotically, since the comparison equation $\dot v=-\Psi(v)+\Delta$ has an equilibrium there (for $\dot V=-\sqrt V+1$ and $V(0)>1$, $V$ tends to $1$ without reaching it). Finite-time entry is certified for any strictly larger set: retaining a fraction $\sigma\in(0,1)$ of the dissipation, the inequality $\dot V\le-\sigma\Psi(V,t)$ holds outside $\{(1-\sigma)\Psi(V,t)\le\Delta\}$, and the entrance time into that set is bounded by the nominal settling-time integral evaluated with $\sigma\Psi$ in place of $\Psi$ and the target level as lower limit. When $\Psi$ depends on $t$, forward invariance of the target set must be established separately. The catalogue tables record, for each certificate, the residual-affected form and its $\sigma$-margin entrance-time bound; in the tables, coefficients are positive unless specified otherwise, $d_V$ denotes a nonnegative additive residual bound, and the displayed formulas use common notation and sufficient comparison bounds that do not replace the model-specific hypotheses of the cited results.

For a time-dependent certificate $V(x,t)$, decay of $V$ implies convergence of the state only when the comparison with the state is uniform over the relevant time interval. In particular, a time-independent class-$\mathcal K$ function $\alpha_1$ must satisfy $\alpha_1(|x|)\leq V(x,t)$; stability estimates additionally require an appropriate uniform upper bound $V(x,t)\leq\alpha_2(|x|)$. For global statements, the comparison functions are taken in class $\mathcal K_\infty$. A weight that vanishes as the terminal time approaches can make $V(x,t)\to0$ without forcing $x(t)\to0$. These conditions, together with existence and continuation, are part of the certificate rather than consequences of its scalar differential inequality.

\subsection{State-Dependent Shaping: Fractional Powers and Homogeneity}
\label{subsec:homogeneous_mechanisms}

Finite-time convergence can be generated by modifying the local scaling of the closed-loop dissipation near the equilibrium. Rather than a locally linear decay, finite-time designs employ nonlinear terms whose relative strength increases as $V$ approaches zero. A representative condition is \citep{bhat2000finite}
\begin{equation}
    \Psi(V,t)=cV^{\alpha},
    \qquad
    0<\alpha<1.
    \label{eq:sec3_ft_decay}
\end{equation}
Its finite-time character follows from
\begin{equation}
    \int_{0}^{V_0}
    \frac{d\xi}{c\xi^{\alpha}}
    =
    \frac{V_0^{1-\alpha}}
         {c(1-\alpha)}
    <\infty,
    \qquad
    0<\alpha<1,
    \label{eq:sec32_integral_ft}
\end{equation}
whereas exponential decay gives $\int_{0}^{V_0}d\xi/(c\xi)=\infty$. Finite settling thus results from a change in the state-dependent scaling of the dissipation rather than from increasing its coefficient. At the feedback level this mechanism appears through signed fractional-power nonlinearities: with $\lceil z\rceil^{\rho}=|z|^{\rho}\operatorname{sgn}(z)$ \citep{bhat2000finite}, a representative scalar term is
\begin{equation}
    u
    =
    -k\lceil e\rceil^{\rho},
    \qquad
    0<\rho<1.
    \label{eq:sec32_fractional_feedback}
\end{equation}
Compared with linear damping $u=-ke$, the ratio $|e|^{\rho}/|e|=|e|^{\rho-1}$ increases as $e\rightarrow0$, so the fractional-power term retains a stronger relative action close to the equilibrium and prevents the progressive slowing of linear asymptotic decay. The mechanism appears in homogeneous feedback \citep{polyakov2018homogeneous}, finite-time observers \citep{razmjooei2023design}, backstepping constructions \citep{xue2022adaptive}, and sliding-mode reaching laws \citep{hua2026finite}; it is independent of the architecture used to realize it.

A systematic interpretation of this local scaling is provided by homogeneity. Consider the weighted dilation \citep{bhat2005geometric}
\begin{equation}
    \Delta_{\lambda}^{r}x
    =
    \left(
        \lambda^{r_1}x_1,
        \ldots,
        \lambda^{r_n}x_n
    \right)^{T},
    \qquad
    \lambda>0,
    \label{eq:weighted_dilation}
\end{equation}
with dilation weights $r_i>0$. A vector field $f$ is homogeneous of degree $q$ with respect to this dilation when
\begin{equation}
    f_i(\Delta_{\lambda}^{r}x)
    =
    \lambda^{q+r_i}f_i(x).
    \label{eq:homogeneous_vector_field}
\end{equation}
For an asymptotically stable homogeneous system, a negative homogeneity degree provides a fundamental route to finite-time convergence: the characteristic time scale of the dynamics contracts as the state is scaled toward the equilibrium, so the trajectory reaches the origin in finite time. Homogeneity is a structural design principle rather than a controller-specific construction; the same scaling argument appears in direct nonlinear feedback \citep{liu2019direct}, terminal manifolds \citep{liu2022finite}, homogeneous observers \citep{xiang2025observer}, recursive virtual controls \citep{wang2022robust}, and disturbance-estimation dynamics \citep{yang2025finite}. It also clarifies a regularity issue: for continuous autonomous systems, exact finite-time convergence is generally associated with a loss of local Lipschitz regularity at the equilibrium \citep{shen2009uniformly}, so fractional-power and related nonsmooth terms are structural components of the finite-time mechanism rather than high-gain corrections.

The exponent $\alpha$ also shapes the local transient. Under \eqref{eq:sec3_ft_decay}, the settling-time estimate is \citep{bhat2000finite}
\begin{equation}
    T(V_0)
    \leq
    \frac{V_0^{1-\alpha}}
         {c(1-\alpha)}.
    \label{eq:sec32_settling_dependence}
\end{equation}
Reducing $\alpha$ strengthens the relative decay near the origin, while $c$ scales the overall dissipation; but the bound remains dependent on $V_0$. Fractional powers and negative-degree homogeneity therefore solve the problem of infinite settling time, not the problem of uniformity with respect to the initial condition. With an uncompensated residual, $\dot V\leq-cV^{\alpha}+\delta$ \citep{zhu2011attitude,wang2017adaptive,li2019finite}, the qualification of Section~\ref{subsec:specification_to_design} applies: $V$ is ultimately bounded by the level $(\delta/c)^{1/\alpha}$ and enters $\{V\le(\delta/[(1-\sigma)c])^{1/\alpha}\}$ in finite time; this is the situation in neural \citep{li2019adaptive_2}, fuzzy \citep{wang2017finite}, and observer-based \citep{li2020observer} implementations, where approximation or estimation errors persist. The mechanism also carries an implementation trade-off: fractional powers below unity have steep local slopes, sign-type realizations amplify measurement noise, and smooth regularizations improve practical behavior at the price of replacing exact finite-time convergence by convergence to a neighborhood.

In summary, state-dependent nonlinear dissipation and negative-degree homogeneity provide the principal route from infinite-time to finite-time convergence. Their defining feature is stronger relative dissipation near the equilibrium without explicit dependence on physical time; their central limitation is that the settling time still depends on the initial state. Removing this dependence requires a complementary mechanism acting on large-state trajectories, which leads to mixed-power and multi-region decay. Table~\ref{tab:finite_catalogue} collects the certificates of this class.

\noindent State-dependent finite-time and practical convergence certificates. Here $V_0=V(x_0)$, coefficients are positive, and inequalities apply before entry into the target set. Bounds are sufficient, not necessarily exact.

{
\small
\setlength{\tabcolsep}{4pt}
\renewcommand{\arraystretch}{1.22}
\begin{longtable}{@{}>{\raggedright\arraybackslash}p{0.34\textwidth} >{\raggedright\arraybackslash}p{\dimexpr0.47\textwidth-16pt\relax} >{\raggedright\arraybackslash}p{0.19\textwidth}@{}}
\caption{Finite-time and practical convergence certificates.}\label{tab:finite_catalogue}\\
\toprule
\textbf{Representative structure} & \textbf{Conditions and temporal interpretation} & \textbf{References}\\
\midrule
\endfirsthead
\multicolumn{3}{@{}l}{\footnotesize\textit{Table~\ref{tab:finite_catalogue} (continued)}}\\
\toprule
\textbf{Representative structure} & \textbf{Conditions and temporal interpretation} & \textbf{References}\\
\midrule
\endhead
\midrule\multicolumn{3}{r@{}}{\footnotesize\textit{continued on next page}}\\
\endfoot
\bottomrule\endlastfoot

$\dot V\leq-cV^\alpha$ & $0<\alpha<1$; $T\leq V_0^{1-\alpha}/[c(1-\alpha)]$. The stochastic counterpart of \citet{yin2011finite} replaces $\dot V$ by the infinitesimal generator $\mathcal LV$ and bounds the expected settling time. & \citealp{bhat2000finite,huang2005global,hong2006adaptive,xiao2009finite,yin2011finite}\\
$\dot V\leq-aV^\beta+d_V$ & $0<\beta<1$, $0<\sigma<1$. For $r=(d_V/[(1-\sigma)a])^{1/\beta}$, entry into $V\leq r$ occurs within $\dfrac{[V_0^{1-\beta}-r^{1-\beta}]_+}{\sigma a(1-\beta)}$. & \citealp{li2019adaptive,xia2019finite}\\
$\dot V\leq-k_1V^\mu-k_2V^\nu$ & $0<\mu<1<\nu$. The bound $V_0^{1-\mu}/[k_1(1-\mu)]$ is valid but ignores the uniform fixed-time certificate; see Table~\ref{tab:fixed_catalogue}. & \citealp{yang2011nonsingular,khoo2014multi,sun2017fast,sun2019fast}\\
$\dot V\leq-aV-bV^p+d_V$ & $0<p<1$. For $0<\sigma<1$ and $r=(d_V/[(1-\sigma)b])^{1/p}$, a sufficient entry bound is $\int_r^{V_0}dv/(av+\sigma bv^p)$ when $V_0>r$; it is zero otherwise. & \citealp{yu2018finite,li2019finite}\\
$\dot V\leq-\beta V^\delta$ during control, $\dot V\leq-\vartheta V$ during rest & $0<\delta<1$, $\vartheta\geq0$. A sufficient finite-prefix condition is $A(t)\geq\Lambda t-b$, where $A(t)$ is accumulated control time, $\Lambda>0$, $b\geq0$. Then $T\leq\bigl[V_0^{1-\delta}/(\beta(1-\delta))+b\bigr]/\Lambda$. An asymptotic time fraction alone does not give this finite-horizon bound. Residual variants require a larger invariant target set and strictly positive retained decay. & \citealp{Wang2026MemristiveFiniteTimeQuasiSync}\\
$\mathcal LV\leq K(V)$, $K'(V)|V_xg|^2\geq2(c(t)+1)K^2(V)$ & $K>0$, $K'\geq0$, $F(s)=\int_0^s dr/K(r)<\infty$, $\alpha(t)=\int_0^tc$: then $\mathbb E\,\alpha(T_0)\leq\mathbb E\,F(V_0)$, with $\mathcal LV$ allowed to be positive because the diffusion supplies the decay. $F(\infty)\leq C$ and convex $\alpha$ give $\mathbb ET_0\leq\alpha^{-1}(C)$, a fixed-time and, if $\alpha^{-1}(C)\leq T$, a predefined bound in expectation. & \citealp{Peng2026NoiseControlFiniteTime}\\
$\sup_u\dot H\geq-\eta H-\zeta H^{1/p}$ (barrier, $H<0$ outside the safe set) & $p>1$ an odd ratio; an unsafe state enters $\{H\geq0\}$ within $\frac{p}{\eta(p-1)}\ln\frac{\eta|\bar H_0|^{1-1/p}+\zeta}{\zeta}$ ($\bar H$ includes the NN-weight correction) and the set is forward invariant; $B_i=\dot B_{i-1}+\lambda_iB_{i-1}+\xi_iB_{i-1}^{1/p}$ extends the condition to relative degree $\kappa$. & \citealp{Wang2026HFNCBFRoboticSafeTracking}\\
\end{longtable}
}

Positive residuals distinguish practical convergence from exact settling; the sets in the table are the $\sigma$-margin sets of Section~\ref{subsec:specification_to_design}, whose entry is finite-time, not the boundary sets, whose approach is asymptotic. The target set and its forward invariance must be included in the certificate. Rows containing a constant negative term, in this table and in Tables~\ref{tab:fixed_catalogue} and~\ref{tab:assigned_catalogue}, cannot hold at the origin for a $C^1$ positive-definite $V$ and locally bounded dynamics; they are to be read on the complement of a neighborhood of the target or under the nonsmooth or set-valued derivative stated in the cited result. Intermittent and barrier versions certify, respectively, settling under aperiodic control-rest intervals \citep{Wang2026MemristiveFiniteTimeQuasiSync} and finite-time entry into a safe set rather than into the origin \citep{Wang2026HFNCBFRoboticSafeTracking}. Finite-horizon boundedness estimates of the form $D^+E[V]\leq-\lambda E[V]+\varphi(t)$ \citep{cao2020finite,min2021adaptive}, including recent finite-time stabilization of networked control systems and finite-time bounded consensus \citep{Liu2026MultiLoopNCSFiniteTime,Chen2026FTBCMultiAgentAttacks}, belong to the short-time stability perspective of Remark~\ref{rem:two_fts}; they are not exact settling-time certificates. Likewise, a weighted inequality of the displayed form $\dot V\leq e^{-\lambda t}(a_0V^p+b_0)$ does not establish finite settling from its form alone: signs, parameter restrictions, and the target set in the underlying result must be specified \citep{li2022event}.

\subsection{Multi-Region Shaping: Mixed Powers and Bi-Limit Homogeneity}
\label{subsec:mixed_power_mechanisms}

The initial-condition dependence of a single sublinear decay law can be removed by shaping the dissipation differently in different regions of the state space. The canonical realization combines a sublinear term governing the dynamics near the equilibrium with a superlinear term that accelerates trajectories at large Lyapunov levels \citep{polyakov2011nonlinear}:
\begin{equation}
    \dot V
    \leq
    -aV^{\alpha}
    -
    bV^{\beta},
    \qquad
    a,b>0,
    \qquad
    0<\alpha<1<\beta.
    \label{eq:sec33_dual_power}
\end{equation}
Write $\Phi(V)=aV^{\alpha}+bV^{\beta}$ for the dissipation on the right-hand side. For $V\rightarrow0^{+}$, $\Phi(V)\sim aV^{\alpha}$, so the sublinear component preserves the finite-time collapse of Section~\ref{subsec:homogeneous_mechanisms}; for $V\rightarrow\infty$, $\Phi(V)\sim bV^{\beta}$, so the superlinear term compresses trajectories with large initial energy and prevents the time to enter an intermediate region from growing without bound. The mechanism may be summarized as
\begin{equation*}
    \underbrace{\text{large-state compression}}
                _{V^{\beta},\,\beta>1}
    \;+\;
    \underbrace{\text{near-origin finite-time collapse}}
                _{V^{\alpha},\,0<\alpha<1}.
\end{equation*}
A standard decomposition of the settling-time integral yields
\begin{equation}
\begin{aligned}
    T(V_0)
    &\leq
    \int_{0}^{\min\{1,V_0\}}
    \frac{dV}{aV^{\alpha}}
    +
    \int_{1}^{\max\{1,V_0\}}
    \frac{dV}{bV^{\beta}}
    \\
    &\leq
    \frac{1}{a(1-\alpha)}
    +
    \frac{1}{b(\beta-1)},
\end{aligned}
\label{eq:sec33_uniform_bound}
\end{equation}
which is independent of the initial condition. The essential innovation is not the presence of a second nonlinear term but the use of complementary scaling regimes to control both local and global convergence.

This interpretation is closely related to homogeneity in the bi-limit \citep{andrieu2008homogeneous}. Negative-degree homogeneity near the origin provides the local mechanism for finite settling \citep{bhat2005geometric}; complementing it with a homogeneous approximation at infinity of positive degree accelerates large-state trajectories and prevents the settling time from increasing without bound. Under the corresponding stability conditions, the combination provides a systematic structural route to fixed-time stability, of which \eqref{eq:sec33_dual_power} is the simplest instance. Bi-limit homogeneity should be viewed as a broad framework underlying many fixed-time designs rather than as a necessary characterization of fixed-time stability.

More general formulations add an intermediate decay component, for example \citep{barman2025novel}
\begin{equation}
    \dot V
    \leq
    -aV^{\alpha}
    -
    cV
    -
    bV^{\beta},
    \qquad
    0<\alpha<1<\beta,
    \label{eq:sec33_three_region}
\end{equation}
a three-region mechanism in which the fractional-power term governs the near-origin dynamics, the linear term supplies exponential-type damping over an intermediate range, and the superlinear term dominates at large $V$. Such composite structures provide additional freedom for transient shaping while retaining a uniform convergence-time bound. The multi-region interpretation is useful when comparing apparently different fixed-time formulations: although published decay laws differ in algebraic form \citep{kong2020new,barman2025novel,zhang2022novel, hu2020fixed,sun2020fixed,polyakov2016robust}, one component prevents slow convergence near the equilibrium while another prevents the settling time from growing for large initial states. The principle extends beyond explicit sums of powers: any composite $\Phi$ with $\Phi(V)\sim V^{\alpha}$ as $V\rightarrow0$ and $\Phi(V)\sim V^{\beta}$ as $V\rightarrow\infty$, $0<\alpha<1<\beta$, realizes it, which is why power-law \citep{wei2022fixed}, exponential \citep{wang2026fixed}, saturation-like \citep{zheng2024saturation}, state-dependent-exponent \citep{Dong2026DynamicPowersFixedTimeSpacecraft,Pang2026NeuralFixedTimeSteerByWire,Trinh2026ALRNNFixedTimePMSM,Shen2026VECFixedTimeDCDC}, and other nonlinear functions yield closely related fixed-time mechanisms.

An important distinction remains between uniform boundedness and direct time assignability. For \eqref{eq:sec33_dual_power}, the bound \eqref{eq:sec33_uniform_bound} is independent of $V(0)$ but is determined jointly by the gains and exponents,
\begin{equation}
    T_{\max}
    =
    \mathcal{F}
    \left(
        a,b,\alpha,\beta,\ldots
    \right),
    \label{eq:sec33_implicit_time}
\end{equation}
and can be inverted or normalized to meet a selected bound when the gain dependence permits it. Its direct expression may nevertheless be conservative relative to the actual settling time \citep{jimenez2017optimal}. The same parameters that reduce the nominal bound increase the required dissipation: increasing $a$ or $b$ decreases \eqref{eq:sec33_uniform_bound}, whereas increasing the superlinear contribution produces rapidly growing feedback demands at large state magnitudes, which matters when the actuator is bounded \citep{jiang2016fixed}. With a residual, $\dot V\leq-aV^{\alpha}-bV^{\beta}+\delta$ \citep{ni2016fixed, zheng2018adaptive, wang2020fixed}, the qualification of Section~\ref{subsec:specification_to_design} applies: the trajectory approaches $\{aV^{\alpha}+bV^{\beta}=\delta\}$ and enters $\{(1-\sigma)(aV^{\alpha}+bV^{\beta})\le\delta\}$ in finite time; the nominal mixed-power form alone does not establish exact fixed-time convergence for an uncertain closed loop.

Overall, mixed-power and multi-region decay make finite-time convergence uniform with respect to the initial condition. The resulting bound depends on the selected gains and nonlinearities and may be conservative; fixed-time stability itself does not preclude calibrating that bound to a desired value. This motivates the next transition: instead of accepting the convergence-time bound as an emergent quantity, the desired bound or horizon is introduced directly into the design. Table~\ref{tab:fixed_catalogue} collects the certificates of this class.

\noindent Mixed-power and composite fixed-time certificates. Residual-free formulas concern exact convergence under the required solution and invariance assumptions; residual-affected formulas concern a stated target set.

{
\small
\setlength{\tabcolsep}{4pt}
\renewcommand{\arraystretch}{1.22}
\begin{longtable}{@{}>{\raggedright\arraybackslash}p{0.33\textwidth} >{\raggedright\arraybackslash}p{\dimexpr0.44\textwidth-16pt\relax} >{\raggedright\arraybackslash}p{0.23\textwidth}@{}}
\caption{Mixed-power and composite fixed-time certificates.}\label{tab:fixed_catalogue}\\
\toprule
\textbf{Representative structure} & \textbf{Conditions and temporal interpretation} & \textbf{References}\\
\midrule
\endfirsthead
\multicolumn{3}{@{}l}{\footnotesize\textit{Table~\ref{tab:fixed_catalogue} (continued)}}\\
\toprule
\textbf{Representative structure} & \textbf{Conditions and temporal interpretation} & \textbf{References}\\
\midrule
\endhead
\midrule\multicolumn{3}{r@{}}{\footnotesize\textit{continued on next page}}\\
\endfoot
\bottomrule\endlastfoot

$\dot V\leq-aV^\alpha-bV^\beta$ & $0<\alpha<1<\beta$; $T\leq1/[a(1-\alpha)]+1/[b(\beta-1)]$. & \citealp{polyakov2011nonlinear,polyakov2015finite,corradini2018nonsingular,ji2018finite,liu2019fixed,hu2017fixed,li2018fixed,wang2019disturbance,zuo2016distributed,ni2016fast,hua2016finite,yang2017fixed,polyakov2016robust,parsegov2013fixed}\\
$\dot V\leq-c_1V^p-c_2V^q+d_V$ & $0<p<1<q$, $0<\sigma<1$. Outside $(1-\sigma)(c_1V^p+c_2V^q)\leq d_V$, entry is bounded by $T\leq\frac{1}{\sigma c_1(1-p)}+\frac{1}{\sigma c_2(q-1)}$. & \citealp{ni2016fixed,zheng2018adaptive,wang2020fixed,ba2019fixed,li2021appointed,liu2022distributed,sun2020fixed,zhang2020fuzzy,zhang2021observer}\\
$\dot V\leq-aV^\xi-bV^\eta-cV$ & $0<\xi<1<\eta$. A sufficient bound is $\begin{aligned}T\leq{}&\frac{\ln(1+c/a)}{c(1-\xi)}\\&+\frac{\ln(1+c/b)}{c(\eta-1)}.\end{aligned}$ & \citealp{chen2020new,hu2020fixed}\\
$\begin{aligned}\dot V\leq{}&-aV\\&-(b_1V^{d_1}+b_2V^{d_2})^c.\end{aligned}$ & $c>0$, $0<d_2c<1<d_1c$. A sufficient bound is $\begin{aligned}T\leq{}&\frac{\ln(1+a/b_2^c)}{a(1-d_2c)}\\&+\frac{\ln(1+a/b_1^c)}{a(d_1c-1)}.\end{aligned}$ & \citealp{zhang2022novel}\\
$\dot V\leq-\Phi(V)+d_V$; $\Phi(v)=k_1v+(k_2v^p+k_3v^q)^l$ & $l>0$, $0<pl<1<ql$. For $(1-\sigma)\Phi(r)\geq d_V$, $0<\sigma<1$, entry into $V\leq r$ has uniform bound $\sigma^{-1}\int_r^\infty dv/\Phi(v)$. & \citealp{barman2025novel}\\
$\dot V\leq-\Phi(V)+d_V$; $\Phi(v)=(\mu_1v^\alpha+\mu_2v^\beta)^\chi$ & $0<\alpha\chi<1<\beta\chi$. For $0<\sigma<1$ and $(1-\sigma)\Phi(r)\geq d_V$, a uniform entry bound is $\sigma^{-1}\int_r^\infty dv/\Phi(v)$. & \citealp{wang2015decentralized,wang2021event,cao2020robust}\\
$\begin{aligned}\dot V\leq{}&\rho(t)V-aV^\delta\\&-bV^\theta-c.\end{aligned}$ & $\rho(t)\leq0$, $0\leq\theta<1<\delta$. On $V>0$, a sufficient comparison bound is $\displaystyle\int_0^\infty\frac{dv}{av^\delta+bv^\theta+c}$. The inequality with $c>0$ is not imposed at the invariant origin. & \citealp{kong2020new}\\
$\begin{aligned}\dot V\leq{}&-M_1^\mu-M_2^\nu+c\\&+\textstyle\sum_m[g_mN(\chi_m)+1]\dot\chi_m\end{aligned}$ & $M_1\geq aV$, $M_2\geq bV$, $0<\mu<1<\nu$. The Nussbaum terms need a separate lemma with the stated assumptions on the unknown control directions, adaptive variables, and $N$. This displayed inequality alone does not yield the ordinary mixed-power settling bound; the practical target and uniform time must follow from the full closed-loop argument. & \citealp{Niu2026SecurityPracticalFixedTimeCPS}\\
$\dot V\leq-a_1V^m-a_2V^l-a_3V$ during control, $\dot V\leq0$ during attacks & $m>1$, $0<l<1$. Let $B$ be the sum of the two logarithmic mixed-power bounds and let active time satisfy $A(t)\geq(1-\rho)t-b$, $\rho<1$, $b\geq0$. Then $T\leq(B+b)/(1-\rho)$. A bound using only $1/(1-\rho)$ presupposes $b=0$ or an equivalent finite-prefix hypothesis. & \citealp{Yang2026FixedTimeMicrogridDoS}\\
$\dot y=-\frac{\kappa}{\nu(y)}\lceil y\rceil^{r(y)}$; $r=r_1$ for $|y|>1$, $r=r_2$ for $|y|<1$ & $r_1>1>r_2>0$, $\nu(y)=\alpha+(1-\alpha)e^{-\beta|y|^\gamma}\in(\alpha,1]$ shapes the rate; $T\leq\frac{\nu(1)}{\kappa(r_1-1)}+\frac{1}{\kappa(1-r_2)}$. A scalar generator, not a Lyapunov criterion. & \citealp{Pang2026NeuralFixedTimeSteerByWire,Trinh2026ALRNNFixedTimePMSM}\\
$\dot s=-k(1+\sigma)^n\lceil s\rceil^{\sigma}+e$, $\sigma(s)=\frac{2|s|}{|s|+\mu-1}$ & $\mu>2$, $|e|\leq\vartheta$, $(1+\sigma)^n\geq N_m$, $k>\vartheta e^{2/e}/N_m$; the exponent tends to $2$ far from and to $0$ near the origin, and two regional inequalities give $T\leq T_1+T_2$ independent of $s(0)$, after the observer transient. & \citealp{Shen2026VECFixedTimeDCDC}\\
\end{longtable}
}

\subsection{Designer-Assigned-Time Design Mechanisms}
\label{subsec:designer_assigned_mechanisms}

Designer-assigned-time control changes the role of the convergence horizon in the synthesis. In fixed-time designs, the uniform bound is obtained after the controller parameters have been chosen, $T_{\max} = \mathcal{F}(k_1,k_2,\alpha,\beta,\ldots)$: independent of the initial condition, but an emergent property of the chosen decay law. Designer-assigned-time approaches reverse this relationship,
\begin{equation*}
    T_{\max}
    =
    \mathcal{F}(\text{design parameters})
    \quad\longrightarrow\quad
    T_c\;\text{or}\;T_p
    =
    \text{explicit design specification},
\end{equation*}
through two branches that share the objective but emphasize different mechanisms (Remark~\ref{rem:no_hierarchy}): predefined-time design normalizes a state-dependent settling-time law \citep{jimenez2017optimal}, whereas the dominant prescribed-time branch considered here embeds the terminal horizon into explicitly time-dependent scaling or transformation \citep{song2017time}.

\subsubsection{Predefined-Time Design Through Settling-Time Normalization}
\label{subsubsec:time_normalized_design}

A natural route from fixed-time to predefined-time design is to normalize the dissipation rate with respect to a selected upper bound. A general structure is \citep{jimenez2017optimal}
\begin{equation}
    \dot V
    \leq
    -\frac{G_c}{T_c}\Phi(V),
    \label{eq:sec34_pdt_general}
\end{equation}
where $T_c>0$ is the designer-selected settling-time bound, $G_c>0$ a normalization constant, and $\Phi(V)>0$ for $V>0$. Integration gives
\begin{equation}
    T(V_0)
    \leq
    \frac{T_c}{G_c}
    \int_{0}^{V_0}
    \frac{d\xi}{\Phi(\xi)},
    \label{eq:sec34_pdt_integral}
\end{equation}
and if $\Phi$ and $G_c$ are selected such that
\begin{equation}
    \sup_{V_0\geq0}
    \frac{1}{G_c}
    \int_{0}^{V_0}
    \frac{d\xi}{\Phi(\xi)}
    \leq1,
    \label{eq:sec34_normalization}
\end{equation}
then $T(V_0)\leq T_c$ uniformly over the considered domain. To relate this convention to Section~\ref{subsec:predefined_time_stability}, let
\begin{equation}
I_\Phi:=\int_0^\infty\frac{dv}{\Phi(v)},\qquad
\widehat\Phi(v):=G_c\Phi(v).
\label{eq:normalization_bridge}
\end{equation}
Then $G_c\geq I_\Phi$ gives
\begin{equation}
\int_0^\infty\frac{dv}{\widehat\Phi(v)}
=\frac{I_\Phi}{G_c}\leq1,
\qquad
-\frac{G_c}{T_c}\Phi(V)=-\frac{1}{T_c}\widehat\Phi(V),
\label{eq:normalization_equivalence}
\end{equation}
so the two conventions are equivalent, with $G_c=1$ the case in which $\Phi$ is already normalized. Even if the scalar comparison integral is tight, a Lyapunov inequality may give a conservative estimate for the physical closed loop; assignability and tightness must be assessed separately.

The key mechanism is therefore not the isolated multiplication of a conventional controller by $1/T_c$: the complete nonlinear decay function must possess a finite state-independent integral, normalized so that this integral is compatible with the selected horizon. A mixed-power realization is \citep{jimenez2020lyapunov}
\begin{equation}
    \dot V
    \leq
    -\frac{G_c}{T_c}
    \left(
        aV^{\alpha}
        +
        bV^{\beta}
    \right)^{k},
    \label{eq:sec34_pdt_mixed}
\end{equation}
with the powers and normalization selected so that the settling-time integral satisfies the required bound; other nonlinear functions serve the same purpose provided their global integral properties are compatible \citep{wang2020attitude, chen2023predefined, zhang2022new, sun2024novel, cheng2024novel, zhang2025new, chen2022new, zhang2022practically}. Recent families give the comparison integral a perfect-square structure, $\dot V\leq-(\theta T_c)^{-1}(1+V^\theta)^2V^{1-\theta}$ \citep{Lai2026PredefinedCMNN,Ding2026GeneralizedPTUAV}, or add a linear term whose weight shapes the actual settling time below the assigned bound \citep{Liu2026AdaptivePTPPCQuadrotor}. With an additive residual, the related kernel $V^{1-\delta/2}(\sqrt\sigma+\sqrt\omega V^{\delta/2})^2$ certifies entry into the residual set only within $\sqrt2\,T$, so the time parameter must be rescaled, $T\le T_d/\sqrt2$, before it is reported as a deadline $T_d$ \citep{Liu2026ElectromagneticTakeover}. The relation with fixed-time design is
\begin{equation}
    \underbrace{
    -aV^{\alpha}-bV^{\beta}
    }_{\text{uniform but indirectly determined bound}}
    \quad\longrightarrow\quad
    \underbrace{
    -\dfrac{G_c}{T_c}\Phi(V)
    }_{\text{designer-selected upper bound}}:
    \label{eq:sec34_fxt_to_pdt}
\end{equation}
the multi-region mechanism may remain, but the overall dissipation is normalized so that the bound is referenced directly to $T_c$.

This explicit dependence reveals a feasibility trade-off. For a fixed Lyapunov level, a smaller assigned bound requires stronger nominal dissipation; under a finite control-authority constraint
\begin{equation}
    \|u\|\leq u_{\max},
    \label{eq:sec34_input_bound}
\end{equation}
there may exist states or requested values of $T_c$ for which the nominal predefined-time inequality cannot be enforced (Section~\ref{subsec:feasible_decay}). With a residual, $\dot V\leq-(G_c/T_c)\Phi(V)+\Delta$, the qualification of Section~\ref{subsec:specification_to_design} applies: the explicit presence of $T_c$ in a nominal decay inequality must not be confused with preservation of the guarantee under uncompensated perturbations, and the appropriate result is practical convergence to a residual set within a designer-related interval.

Overall, time normalization converts a uniform convergence-time bound from an implicit consequence of the decay parameters into an explicit synthesis variable, while the construction can remain autonomous. Table~\ref{tab:assigned_catalogue} collects the certificates of this class.

\noindent Representative designer-assigned-time decay families, grouped by their analytical structure. Original terminology is retained in the cited literature; the mechanism and the exact/practical guarantee must be identified separately.

{
\small
\setlength{\tabcolsep}{4pt}
\renewcommand{\arraystretch}{1.22}
\begin{longtable}{@{}>{\raggedright\arraybackslash}p{0.33\textwidth} >{\raggedright\arraybackslash}p{\dimexpr0.44\textwidth-16pt\relax} >{\raggedright\arraybackslash}p{0.23\textwidth}@{}}
\caption{Normalized designer-assigned-time certificates.}\label{tab:assigned_catalogue}\\
\toprule
\textbf{Representative structure} & \textbf{Conditions and temporal interpretation} & \textbf{References}\\
\midrule
\endfirsthead
\multicolumn{3}{@{}l}{\footnotesize\textit{Table~\ref{tab:assigned_catalogue} (continued)}}\\
\toprule
\textbf{Representative structure} & \textbf{Conditions and temporal interpretation} & \textbf{References}\\
\midrule
\endhead
\midrule\multicolumn{3}{r@{}}{\footnotesize\textit{continued on next page}}\\
\endfoot
\bottomrule\endlastfoot

$\dot V\leq-(G_c/T_c)\Phi(V)$; $\Phi(v)=(\alpha v^\delta+\beta v^\theta)^k$ & $k>0$, $0<k\theta<1<k\delta$; choose $G_c\geq\int_0^\infty dv/\Phi(v)$. & \citealp{jimenez2020lyapunov,ni2019predefined,munoz2019predefined}\\
$\dot V\leq-(G_c/T_c)(\alpha V^\delta+\beta V^\theta)+d_V$ & $0<\theta<1<\delta$. A positive residual requires a target-set and time certificate accounting for $d_V$; nominal normalization alone is insufficient. & \citealp{wang2022adaptive,xie2022adaptive,xie2021adaptive,xu2022event,sui2024command,wu2021predefined,munoz2021predefined}\\
$\dot V\leq-\dfrac{e^{V^p}V^{1-p}}{pT_c}$ & $0<p<1$; $T\leq T_c(1-e^{-V_0^p})\leq T_c$. & \citealp{wang2020attitude,jimenez2020lyapunov,pan2023improved,jimenez2017optimal}\\
$\dot V\leq-\dfrac{e^{V^p}V^{1-p}}{pT_c}+d_V$ & $0<p<1$, $d_V>0$; a residual-set certificate is needed. & \citealp{lv2023neural}\\
$\dot V\leq-(G_c/T_c)\Phi(V)$; $\begin{aligned}\Phi(v)={}&\alpha v^\delta+\beta v^\theta\\&+cv+\rho.\end{aligned}$ & $0<\theta<1<\delta$, $\rho\geq0$; normalize using $\int_0^\infty dv/\Phi(v)$. If $\rho>0$, apply the inequality only before reaching the target. & \citealp{liu2022new,kong2023fixed,zhang2022new,zhao2024new,wu2025new,yang2025chaotic,jia2023new,guo2023predefined,jia2023predefined}\\
$\begin{aligned}\dot V\leq{}&-\frac{G_c}{T_c}\Phi(V)+d_V,\\\Phi(v)={}&\alpha v^\delta+\beta v^\theta+cv.\end{aligned}$ & $d_V>0$; specify the residual target and its entrance-time bound. & \citealp{shihao2024neural,xie2023predefined}\\
$\dot V\leq-(G_c/T_c)(e^{-V}+e^V)$ & For $V>0$, $T\leq (T_c/G_c)\int_0^{V_0}dv/(e^{-v}+e^v)$. The integral over $[0,\infty)$ is $\pi/4$; continuation at zero is separate. & \citealp{zhang2025new}\\
$\dot V\leq-\dfrac{G_c}{T_c\ln a}a^VV^{1-\beta}$ & $a>1$, $\beta>0$ for integral convergence. The scalar comparison gives $T\leq T_c(\ln a)G_c^{-1}\int_0^{V_0}a^{-v}v^{\beta-1}dv$; normalize and check regularity at zero. & \citealp{chen2022new}\\
$\dot V\leq-(G_c/T_c)V^ae^V+d_V$ & $0<a<1$, $d_V>0$; practical convergence requires an explicit residual target and normalization appropriate to its entrance time. & \citealp{zhang2022practically}\\
$\dot V\leq-\dfrac{(1+V^\theta)^2V^{1-\theta}}{\theta T_c}\,(+\bar\Delta)$ & $0<\theta<1$; $T\leq T_c[1-(1+V_0^\theta)^{-1}]\leq T_c$; $\theta=(\alpha+1)/2$ gives $2V+V^{\frac{1-\alpha}{2}}+V^{\frac{\alpha+3}{2}}$. With a persistent $\bar\Delta>0$ and $0<\theta<\frac12$, the generalized characterization gives $T\leq T_c/(2^{1-\theta}-1)\leq(\sqrt2+1)T_c$ under its conditions on $\bar\Delta$. & \citealp{Lai2026PredefinedCMNN,Ding2026GeneralizedPTUAV}\\
$\begin{aligned}\dot V\leq{}&-\tfrac{\pi}{\eta\beta T_c\sqrt{ab}}\bigl(2\beta\sqrt{ab}V\\&+aV^{1-\eta/2}+\beta^2bV^{1+\eta/2}\bigr)\end{aligned}$ & $0<\eta<1$, $a,b,\beta>0$; $T\leq\frac{2T_c}{\pi}\arctan(\beta\sqrt{b/a}\,V_0^{\eta/2})\leq T_c$ (the linear term only strengthens the decay); $\beta$ shapes the actual settling time without changing the bound. & \citealp{Liu2026AdaptivePTPPCQuadrotor}\\
$\dot V\leq-\rho\lambda_c(1+V^2)$, $V=\|x\|^\rho$ & $T<\arctan(V_0)/(\rho\lambda_c)$; the gain $\lambda_c=\arctan(\|x_0\|^\rho)/(\rho t_c)$ meets $t_c$ but is calibrated from $x_0$, whereas $\lambda_c=\pi/(2\rho t_c)$ gives an $x_0$-independent bound. A normalized-sign activation makes all components arrive together. & \citealp{Yu2026ContinuumPTSS}\\
$\mathcal LV\leq-\frac{\pi}{\varpi T_c}(V^{1+\varpi/2}+V^{1-\varpi/2})+C$ & $0<\varpi<1$, $C>0$; stochastic practical counterpart in probability; residual and constant to be read from the cited proof. & \citealp{Liu2026DynamicMemoryPT}\\
\end{longtable}
}

\subsubsection{Prescribed-Time Design Through Temporal Scaling and Transformation}
\label{subsubsec:time_varying_pt}

Prescribed-time design introduces the selected terminal horizon directly into the temporal evolution of the closed loop \citep{song2017time,song2023prescribed}. A representative mechanism employs a positive scaling function $\mu(t)$ with
\begin{equation}
    \mu(t)\rightarrow\infty
    \qquad
    \text{as}
    \qquad
    t\rightarrow T_p^{-},
    \label{eq:sec34_mu_divergence}
\end{equation}
together with a Lyapunov inequality such as
\begin{equation}
    \dot V
    \leq
    -k\mu(t)V,
    \label{eq:sec34_pt_lyap}
\end{equation}
so that
\begin{equation}
    V(t)
    \leq
    V(0)
    \exp
    \left[
        -k
        \int_{0}^{t}\mu(\tau)d\tau
    \right],
    \label{eq:sec34_pt_integral}
\end{equation}
and, when $\int_{0}^{T_p}\mu(\tau)d\tau=\infty$ as in \eqref{eq:pt_integral}, the cumulative dissipation becomes unbounded as the horizon is approached, forcing $V(t)$ toward zero. A closely related realization is state scaling,
\begin{equation}
    z
    =
    \mu(t)x,
    \label{eq:sec34_state_scaling}
\end{equation}
for which boundedness of the transformed state implies
\begin{equation}
    x(t)
    =
    \frac{z(t)}{\mu(t)}
    \rightarrow0
    \qquad
    \text{as}
    \qquad
    t\rightarrow T_p^{-};
    \label{eq:sec34_state_scaling_result}
\end{equation}
the convergence problem is converted into a boundedness or stabilization problem in the scaled coordinates. A complementary mechanism is nonlinear time transformation: a monotonic map
\begin{equation}
    \tau=a(t),
    \qquad
    a(0)=0,
    \qquad
    a(t)\rightarrow\infty
    \quad
    \text{as}
    \quad
    t\rightarrow T_p^{-},
    \label{eq:sec34_time_transform}
\end{equation}
sends the finite physical-time interval $[0,T_p)$ onto the infinite transformed-time interval $[0,\infty)$, so that an asymptotic stabilization mechanism in the transformed coordinate is compressed into a finite physical-time horizon. From this viewpoint, prescribed-time control is temporal reshaping of an infinite-time convergence process. More elaborate formulations combine state \citep{su2025adaptive} or time transformations \citep{zhang2025adaptive_2}, dynamic high-gain scaling \citep{krishnamurthy2020dynamic}, parametric Lyapunov equations \citep{zhou2021prescribed}, time-dependent auxiliary dynamics \citep{tan2026flexible}, or recursive scaling \citep{wang2023prescribed}; their common characteristic is that the selected horizon enters the evolution of the closed loop explicitly rather than only through an autonomous state-dependent settling-time bound.

A rapidly increasing scaling function does not, however, imply that the physical input must diverge:
\begin{equation}
    \mu(t)\rightarrow\infty
    \quad\not\Rightarrow\quad
    \|u(t)\|\rightarrow\infty.
    \label{eq:sec34_false_implication}
\end{equation}
Increasing scaling terms may be multiplied by simultaneously vanishing states or compensated by the transformed dynamics. Conversely, plant-state convergence alone is not an admissibility certificate: boundedness of the actual input and of all internal signals must be established independently, and this is the form in which the issue recurs in every architecture of Section~\ref{sec:architectures}. Operation at and after the horizon is a second distinctive issue \citep{zuo2022adaptive}: a feedback expression containing factors such as $(T_p-t)^{-1}$ is defined only for $t<T_p$, so a complete construction must establish a continuation beyond the terminal time and preserve equilibrium invariance,
\begin{equation}
    x(t)=0,
    \qquad
    t\geq T_p.
    \label{eq:sec34_post_tp}
\end{equation}
Bounded-gain \citep{su2025prescribed,Zhou2026PTQuasiSynchronization,Kim2026EventImpulsivePrescribedTimeUUV,Cai2026MSVPracticalPsT,Golestani2026PracticalPrescribedTimePPC}, nonsingular \citep{shi2024prescribed}, switching \citep{ding2026switching}, regularized time-scaling \citep{Sun2026PredefinedUAVSpatiotemporal,Li2026TeleoperationPT,Li2026PsTClusterConsensusHMAS}, and practical prescribed-time formulations \citep{cao2022practical} have been developed partly to reduce the implementation difficulties associated with increasingly aggressive behavior near $T_p$. Most of them trade exact arrival for a residual; exactness with bounded gains has been retained by freezing the scaling after a certified arrival before $T_p$ \citep{Cui2026PrescribedTimeMorphingHMV} and by periodic delayed feedback inside a super-twisting algorithm \citep{Ding2026PDFSuperTwistingPsT}. The scaling certificate has also been extended to delayed \citep{Zhu2026PrescribedTimeDelayDoS} and DoS-interrupted \citep{Zhang2026PrescribedTimeConsensusDelayDoS} dynamics and, through a prescribed-time Nussbaum lemma, to unknown control directions \citep{Li2026EventTriggeredUnknownCoefficientsPsT}.

In summary, temporal scaling and transformation supply the horizon as an explicit time-dependent quantity, at the price of two verification obligations that autonomous normalization does not create in the same form: admissibility of the input and internal signals near $T_p$, and continuation at and after $T_p$. Table~\ref{tab:temporal_catalogue} collects the certificates of this class.

\noindent Temporal-scaling and auxiliary-dynamics mechanisms. A terminal-limit certificate must be distinguished from bounded physical input and a well-defined continuation after the selected horizon.

{\small
\setlength{\tabcolsep}{4pt}
\renewcommand{\arraystretch}{1.22}
\begin{longtable}{@{}>{\raggedright\arraybackslash}p{0.34\textwidth} >{\raggedright\arraybackslash}p{\dimexpr0.47\textwidth-16pt\relax} >{\raggedright\arraybackslash}p{0.19\textwidth}@{}}
\caption{Temporal-scaling and auxiliary-dynamics mechanisms.}\label{tab:temporal_catalogue}\\
\toprule
\textbf{Representative structure} & \textbf{Conditions and temporal interpretation} & \textbf{References}\\
\midrule
\endfirsthead
\multicolumn{3}{@{}l}{\footnotesize\textit{Table~\ref{tab:temporal_catalogue} (continued)}}\\ \toprule
\textbf{Representative structure} & \textbf{Conditions and temporal interpretation} & \textbf{References}\\
\midrule
\endhead
\midrule\multicolumn{3}{r@{}}{\footnotesize\textit{continued on next page}}\\ \endfoot
\bottomrule\endlastfoot
$z=\mu(t)x$, $\mu(t)=T_p/(T_p-t)$ & Bounded transformed state $z$ gives $x(t)\to0$ as $t\to T_p^-$. Input boundedness is a separate condition. & \citealp{song2017time,song2023prescribed}\\
$\dot V\leq-k\mu(t)V+|d(t)|$ & The integrating-factor estimate gives $V(t)\le e^{-k\int_0^t\mu}\bigl[V_0+\int_0^t e^{k\int_0^s\mu}|d(s)|ds\bigr]$; for $\mu=T_p/(T_p-t)$ and bounded $d$ the terminal limit $V(t)\to0$ survives (bounded-disturbance case of Section~\ref{subsec:numerical_constraints}), whereas retention after $T_p$ depends on the continuation. & \citealp{song2017time,song2023prescribed}\\
$\tau=a(t)\to\infty$ as $t\to T_p^-$; auxiliary high-gain state $r=r(\tau)$ & A finite physical horizon is mapped to an infinite transformed horizon; auxiliary-state and physical-input behavior must also be checked; truncating the map before $T_p$ gives practical entry followed by asymptotic convergence. & \citealp{song2023prescribed,krishnamurthy2020dynamic,Shao2026PsTSpacecraftSafety}\\
$u=-b^TP(\gamma(t))x$, $\gamma(t)\to\infty$ & Parametric-Lyapunov-equation gain generation; its closed-loop analysis must establish the stated terminal property. & \citealp{zhou2021prescribed}\\
$\dot V\leq-\beta_1(\alpha m/T_c)\omega(t)V+\beta_2\omega(t)^{\beta_3}$; $\omega(t)=T_c/(T_c-\alpha t)$ on $[0,T_c]$ & For $0<\alpha<1$, the gain remains finite at $T_c$. This displayed inequality alone is not an exact zero-error terminal certificate. & \citealp{chen2023predefined,Zhou2026PTQuasiSynchronization}\\
$\dot V\le-p_1\omega(t)V+p_2$, $\omega(t)=\dfrac{T^2}{(1-\iota)(T-t)^2+\iota T^2}$ on $[0,T)$, $1/\iota$ afterwards & Bounded scaling, $\omega\le1/\iota$; the residual at $T$ decreases with $\iota$. Bounded high-gain and performance-rate profiles behave alike. & \citealp{Kim2026EventImpulsivePrescribedTimeUUV,Zhang2026FNNPTESOUSV,Cai2026MSVPracticalPsT,Golestani2026PracticalPrescribedTimePPC}\\
$\dot x=-\dfrac{\dot\xi(t)}{1-\xi(t)+\delta}x$, $\xi$ a polynomial time-base generator, $\xi(0)=0$, $\xi(t_f)=1$ & $x(t_f)=\frac{\delta}{1+\delta}x_0$: the regularization $\delta>0$ bounds the gain and makes the arrival practical; coefficient conditions on $\xi$ limit the initial input. & \citealp{Sun2026PredefinedUAVSpatiotemporal}\\
$\dot V\le-\Bigl[a_0+\dfrac{a_1\dot\xi(t)}{1-\xi(t)+\delta}\Bigr]V+\Delta$, $\xi$ any admissible time-base generator & $a_0>0$, $a_1>1$, $0<\delta\ll1$: $V(t_0+t_b)<e^{-a_0t_b}\bigl(\frac{\delta}{1+\delta}\bigr)^{a_1}V_0+\Delta/a_0$; practical, and not tied to a particular $\xi$. & \citealp{Li2026PsTClusterConsensusHMAS}\\
$\dot V=-\alpha_1V+\dfrac{\alpha_2\dot\phi(t,T_f)}{\alpha_2\phi(t,T_f)+\varepsilon}V+\delta$ & $\phi$ decreasing from $1$ to $0$ on $[0,T_f]$; $V(T_f)\leq\frac{\varepsilon}{\alpha_2+\varepsilon}V_0e^{-\alpha_1T_f}+\frac{\delta}{\alpha_1}$. $\varepsilon=0$ gives exact arrival with a singular gain; $\varepsilon>0$ trades it for a residual. & \citealp{Li2026TeleoperationPT}\\
Time gain set to zero during attacks, $\varepsilon\bigl(\frac{\dot r(t^N)}{1-r(t^N)+\sigma_1}+\sigma_2\bigr)$ otherwise, $t^N=t-|\Pi_D(t_0,t)|$ & The clock advances in attack-free time. If every prefix satisfies $A(t)\geq\eta_N(t-t_0)-b$, $\eta_N>0$, then the wall-clock budget is at most $(t_f-t_0+b)/\eta_N$. The practical target and prefix constants must be specified. & \citealp{Zhang2026DCMicrogridPT}\\
$\dot V\le-\eta V-2\dfrac{\dot\mu_s}{\mu_s}V$ attack-free, $\dot V\le\zeta V$ under DoS & Piecewise estimates; with the horizon placed according to the attack pattern, $V\equiv0$ after $T$. The ratio $\dot\mu_s/\mu_s$ is unbounded. & \citealp{Zhang2026PrescribedTimeConsensusDelayDoS}\\
$\dot Q\leq-\varepsilon\dfrac{\dot\phi}{\phi}Q+\gamma Q(t-\tau)$, $\phi=\dfrac{T^h}{(T+t_s-t)^h}$, $h>2$ & A delay-dependent comparison must account for the initial history, for example $Q_{\rm hist}=\sup_{s\in[t_s-\tau,t_s]}Q(s)$, and compatibility of the scaling with the delayed term. The displayed inequality alone does not establish the terminal limit or invariance after $t_s+T$; those require the cited model's additional hypotheses. & \citealp{Zhu2026PrescribedTimeDelayDoS}\\
$\dot V\le-c\mu(t)\tilde V+[g(t)E(\nu)+a]\dot\nu+\beta$, $\dot\nu\ge0$, $\tilde V\le V$; $\mu=1/(t_f-t)$, then $\sigma$ & $E$ a Nussbaum function, $g$ of unknown sign: $\nu$, $V$ bounded and $\tilde V\to0$ as $t\to t_f^-$; with the switch, $x=0$ and $u=0$ after $t_f$. & \citealp{Li2026EventTriggeredUnknownCoefficientsPsT}\\
$\dot V\le-\dfrac{\mu(t)}{\alpha k_1k_2}V^{1-\alpha}\cosh^2(k_1V^{\alpha})$, $\mu=T/(T-t)$ frozen at $\bar T<T$ & $0<\alpha\le\frac12$: $T(x_0)\le T\bigl[1-e^{-(k_2/T)\tanh(k_1V_0^{\alpha})}\bigr]\le T(1-e^{-k_2/T})<T$; exact, bounded gain, uniform margin. Adding $\mu\Theta$ gives a residual set. & \citealp{Cui2026PrescribedTimeMorphingHMV}\\
$\dot x=-kx-K_{k,h}(t)x(t-h)$, $K_{k,h}$ $2h$-periodic, inside a super-twisting algorithm with $h=T/2$ & Bounded gains; $x\equiv0$ for $t\ge2h$ in the scalar case and exact settling at $T$ in closed loop. Measurement noise leaves a residual. & \citealp{Ding2026PDFSuperTwistingPsT}\\
$w_i=-\dfrac{N}{t_f-t}s_i$, nominal reference $r(t)\propto(t_f-t)^N$ & Nominal arrival exactly at $t_f$, not earlier; under disturbance, entry into the envelope at $t_f$. & \citealp{Shao2026TTRPrescribedPerformanceSMC}\\
$\dot V\leq-\dfrac{G_c(1-e^{-V})}{T_c-t}$ & The terminal horizon enters through explicit temporal scaling; behavior at and after $T_c$ must be specified. & \citealp{cheng2024novel,jiang2024novel}\\
$\dot V\leq-\Theta$, $\dot\Theta=\dfrac{\pi^2}{4T_c^2}V$ & An auxiliary dynamic variable shapes decay. Its initialization, admissible interval, and continuation belong to the certificate. & \citealp{sun2024novel}\\
\end{longtable}
}

\subsection{From Desired Decay to Feasible Control}
\label{subsec:feasible_decay}

\begin{figure*}[pos=!t]
\centering
\reviewfigure{\textwidth}{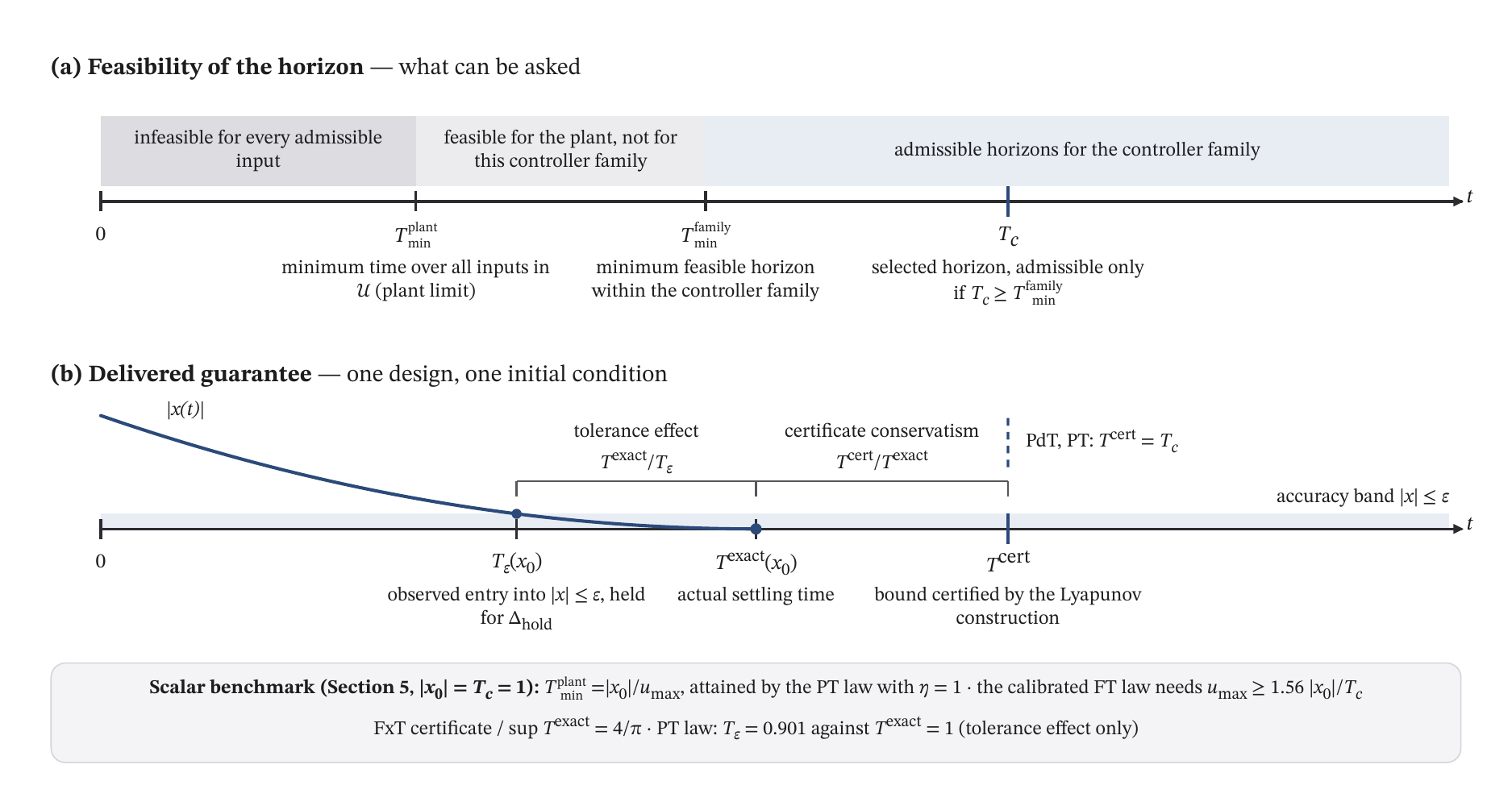}
\caption{Horizon feasibility versus the delivered guarantee.}
\label{fig:time_quantities}
\end{figure*}

A decay law has practical meaning only if it can be generated by an admissible physical input. Denoting the available control set by $u(t)\in\mathcal{U}$, the synthesis problem reads
\begin{equation}
    \exists\,u(t)\in\mathcal{U}
    \quad\text{such that}\quad
    \dot V(x,t)
    \leq
    -\Psi(V,t)+\Delta(x,t).
    \label{eq:admissible_decay_problem}
\end{equation}
This consideration is particularly important in designer-assigned-time control \citep{li2023adaptive,vazquez2024prescribed, yang2025adaptive,tan2026flexible}. Reducing the selected horizon requires the closed loop to dissipate $V$ more rapidly, while the achievable dissipation remains limited by the physical control channel. A sufficiently constrained system may therefore possess a minimum feasible convergence horizon
\begin{equation}
    T_{\min}
    =
    T_{\min}
    \left(
        \mathcal X_0,
        u_{\max},
        \dot u_{\max},
        d,
        \mathcal{X},
        \ldots
    \right),
    \label{eq:min_feasible_time}
\end{equation}
below which the temporal specification cannot be enforced over the initial set $\mathcal X_0$ without violating one or more admissibility conditions; whether the minimum is taken over open-loop inputs for each initial condition or over a single feedback law for the whole set is made precise in Section~\ref{subsec:open_temporal}. The systematic characterization of such limits remains substantially less developed than the nominal convergence theory \citep{song2023prescribed}. Three quantities must be kept apart: the minimum time achievable by the plant over all admissible inputs, the minimum feasible horizon within a selected controller family, and a sufficient horizon certified by a particular Lyapunov construction. A controller-family feasibility calculation does not establish the plant's minimum-time reachability limit. Figure~\ref{fig:time_quantities} separates the three quantities and relates them to the guarantee that a given design delivers. Its panel (a) orders the minimum time achievable by the plant, the minimum feasible horizon within a controller family, and the selected horizon; panel (b) shows, for one design and one initial condition, the observed time $T_\varepsilon$, the exact settling time, and the certified bound, with the decomposition \eqref{eq:settling_time_conservatism_index}, using the values of the benchmark of Section~\ref{sec:numerical}. This distinction connects the mechanism-level analysis to the architectural and actuator constraints considered next.

\section{Architectural Realizations, Implementation Constraints, and Design Guidance}
\label{sec:architectures}

The mechanisms of Section~\ref{sec:mechanisms} do not prescribe a unique controller structure. Fractional-power, mixed-power, time-normalized, and explicitly time-varying mechanisms can be embedded into substantially different feedback architectures, and a single controller may combine several of them \citep{zhang2012finite,hou2019finite,polyakov2018homogeneous, yu2023finite,moulay2022fixed,yu2017design,wang2020fixed, song2017time}. The emphasis of this section therefore shifts from the mathematical form of the decay law to the structural means by which the required dissipation is generated in the closed loop. This matters because practical realizations introduce additional states \citep{toth2011state}, estimation or adaptation dynamics \citep{wang2020parameter}, optimization variables \citep{parisio2014model}, switching mechanisms \citep{jia2025adaptive}, and implementation constraints \citep{zheng2019simple} that are not visible in the nominal Lyapunov inequality.

The realizations are organized into three groups. Primary feedback architectures (Section~\ref{subsec:primary_architectures}) determine how the available plant variables are converted into the physical input: direct state feedback, sliding-mode control, and backstepping. Augmented and information-structured architectures (Section~\ref{subsec:augmented_architectures}) add estimation, adaptation, learning, or a distributed information structure to a primary architecture. Constraint- and implementation-aware realizations (Section~\ref{subsec:implementation_architectures}) impose performance, safety, optimization, actuation, sampling, switching, or communication requirements. The classification is functional rather than mutually exclusive: a predefined-time backstepping design may include adaptive approximation \citep{pan2023improved}, prescribed-performance shaping \citep{zhang2026predefined}, and an event-triggered implementation \citep{pan2024event}, and a methodology may occupy more than one role depending on whether it generates the physical feedback or is superimposed on another controller.

\begin{figure*}[pos=!t]
\centering
\reviewfigure{\textwidth}{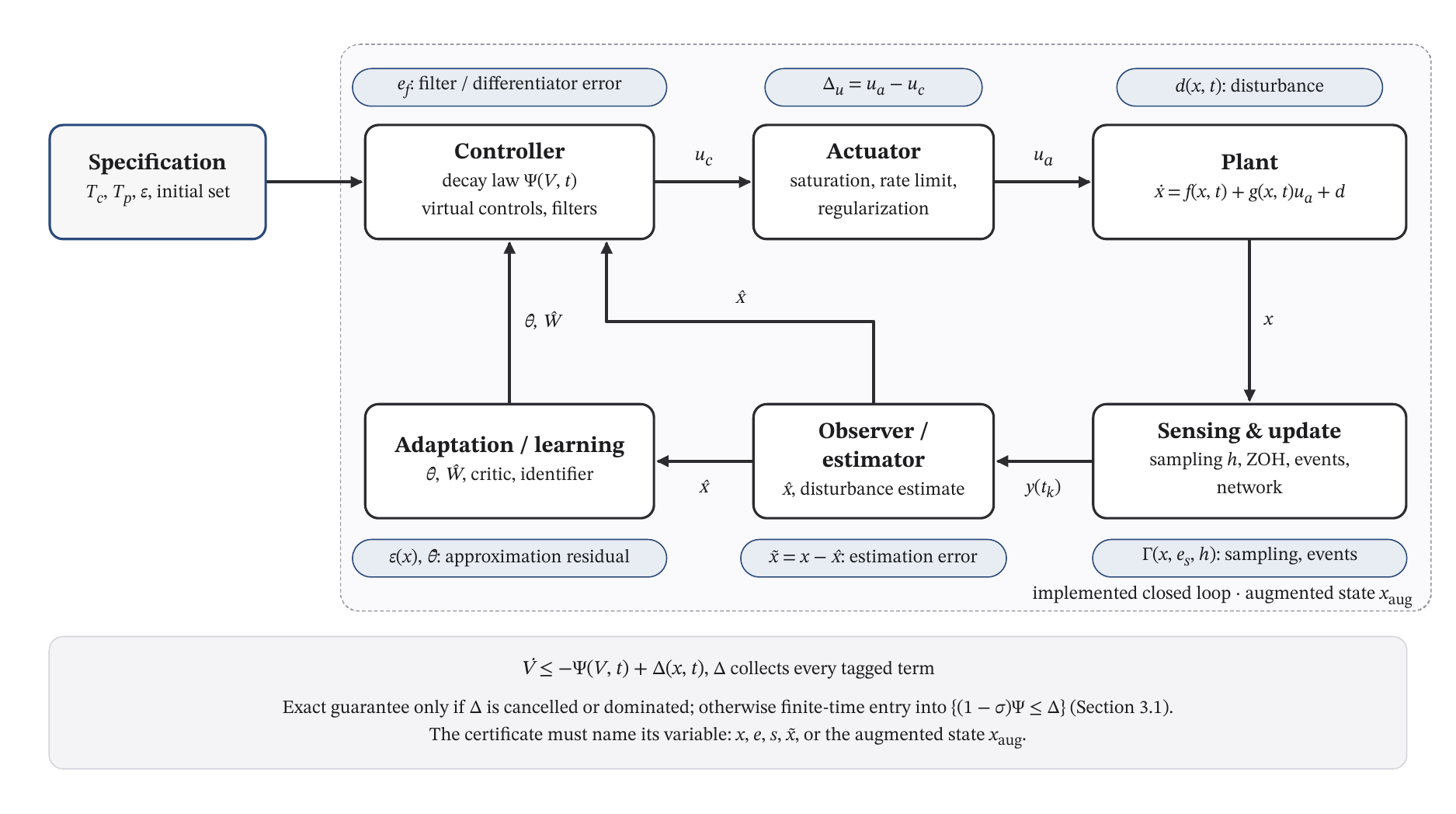}
\caption{Implemented closed loop and the residual terms contributed by its layers.}
\label{fig:implemented_loop}
\end{figure*}

Each architecture is examined through the same questions: its structural role, the variable on which its certificate is established, the difficulty specific to the architecture, how representative audited studies address that difficulty, and the design consequence. The central question is not only whether an architecture can reproduce a mechanism of Section~\ref{sec:mechanisms}, but whether the temporal guarantee is preserved after all architecture-specific dynamics and implementation requirements are included in the complete closed loop. Figure~\ref{fig:implemented_loop} shows where each layer of an implemented loop contributes to the residual $\Delta$ of \eqref{eq:general_decay_sec3}. These contributions are filter and differentiator errors, the saturation deficiency $\Delta_u$, disturbance, approximation residual, estimation error, and sampling- or event-induced error; a temporal certificate is exact only if every tagged term is cancelled or dominated, and it must name the variable it concerns. The study-level classification is collected in Table~\ref{tab:audit_full} (Appendix~\ref{app:audit}) and summarized in Table~\ref{tab:audit} (Section~\ref{subsec:design_guidance}); the prose retains only the studies that mark the boundaries of each certificate.

\subsection{Primary Feedback Architectures}
\label{subsec:primary_architectures}

Primary feedback architectures determine the fundamental route through which the available plant variables become the physical control input. Their defining feature is the feedback synthesis mechanism itself, before estimation, adaptation, learning, communication, or performance layers are added. Three families are considered: direct state feedback, which provides the most transparent connection between a target dissipation law and the input; sliding-mode realizations, which introduce reaching and manifold dynamics to combine convergence-time shaping with robustness against matched uncertainties; and backstepping, which propagates the same mechanisms to strict-feedback systems through virtual controls and composite Lyapunov constructions.

\subsubsection{Direct State-Feedback Realizations}
\label{subsubsec:direct_feedback}

Direct state feedback constructs the physical input from the available plant state \citep{zhang2012finite}, without intrinsically requiring observers, filters, adaptation laws, or other auxiliary dynamics. For
\begin{equation}
    \dot{x}
    =
    f(x,t)+g(x,t)u+d(x,t),
    \label{eq82}
\end{equation}
the controller has the form
\begin{equation}
    u=\kappa(x,t),
    \label{eq:direct_general_feedback}
\end{equation}
where $\kappa$ is designed so that the closed loop generates the dissipation required by the temporal specification. The linear law $u=-Kx$ is the asymptotic or exponential baseline when the closed-loop matrix is Hurwitz \citep{slotine1991applied}. Stronger guarantees are obtained within the same information architecture by embedding nonlinear powers, homogeneous terms, normalized decay laws, or time-varying gains in $\kappa$; finite-time \citep{zhang2012finite}, fixed-time \citep{hao2020bipartite}, predefined-time \citep{kokolakis2024safe}, and prescribed-time \citep{hua2021adaptive} formulations are not different architectures, the distinction lying in the mechanism embedded in $\kappa(x,t)$.

The certificate concerns the plant state directly, through $V$ evaluated along \eqref{eq82}. Control Lyapunov functions (CLFs) \citep{artstein1983stabilization,sontag1989universal} make this explicit: the temporal property is represented by the admissible control set \citep{bertsekas1995dynamic}
\begin{equation}
    \mathcal{U}_{\Psi}(x,t)
    =
    \left\{
        u\in\mathcal{U}:
        \dot V(x,t,u)
        \leq
        -\Psi(V,t)
    \right\},
    \label{eq:direct_clf_set}
\end{equation}
where $\Psi$ is the convergence-shaping law of the selected specification. Any pointwise selection from $\mathcal{U}_{\Psi}$ satisfies the nominal control-Lyapunov-function (CLF) condition and remains a direct realization as long as no controller state is introduced. This separates the required decay law from the control-selection rule and connects to the optimization- \citep{jiang2025safety}, constraint- \citep{si2023event}, and safety-aware \citep{jang2024safe} frameworks of Section~\ref{subsec:implementation_architectures}.

Direct feedback requires adequate state measurements and sufficient knowledge of the plant structure and control effectiveness \citep{tegling2019fundamental}. Its specific difficulty is admissibility: the nominal command $u^{*}=\kappa(x,t)$ is physically meaningful only if
\begin{equation}
    u^{*}(x,t)\in\mathcal{U}
    \label{eq:direct_admissibility}
\end{equation}
throughout the operating domain, which becomes restrictive for large-state fixed-time feedback \citep{zuo2026fixed} and for designer-assigned-time controllers with aggressive horizons \citep{tan2026flexible}. The feasibility limit of Section~\ref{subsec:feasible_decay} therefore appears here in its purest form, uncomplicated by auxiliary time scales.

Three audited designs mark the scope and the boundary of the category. \citet{nezhad2026finite} develop a voltage-based robust state-feedback controller for finite-time tracking of flexible-joint manipulators under matched and mismatched uncertainties, with hardware-in-the-loop validation. Two prescribed-time designs illustrate distinct routes within the same architecture: \citet{ning2025global} construct the law from a prescribed-time adjustment function and a solution-parameter Lyapunov construction for fully actuated systems, whereas \citet{zhang2026prescribed} generate linear time-varying feedback from a parametric Lyapunov equation and establish boundedness both on $[0,T_p)$ and after $T_p$. The intermittent and impulsive controller of \citet{you2024finite} is a hybrid state-feedback realization whose invoked property is the short-time stability of Remark~\ref{rem:two_fts}, not finite settling. Designs that add neural adaptation and command filters \citep{wang2025predefined}, observer-based hysteresis compensation \citep{lai2026output}, linear time-varying observers \citep{wang2026global}, or dual dynamic-gain observers \citep{mao2025fixed} fall outside strict direct feedback and are audited under Section~\ref{subsec:augmented_architectures} in Table~\ref{tab:audit_full}.

Direct state feedback is structurally the simplest realization \citep{berberich2020robust}: when state and model information are available, the desired dissipation can be imposed without observer errors or auxiliary convergence time scales. It does not by itself guarantee input admissibility, robustness, or practical implementability, and the two conditions that carry over to every other architecture are already visible here: the decay law must be realizable within $\mathcal{U}$, and the certificate must concern exactly the variable that the architecture regulates.

\subsubsection{Sliding-Mode Realizations}
\label{subsubsec:smc}

Sliding-mode control (SMC) embeds the temporal mechanism in the reaching dynamics, in the reduced-order sliding dynamics, or in both, and its discontinuous or high-gain structure provides inherent robustness against matched uncertainties \citep{young1999control}. For system \eqref{eq82} with a possibly time-dependent sliding variable $s=s(x,t)$, the sliding dynamics may be written as
\begin{equation}
    \dot{s}
    =
    a_s(x,t)+b_s(x,t)u+\delta_s(x,t),
    \label{eq:smc_s_dynamics}
\end{equation}
where $b_s(x,t)\neq0$ and $\delta_s$ is the disturbance component entering the sliding dynamics, and the control is decomposed as
\begin{equation}
    u=u_{\mathrm{eq}}+u_{\mathrm{r}},
    \label{eq:smc_decomposition}
\end{equation}
where $u_{\mathrm{eq}}$ determines the nominal manifold dynamics and $u_{\mathrm{r}}$ enforces reaching and robustness. At the classical baseline, the linear reaching law
\begin{equation}
    \dot{s}=-k_s s,
    \qquad
    k_s>0,
    \label{eq:smc_exp_reaching}
\end{equation}
gives exponential convergence of $s$, and with the linear manifold
\begin{equation}
    s=\dot e+\lambda e,
    \qquad
    \lambda>0,
    \label{eq:smc_linear_manifold}
\end{equation}
the ideal sliding motion $\dot e=-\lambda e$ is also exponential. The progression from asymptotic to designer-assigned-time SMC retains this reaching-sliding structure and changes only the mechanism embedded in it: linear damping, non-Lipschitz, mixed-region, normalized, or explicitly time-dependent shaping.

The two-phase structure permits the temporal properties of the phases to be shaped separately, but both must appear in the closed-loop certificate, and for discontinuous realizations the adopted solution concept and the post-reaching invariance of the manifold form part of it. Conventional discontinuous reaching drives $s$ to zero in finite time while the motion on a linear manifold remains exponential; finite-time arrival at $s=0$ must therefore not be identified with finite-time stabilization of the plant state. For the reaching law \citep{song2016finite,hou2019finite}
\begin{equation}
    \dot s
    =
    -k_F|s|^\alpha\operatorname{sgn}(s),
    \qquad
    0<\alpha<1,
    \label{eq:smc_ft_reaching}
\end{equation}
setting $V_s=s^2/2$, $\gamma=(1+\alpha)/2$, and $c_F=k_F2^{(1+\alpha)/2}$ gives $\dot V_s=-c_FV_s^\gamma$ and
\begin{equation}
    T_{\mathrm{reach}}
    \leq
    \frac{V_s(0)^{1-\gamma}}
         {c_F(1-\gamma)}
    =
    \frac{|s(0)|^{1-\alpha}}
         {k_F(1-\alpha)}.
    \label{eq:smc_ft_reaching_time}
\end{equation}
Since finite-time reaching alone does not establish finite-time stabilization \citep{polyakov2009lyapunov}, terminal SMC replaces the linear manifold with a nonlinear finite-time manifold \citep{yang2011nonsingular},
\begin{equation}
    s
    =
    \dot e
    +
    \lambda|e|^\rho\operatorname{sgn}(e),
    \qquad
    0<\rho<1,
    \label{eq:smc_terminal_surface}
\end{equation}
on which $\dot e=-\lambda|e|^\rho\operatorname{sgn}(e)$ and
\begin{equation}
    T_{\mathrm{slide}}
    \leq
    \frac{|e(t_r)|^{1-\rho}}
         {\lambda(1-\rho)},
    \label{eq:smc_ft_sliding_time}
\end{equation}
where $t_r$ is the reaching instant. Once both phase bounds are established, a sequential certificate follows,
\begin{equation}
    T_{\mathrm{total}}
    \leq
    T_{\mathrm{reach}}
    +
    T_{\mathrm{slide}}
    <\infty.
    \label{eq:smc_ft_total_time}
\end{equation}
The additive estimate is valid once both bounds are well defined, but an a priori closed-loop bound additionally requires control of $|e(t_r)|$: the tracking error need not decrease while $s\neq0$ and may exceed its initial value, so bounded reaching time alone does not certify the state transient over the reaching interval. Fixed-time SMC removes the initial-condition dependence of the reaching bound by combining small- and large-error actions \citep{hu2022robust},
\begin{equation}
    \dot{s}
    =
    -k_1|s|^\alpha\operatorname{sgn}(s)
    -
    k_2|s|^\beta\operatorname{sgn}(s),
    \qquad
    0<\alpha<1<\beta,
    \label{eq:smc_fxt_reaching}
\end{equation}
producing a uniform reaching-time bound \citep{moulay2022fixed}; nonsingular \citep{yang2011nonsingular} and fast terminal \citep{ni2016fast} constructions modify the manifold to avoid singular control terms or improve large-error transients. In every case the certified variable must be named: the sliding variable, the reduced-order state, or the complete closed loop including the reaching transient.

\begin{figure*}[pos=!t]
\centering
\reviewfigure{\textwidth}{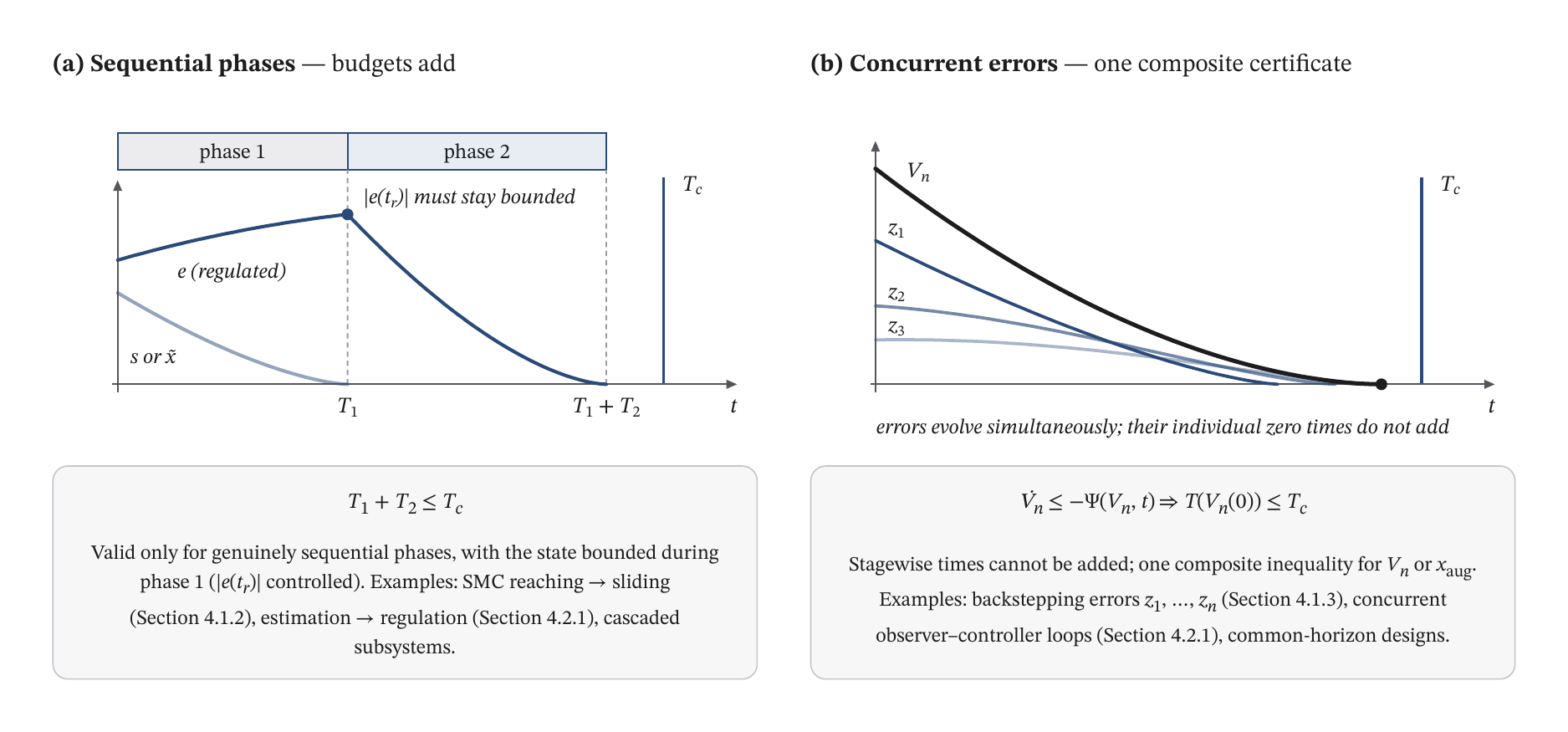}
\caption{Horizon allocation: sequential budgets versus concurrent errors (schematic).}
\label{fig:horizon_budget}
\end{figure*}

Two issues are specific to the two-phase structure. The first is horizon allocation. Predefined-time SMC \citep{munoz2022second} requires either a sequential budget,
\begin{equation}
    T_{\mathrm{reach}}\leq T_r,
    \qquad
    T_{\mathrm{slide}}\leq T_s,
    \qquad
    T_r+T_s\leq T_c,
    \label{eq:smc_pdt_budget}
\end{equation}
where $T_c$ is the desired closed-loop bound, or coordination of both phases with respect to a common horizon \citep{lai2025predefined}, which may reduce the conservatism of additive allocation; assigning $T_c$ to the reaching dynamics alone does not establish predefined-time stabilization of the plant. The distinction between a sequential budget and a common composite certificate, which recurs for backstepping and observer-based designs, is summarized in Fig.~\ref{fig:horizon_budget}. Additive budgets such as \eqref{eq:smc_pdt_budget}, \eqref{eq:recursive_pdt_budget}, and \eqref{eq:observer_controller_budget} apply to a valid sequential analysis, with bounded evolution during the first budget and a second-stage bound valid from the reached state; physical switching is not necessary. Concurrent errors instead require an interconnection argument, such as a composite or vector Lyapunov estimate, a cascade bound, or ISS/small-gain analysis, and their isolated settling times cannot simply be added. Prescribed-time SMC \citep{chen2022prescribed} introduces the horizon into the reaching law, for example
\begin{equation}
    \dot{s}
    =
    -\mu_r(t)s,
    \qquad
    \mu_r(t)
    =
    \frac{k_r}{T_{p,r}-t},
    \qquad
    0\leq t<T_{p,r},
    \label{eq:smc_pt_reaching}
\end{equation}
with $k_r>0$ and $T_{p,r}$ the horizon assigned to reaching, or into the manifold \citep{shi2024prescribed}, and the two phases again need compatible budgets or a common-horizon construction. As stated in \eqref{eq:sec34_false_implication}, divergence of $\mu_r(t)$ does not by itself make the input unbounded; boundedness depends on how fast the sliding, state, and auxiliary signals decay and on how the scaling enters the implemented law, so convergence, internal-signal boundedness, input admissibility, and continuation at and after the horizon are verified separately. The second issue is switching. Higher-order sliding modes \citep{shtessel2014sliding} and super-twisting algorithms \citep{chalanga2016implementation} reduce the direct effect of discontinuity on the input at the cost of additional controller states that must enter the temporal certificate; boundary-layer, saturation, and smooth-approximation remedies for chattering alter the dissipation inequality itself and are treated in Section~\ref{subsubsec:actuation_regularization}.

Three audited designs mark the boundaries of the certificate. In \citet{shi2024prescribed}, the reaching and sliding times are assigned separately using bounded time-varying gains, so the horizon covers the complete plant state and the gain issue of \eqref{eq:sec34_false_implication} is avoided by construction. In \citet{yin2025nonsingular}, the predefined-time result concerns the plant state under a disturbance observer that has no predefined-time certificate, so the guarantee does not extend to the augmented system. In \citet{su2025second}, the prescribed-performance envelope has a nonzero terminal width, so the second-order predefined-time result is practical although the sliding dynamics are exact. Designs that coordinate the sliding dynamics with observers or estimators having assignable horizons \citep{kong2025predefined,kumar2025prescribed}, the observer-free dual sliding-mode design of \citet{lei2024fixed}, adaptive, approximation-based, and regularized variants \citep{hu2024finite,li2025fastfinite,ding2026switching,wu2024adaptive,alsinai2026finite,liu2025practical,jia2025prescribed}, and the free-will arbitrary-time formulation of \citet{asadollahi2024fixed} (Remark~\ref{rem:arbitrary_time}) are classified in Table~\ref{tab:audit_full} together with their validation platforms. Finite-time reaching of a sliding manifold while the state is only confined to a reachable set \citep{Zhang2026SecurityReachableSetSMC}, or a fast simulated transient of an asymptotic Barbalat-type sliding-mode design \citep{Xiong2026RFNNSMCTiltingQuadrotor}, does not place the closed loop in a finite- or fixed-time class.

For SMC the question is whether the guarantee covers only the sliding variable or also the reaching transient, the reduced-order dynamics, the auxiliary controller states, and the post-horizon continuation; for discontinuous realizations the solution concept and manifold invariance belong to the certificate.

\subsubsection{Backstepping Realizations}
\label{subsubsec:recursive}

Backstepping extends convergence-time control to systems in which the input does not act directly on the regulated variable \citep{kokotovic1992joy}: the mechanism must be propagated through a sequence of virtual controls before reaching the physical input \citep{zhou2008adaptive}. For the strict-feedback system
\begin{equation}
\begin{aligned}
    \dot x_1
    &=
    f_1(x_1)+g_1(x_1)x_2,
    \\
    \dot x_2
    &=
    f_2(x_1,x_2)+g_2(x_1,x_2)u,
\end{aligned}
\label{eq:recursive_strict_feedback}
\end{equation}
and more generally for an $n$th-order strict-feedback system, the error coordinates are
\begin{equation}
    z_1=x_1-x_{1d},
    \qquad
    z_i=x_i-\alpha_{i-1},
    \quad
    i=2,\ldots,n,
    \label{eq:recursive_general_coordinates}
\end{equation}
where $\alpha_{i-1}$ is the virtual control generated at the preceding step. A typical composite Lyapunov construction is
\begin{equation}
    V_i
    =
    V_{i-1}
    +
    \frac{1}{2}z_i^2,
    \qquad
    V_1=\frac{1}{2}z_1^2,
    \label{eq:recursive_lyapunov_structure}
\end{equation}
and the complete design seeks
\begin{equation}
    \dot V_n
    \leq
    -\Psi(V_n,t).
    \label{eq:recursive_generic_decay}
\end{equation}
Classical backstepping compensates the interconnection terms and introduces smooth damping \citep{zhou2008adaptive,coron2013local}; finite-time constructions propagate fractional-power or homogeneous terms through the virtual-control hierarchy \citep{yu2023finite,xue2022adaptive,wang2022robust}, fixed-time designs preserve both near-origin and large-error dissipation in the composite dynamics \citep{li2017fixed}, and predefined-time constructions distribute normalized decay terms throughout the recursion \citep{liu2021predefined}. Backstepping does not define a separate convergence mechanism; it is the recursive architecture through which the mechanisms of Section~\ref{sec:mechanisms} reach the physical input.

The relevant certificate is the target inequality applied to the complete function $V_n$, not to an individual virtual-control error. Obtaining it involves a step specific to the recursive architecture. Stagewise design produces a sum of stagewise dissipation terms,
\begin{equation}
    \dot V_n
    \leq
    -\sum_{i=1}^{n} k_i|z_i|^{1+\alpha},
    \label{eq:recursive_stagewise_sum}
\end{equation}
whereas the target inequalities of Section~\ref{sec:mechanisms} are expressed in powers of $V_n$. Converting \eqref{eq:recursive_stagewise_sum} into a bound of the form $-cV_n^{(1+\alpha)/2}$ requires a power-mean inequality: for sublinear exponents, $\sum_i a_i^{p}\ge(\sum_i a_i)^{p}$ with constant one, independent of $n$, whereas for the superlinear branch used in fixed-, predefined-, and prescribed-time designs $\sum_i a_i^{p}\ge n^{1-p}(\sum_i a_i)^{p}$, so the effective gain degrades as $n$ grows. Two consequences follow. Part of the conservatism of a recursive convergence-time bound originates in the architecture rather than in the decay law (Section~\ref{subsec:open_temporal}), and retaining the same assigned horizon for a higher-order system requires larger gains, so the feasibility limits of Section~\ref{subsec:feasible_decay} tighten with the order of the recursion. A further distinction concerns time budgets. For genuinely sequential subsystems, individual budgets may be assigned according to
\begin{equation}
    \sum_{i=1}^{m}T_i
    \leq
    T_c,
    \label{eq:recursive_pdt_budget}
\end{equation}
provided the state remains bounded during every preceding stage. The recursive steps of ordinary backstepping, however, are design steps rather than temporal phases: $z_1,\ldots,z_n$ evolve simultaneously, their individual time estimates cannot be added, and a common composite certificate for $V_n$ is required, which may also reduce the conservatism of an artificial sequential allocation.

Two difficulties are specific to the recursion. The first concerns time-varying designs. Prescribed-time backstepping introduces the horizon through time-varying scaling \citep{ye2022backstepping}, state transformations \citep{zeng2025backstepping}, transformed-time coordinates \citep{hao2026prescribed}, or scaling of the virtual tracking errors in multi-input multi-output (MIMO) designs \citep{ye2022prescribed}; schematically,
\begin{equation}
    \alpha_i
    =
    \alpha_i
    \left(
        z_1,\ldots,z_i,t;T_p
    \right),
    \qquad
    i=1,\ldots,n-1,
    \label{eq:recursive_pt_virtual_structure}
\end{equation}
with the composite behavior
\begin{equation}
    \dot V_n
    \leq
    -\mu(t;T_p)\Psi(V_n),
    \qquad
    0\leq t<T_p.
    \label{eq:recursive_pt_compact}
\end{equation}
Unlike direct feedback, the scaling propagates through the virtual controls and their derivatives, so the certificate must establish that
\begin{equation}
    \alpha_i(t),\;
    \dot{\alpha}_i(t),\;
    u(t)
    \quad
    \text{remain well defined as }
    t\rightarrow T_p^{-},
    \label{eq:recursive_pt_internal_signals}
\end{equation}
and must specify the continuation at and after $T_p$; plant-state convergence alone does not establish these internal-signal properties \citep{portella2026adaptive}. The second difficulty is the explosion of complexity \citep{wang2024adaptive,portella2026adaptive}: the repeated differentiation of virtual controls, aggravated when nonlinear powers, normalized decay laws, adaptive terms, or temporal scaling are propagated through several stages. Dynamic-surface \citep{jiang2026adaptive} and command-filtered methods \citep{hu2026prescribed} introduce auxiliary filtering dynamics,
\begin{equation}
    \tau_f\dot{\alpha}_f+\alpha_f=\alpha,
    \qquad
    e_f=\alpha_f-\alpha,
    \label{eq:recursive_filter}
\end{equation}
and the implemented closed-loop state is enlarged from $z=[z_1,\ldots,z_n]^{T}$ to
\begin{equation}
    z_{\mathrm{aug}}
    =
    [z_1,\ldots,z_n,e_{f,1},\ldots,e_{f,r}]^{T}.
    \label{eq:recursive_augmented_state}
\end{equation}
A temporal property established for $z$ does not automatically extend to $z_{\mathrm{aug}}$: filter, differentiator, observer, and adaptation errors must either satisfy the same specification or enter the composite inequality as explicitly quantified residual terms. Filtering also couples to feasibility: $\tau_f$ must be small relative to the assigned horizon, so reducing $T_c$ or $T_p$ increases the required filter bandwidth and hence the sensitivity to measurement noise and to the sampling period.

Three audited designs mark the boundaries. The two-time-scale construction of \citet{zeng2025backstepping} separates the slow and fast dynamics, assigns both a common prescribed horizon through time-varying scaling, and, under its controllability and small-perturbation conditions, establishes prescribed-time convergence while also examining boundedness of the physical input. In \citet{liu2025auv}, predefined-time backstepping is combined with an adaptive predefined-time disturbance observer and predefined-time filters for an underactuated autonomous underwater vehicle (AUV); the tracking, velocity, and filter errors are collected in one Lyapunov function but converge to a residual set, so the guarantee is practical predefined-time. In \citet{peng2024antisaturation}, barrier Lyapunov functions, fixed-time prescribed-performance boundaries, and antisaturation auxiliary systems keep the constrained variables within specified regions while the closed loop remains uniformly ultimately bounded; this is a fixed-time performance-envelope result, not exact fixed-time stabilization of the plant state, and the distinct roles of envelopes, antisaturation systems, and event-triggered actuation are examined in Sections~\ref{subsubsec:safety_performance}-\ref{subsubsec:digital_implementation}. Designs combining finite-time backstepping with fixed-time observers or differentiators \citep{ding2024rectifier,li2024autonomous}, learning-based variants \citep{shi2024adaptive,hao2026predefined,wang2025optimal}, and the command-filtered fuzzy design of \citet{zhang2025command} are classified in Table~\ref{tab:audit_full}; in each, the guarantee is limited by the component certified only to a neighborhood, and the tuning horizon of \citet{hao2026predefined} is recorded through the proved bound (a multiple of that horizon) rather than as a settling time. Likewise, a finite-time differentiator inside a recursive design whose closed loop is shown only to be bounded \citep{Chu2026UnifiedNeuralEventTriggered} does not make that loop finite-time.

A backstepping guarantee is complete when the target inequality holds for the composite function $V_n$, given the stagewise conversion above; a claim established at one recursive step does not transfer to the complete system without the interconnections and filter dynamics they introduce.

\subsection{Augmented and Information-Structured Architectures}
\label{subsec:augmented_architectures}

The three families of this group modify a primary architecture in different ways:
\begin{equation}
\begin{aligned}
    \text{observer/output feedback}
    &:\quad
    x
    \rightarrow
    \hat{x}
    \rightarrow
    u,
    \\
    \text{adaptation/learning}
    &:\quad
    f,\theta
    \rightarrow
    \hat f,\hat\theta
    \rightarrow
    u,
    \\
    \text{distributed control}
    &:\quad
    \{x_j-x_i\}_{j\in\mathcal N_i}
    \rightarrow
    u_i.
\end{aligned}
\label{eq:augmented_architecture_summary}
\end{equation}
Their common challenge is the preservation of the nominal convergence mechanism after the estimation, adaptation, approximation, or information structure is introduced. Analyzing the complete closed loop does not mean that every auxiliary state must converge within the same horizon as the regulated variable; it means that the certificate states which property is proved for each auxiliary error and shows that this property suffices to preserve the guarantee for the regulated variable.

\subsubsection{Observer-Based and Output-Feedback Realizations}
\label{subsubsec:observer_output}

Output-feedback architectures are required when the states needed by a state-feedback law are not measurable \citep{calise2001adaptive}. For
\begin{equation}
    \dot{x}=f(x,t)+g(x,t)u,
    \qquad
    y=h(x,t),
    \label{eq:output_feedback_system}
\end{equation}
with only $y\in\mathbb{R}^{p}$ available, two realizations are distinguished. Observer-based output feedback \citep{lv2025neural} reconstructs the unavailable states and evaluates the law on the estimates: with $\tilde{x}=x-\hat{x}$,
\begin{equation}
\begin{aligned}
    \dot{\hat{x}}&=\mathcal{F}(\hat{x},y,u,t),\\
    u&=\kappa(\hat{x},t).
\end{aligned}
\label{eq:observer_based_structure}
\end{equation}
Dynamic output feedback \citep{rizvi2019reinforcement} need not reconstruct the state; it includes the static law $u=\kappa_y(y,t)$ and the dynamic realization
\begin{equation}
\begin{aligned}
    \dot{\eta}&=\varphi(\eta,y,t),\\
    u&=\kappa_y(\eta,y,t),
\end{aligned}
\label{eq:dynamic_output_feedback}
\end{equation}
with internal controller state $\eta$; observer-based control is the particular case in which $\eta$ is a state estimate \citep{chen2015disturbance}. The classification adopted here follows the measurement structure: a design is output feedback when its implemented law uses only the measured output $y$ (statically, through an internal state $\eta$, or through a state estimate), and a design that measures the full state and adds a disturbance or extended-state observer remains a state-feedback architecture with an augmentation.

The certificate must apply to the augmented dynamics, not only to the measured or regulated output. Defining
\begin{equation}
    x_{\mathrm{aug}}
    =
    \begin{bmatrix}
        x^{T} & \eta^{T}
    \end{bmatrix}^{T},
    \label{eq:output_feedback_augmented_state}
\end{equation}
the analysis should state whether convergence concerns $y$, $x$, $\tilde{x}$, $\eta$, or the complete augmented state, and whether the result is exact or practical; output convergence alone may leave estimation, zero, or controller dynamics uncertified. This requirement has a structural origin. For linear time-invariant designs a separation principle permits the observer and the controller to be designed independently and combined. Under the non-Lipschitz feedback of finite- and fixed-time designs, or the time-varying feedback of prescribed-time designs, the separation generally fails: the estimation error enters the regulation dynamics through nonsmooth or time-varying terms, and its transient can invalidate the nominal state-feedback certificate. The certificate must therefore address the augmented state explicitly, answering three questions for each design: which regulated variable converges, which property (convergence, boundedness, or entry into a set) is proved for the estimation error, and whether that property suffices to preserve the temporal guarantee of the regulated variable. Convergence of the estimation error within the same horizon is one sufficient answer, not the only one (Section~\ref{subsec:open_augmented}).

Finite- and fixed-time observers employ nonlinear or homogeneous correction mechanisms \citep{liu2024observer,liu2024adaptive}, whereas predefined- and prescribed-time observers assign an estimation horizon \citep{xuan2025predefined,ji2026prescribed}; prescribed-time observer gains, parametric-Lyapunov-equation feedback, and nonscaling stochastic constructions \citep{holloway2019prescribed,zhou2021prescribed,zhang2024global,li2022prescribed} are output-feedback realizations of the mechanisms of Section~\ref{subsubsec:time_varying_pt}, not additional convergence definitions. Separate observer and regulation budgets can be added when the augmented system remains well posed during the observer transient and the regulation bound is valid from every state reached at its end,
\begin{equation}
    T_{\mathrm{obs}}+T_{\mathrm{reg}}\leq T_c,
    \label{eq:observer_controller_budget}
\end{equation}
where $T_{\mathrm{obs}}$ and $T_{\mathrm{reg}}$ are analysis budgets and $T_c$ is the desired closed-loop bound, as in \eqref{eq:smc_pdt_budget}. The controller may operate during observation; physical switching is not necessary if the post-observer regulation estimate applies to the reached state. The total bound concerns the regulated variable unless all augmented states are also certified to converge. An exact observer horizon followed by an exact tracking horizon under structured nonvanishing disturbances realizes this budget \citep{ji2026prescribed}, whereas bounded time-varying high gains shared by observer and controller give a practical common-frame result \citep{Cai2026MSVPracticalPsT,Zhang2026FNNPTESOUSV}. For concurrent dynamics the coupling must be analyzed jointly, using, for example, a composite or vector Lyapunov function, a cascade estimate, or an appropriate ISS/small-gain argument \citep{xu2023observer,wang2026predefined,li2022prescribed}. Implementation effects are especially consequential here: high-gain peaking \citep{khalil2017high}, measurement-noise amplification \citep{lv2025h}, finite sensor resolution \citep{zhao2025disturbance}, and sampling \citep{wang2024sampled} degrade the nominal guarantee, and a short assigned observer horizon aggravates them, since it demands higher correction gains whose transient peaking propagates into the control law during precisely the interval over which the guarantee must hold. Reducing the estimation horizon therefore tightens the feasibility limits of Section~\ref{subsec:feasible_decay} in the same way that a divergent scaling function does for sliding-mode reaching and a small filter constant does for command filtering.

Three audited designs illustrate the three answers to the questions above. The decentralized design of \citet{tan2026decentralized} provides a common-horizon prescribed-time certificate for the plant, the observer error, and the control input of interconnected systems using only local outputs, and thereby exemplifies the augmented-state guarantee in its strongest form; \citet{que2024disturbance} similarly establishes exact prescribed-time convergence of both the estimation and the tracking errors. The unmanned surface vessel design of \citet{sui2025usv} reconstructs unmeasured velocities and lumped disturbances within an assigned observer horizon and drives the tracking error into a prescribed-performance region while the augmented signals remain uniformly ultimately bounded: the regulated variable carries the temporal guarantee, the auxiliary signals carry a boundedness property, and the certificate connects the two. The visual-feedback bolt-tightening controller of \citet{you2025visual} establishes finite-time convergence of the tracking error with experimental validation, and is strict output feedback only to the extent that the reconstructed-state or image-feature estimation error carries its own certificate. Designs with extended-state or disturbance observers augmenting backstepping, sliding-mode, or composite controllers \citep{farhat2024walking,xu2025autotuned,qin2024enhanced,ye2024fdi,liu2026hydraulic,cai2026predefined} are classified in Table~\ref{tab:audit_full} according to whether the observer reconstructs plant states. Sequential budgets of the form \eqref{eq:observer_controller_budget} appear explicitly in a switched fixed-time observer-controller pair validated in field tests \citep{Meng2026FixedTimeAmphibiousVehicle} and in a quadrotor attitude loop whose proof chains observer transient, reaching, and sliding and restarts after each disturbance jump, the sampled implementation being practical \citep{Ren2027CompositeFiniteTimeQuadrotor}. Conversely, a fixed-time disturbance observer within semiglobally uniformly ultimately bounded tracking \citep{Zhang2026USVOptimizedBackstepping} gives a fixed-time property to the estimation error only.

For output feedback, the failure of separation under nonsmooth or time-varying feedback makes the augmented-state statement necessary, and the peaking-horizon coupling makes short estimation horizons costly.

\subsubsection{Adaptive and Learning-Based Realizations}
\label{subsubsec:adaptive_learning}

Adaptive and learning-based methods are cross-cutting augmentations rather than a separate feedback geometry: online estimation \citep{lv2016online}, function approximation \citep{zhang2024fault}, and policy learning can be incorporated into direct-feedback, sliding-mode, backstepping, or output-feedback architectures. For the parameterized plant
\begin{equation}
    \dot{x}
    =
    f(x,\theta)+g(x)u,
    \label{eq:adaptive_system}
\end{equation}
an adaptive realization introduces an estimate $\hat{\theta}$ with error $\tilde{\theta}=\theta-\hat{\theta}$. For neural and fuzzy approximation \citep{farrell2006adaptive}, an unknown nonlinearity is represented as
\begin{equation}
    f(x)
    =
    W^{*T}\phi(x)+\varepsilon(x),
    \label{eq:learning_approx}
\end{equation}
with ideal weights $W^*$ and approximation residual $\varepsilon(x)$, of which only a bound $|\varepsilon(x)|\leq\bar{\varepsilon}$ over the operating domain is usually available.

Plant regulation and parameter identification remain distinct:
\begin{equation}
    x(t)\rightarrow0
    \quad\not\Rightarrow\quad
    \tilde{\theta}(t)\rightarrow0,
    \label{eq:adaptive_parameter_distinction}
\end{equation}
and exact parameter recovery requires excitation or identifiability conditions \citep{wang2020fixed}. A finite-, fixed-, predefined-, or prescribed-time claim must therefore specify whether it applies to the plant error \citep{sun2022adaptive}, the parameter-estimation error \citep{cao2026parameter}, or the complete augmented system \citep{lu2025parallel}. The level at which stability is proved matters as well: a transformed game system that is globally asymptotically stable under the ideal policy becomes only uniformly ultimately bounded once critic learning and event triggering are implemented \citep{Zhang2026HumanMachineDynamicAuthority}, and asymptotic convergence of a learned model error \citep{Xu2026ZeroSumDSTC} or of a policy iteration \citep{Yue2026SMSEventTriggeredFTC} does not transfer to the plant. The approximation residual must appear explicitly in the certificate \citep{li2019adaptive_2,wang2017finite}; unless it is eliminated or dominated by a robust term, exact settling is replaced by practical \citep{nevsic2009extremum} or residual-set convergence \citep{zhu2022neural}, as stated once in Section~\ref{subsec:specification_to_design}.

Two difficulties are specific to this class. The first is that the residual is generated inside the controller: approximation, observation, or adaptation residuals persist in the Lyapunov inequality even when the plant model is otherwise exact, so the certified conclusion concerns a compact target set unless the design adds a robust term or an excitation condition that removes the residual. This is the pattern across finite-time \citep{cai2016fuzzy,cuan2023cps,fu2025inverter,rodriguez2022usv,chen2023uam}, fixed-time \citep{cui2022quav,guo2025hydraulic,liu2025robot}, and designer-assigned-time \citep{ni2021global,cui2025observer,liu2025global,cao2026parameter} designs. The second is that learning for optimization improves performance without certifying time: reinforcement learning and actor-critic structures may reduce transient cost or control effort \citep{wang2022reinforcement,wang2023adaptive}, but an approximately solved Hamilton-Jacobi-Bellman equation does not by itself establish convergence-time stability \citep{recht2019tour}; policy approximation, exploration, triggering errors, and observer dynamics must enter the stability certificate. The finite-time condition can instead be imposed as a constraint on the network weights during policy training \citep{Phothongkum2026ConstrainedRLFiniteTime}, whereas fixed-time convergence of actor-critic weights is compatible with plant states that are only ultimately bounded \citep{Xue2026NAGFixedTimeActorCritic}. Prescribed-time actor-critic designs either attribute the temporal property to the error transformation rather than to weight convergence \citep{tan2026flexible} or certify observer and parameter errors only to compact sets \citep{Mai2026PrescribedTimeRLAUV,Liu2026DistributedPrescribedTimeOptimalFormation}.

Three audited designs mark the boundaries. In \citet{wang2023maglev}, the magnetic-levitation position states are driven to the origin in fixed time under the stated robust condition, whereas the disturbance-observer error retains a residual bound: an exact plant-state result need not imply exact convergence of the complete augmented state. The composite fuzzy scheme of \citet{jiang2025composite} uses current and historical data under an interval-excitation condition and establishes exact preassigned-time tracking while the approximation parameters converge to a neighborhood, showing how an excitation condition converts a practical result into an exact one for the regulated variable; adaptive prescribed-time designs for spacecraft formation flying \citep{chen2023spacecraft}, robotic arm-finger coordination \citep{ye2025armfinger}, and vehicular platoons \citep{li2025platoon}, and the adaptive prescribed-time control of \citet{hua2021adaptive}, likewise certify the regulated variables while maintaining only boundedness of the adaptive signals. In \citet{song2026rl}, an actor-critic-identifier structure is combined with a predefined-time disturbance observer, dynamic-surface control, and event-triggered communication; the policy is approximately optimal while the closed-loop guarantee remains practical predefined-time boundedness. The appointed-time formulation of \citet{sun2023appointed}, which provides practical tracking with fixed-time boundedness of the augmented signals, is mapped onto the classes of Table~\ref{tab:terminology} according to the role of its horizon. The remaining audited designs are classified in Table~\ref{tab:audit_full}.

The items specific to adaptive and learning-based results are the excitation condition and the treatment of approximation residuals; approximation accuracy, parameter boundedness, and learned performance do not substitute for a temporal guarantee of the implemented loop.

\subsubsection{Distributed and Cooperative Realizations}
\label{subsubsec:distributed}

Distributed control changes the information structure of the feedback law: each agent constructs its input from locally available variables and information received from its neighbors \citep{xu2019distributed},
\begin{equation}
    u_i
    =
    \kappa_i
    \left(
        x_i,
        \{x_j-x_i\}_{j\in\mathcal N_i},
        t
    \right),
    \label{eq:distributed_general_feedback}
\end{equation}
where $\mathcal N_i$ is the neighbor set of agent $i$; depending on the realization, the exchanged quantity may instead be an observer state, an optimization variable, a quantized signal, or an estimate of a leader or exosystem state. The mechanisms of Section~\ref{sec:mechanisms} can be embedded into \eqref{eq:distributed_general_feedback} unchanged; distributed time-varying scaling, for example, has been used for prescribed-time consensus and containment \citep{wang2018prescribed}. Two terms must be kept apart. A \emph{distributed} protocol uses local information in the control law but may still require global quantities at the design stage, a Laplacian eigenvalue, the number of agents, a graph-dependent gain. A \emph{fully distributed} realization constructs the local gains without such quantities, typically through adaptation.

The certified variable is a consensus, formation, synchronization, output-regulation, or optimization error, and its convergence property is coupled to the communication graph \citep{zhang2011lyapunov}: connectivity, Laplacian eigenvalues, directed spanning trees, weight balance, and switching topology enter the closed-loop certificate in addition to the local dynamics. The defining question is therefore not the mechanism but whether its temporal guarantee remains valid under the graph-dependent information structure \citep{liu2025spatiotemporal}, and a certificate should state which graph assumption it requires, which graph quantities the bound and the gains depend on, and which information each agent must possess. As in the other augmented architectures, the convergence of an auxiliary distributed observer or optimizer must be separated from the convergence of the regulated agent variables: exact estimation at an assigned time for every node does not by itself establish an assigned time for the formation or regulation error, and the interconnection must be certified.

For designer-assigned-time cooperative control, the specific difficulty is scalability: can an assigned horizon be retained as network size and topology vary \citep{liu2024prescribed}? Maintaining the horizon as the number of agents grows may require larger gains or control amplitudes, depending on the spectral properties of the graph, so that the feasibility limit of Section~\ref{subsec:feasible_decay} acquires a graph-dependent form; prescribed-time transformations may also produce increasingly aggressive transients as the horizon is shortened \citep{chen2024prescribed,zhang2025dptuio}. Fully distributed adaptive gains can remove some global spectral requirements \citep{an2025finite}, but they do not remove bandwidth, saturation, noise, or energy limitations. Directed, switching, or intermittently connected networks add the difficulty that the information structure itself varies with time. Distributed control should finally be distinguished from networked implementation: the architecture determines which information each local controller requires, whereas sampling, delays, quantization, packet losses, and limited data rate determine how and when it is transmitted. The two layers interact, quantized inputs \citep{zhang2024quantized} and limited data rate with collision avoidance \citep{zhao2025collision} change the attainable error set even when the coordination mechanism is unchanged, but they are different structural questions, and the transmission layer is treated in Section~\ref{subsubsec:digital_implementation}.

The cooperative designs examined here separate along the three axes above. In the prescribed-time group, distributed scaling assigns a common horizon to consensus and containment errors over connected undirected graphs and directed graphs containing a spanning tree \citep{wang2018prescribed}; prescribed-time cooperative output regulation of heterogeneous linear agents combines a distributed prescribed-time exosystem observer with state- or output-feedback protocols under directed communication, with bounded inputs under the stated assumptions \citep{chen2024prescribed}; and the distributed prescribed-time unknown-input observer of \citet{zhang2025dptuio} reconstructs the complete plant state at every node at the assigned time under observer-matching and joint strong-observability conditions over a strongly connected digraph, with measurement noise producing a bounded estimation error of characterized size, while remaining distributed but not fully distributed. In the predefined-time group, the resource-allocation algorithm of \citet{yue2025predefined} reaches the time-varying optimizer at the predefined time over a strongly connected weight-balanced digraph while maintaining the global equality constraint, whereas the multiobjective method of \citet{liu2023multiobjective} and the heterogeneous output-allocation schemes of \citet{jin2025practical} reach an assignable neighborhood of the optimum at the selected time and are therefore practical. In the finite- and fixed-time group, \citet{an2025finite} achieves exact finite-time formation tracking of heterogeneous agents over directed networks with a fully distributed adaptive observer gain and a disturbance observer, with an initial-condition-dependent settling time; \citet{zhou2026formation} combines a distributed fixed-time leader observer, a fixed-time solver for the regulator equations, and a formation controller without Laplacian knowledge; \citet{liu2023resource} gives an initial-condition-independent bound for resource management under fixed or switching connected graphs; and the platoon \citep{liu2022distributed} and fault-tolerant unmanned aerial vehicle (UAV) formation \citep{duan2022uav} designs are practical fixed-time, with the error entering a residual set within a uniform bound. Cooperative applications whose guarantees are asymptotic, steady-state, or optimization-based \citep{Zhang2026DynamicEventTriggeredADPConsensus,He2026MarkovianDMPCPlatooning,Xu2026DualMotorSteerByWireConsensus,Zhu2026KConnectedNMSV,liu2022hst,wang2024cutin,xiong2023wind,zeng2024microgrid,sadabadi2023sparse,song2025rmpc,zhou2022satellite} broaden the range of distributed realizations but lie outside the convergence-time classes of Section~\ref{sec:notions} and are not audited. Among recent prescribed-time formation designs, those with envelopes, self-triggering, or time-base-generator clocks report practical residuals \citep{Zhang2026DynamicSelfTriggeredBipartitePsT,Li2026PsTClusterConsensusHMAS,Liu2026DistributedPrescribedTimeOptimalFormation}, and the rigid-formation controller of \citet{Zhou2026PracticalPsTRigidFormation} is one of the few cooperative prescribed-time designs validated experimentally. The classification of the audited designs is recorded in Table~\ref{tab:audit}.

A distributed convergence-time claim is complete only when it states (i) whether convergence is exact or practical, (ii) whether the bound depends on the initial condition and whether it is assignable, (iii) which agent, observer, optimization, and formation errors carry the guarantee, (iv) which graph assumption is required and how the bound and the gains depend on the graph, and (v) which communication assumptions are needed. Convergence of an auxiliary observer or optimizer is not convergence of the cooperative loop unless the interconnection is certified.

The three families of this group modify a primary architecture in the ways summarized in \eqref{eq:augmented_architecture_summary}; their common requirement is that the certificate name the property proved for each auxiliary error and show that it suffices for the regulated variable.

\subsection{Constraint- and Implementation-Aware Realizations}
\label{subsec:implementation_architectures}

The architectures of Sections~\ref{subsec:primary_architectures} and~\ref{subsec:augmented_architectures} describe how a nominal convergence mechanism is embedded in the feedback loop. The implemented controller must also operate within constraints imposed by the plant \citep{franze2017command}, the actuator \citep{su2021velocity}, the communication network \citep{liu2018predictive,jang2023probabilistic}, the sensing architecture \citep{dentler2019model}, and the digital implementation \citep{pichan2016straightforward}. These constraints are not details added after the stability analysis. They alter the admissible trajectories, reduce the effective control authority, introduce actuator-, sampling-, or quantization-induced error dynamics, and, in the sense of the qualification stated once in Section~\ref{subsec:specification_to_design}, may replace exact convergence by convergence to a residual set \citep{ji2023saturation,diFerdinando2024digital,zhang2024quantized}. Typical requirements are
\begin{equation}
    x(t)\in\mathcal X_{\mathrm{safe}},
    \qquad
    u_a(t)\in\mathcal U,
    \qquad
    e(t)\in\mathcal P(t),
    \label{eq:constraint_aware_requirements}
\end{equation}
where $\mathcal X_{\mathrm{safe}}$ is a safe state set, $\mathcal U$ the available control authority, $u_a$ the input actually applied, and $\mathcal P(t)$ an allowable performance envelope \citep{ames2019cbf,sun2022adaptive}; optimization objectives (energy, effort, economic cost) and the digital layer (sampling, quantization, event triggering, computation time, delay, packet loss, data rate) act on the same closed loop \citep{liu2018predictive,diFerdinando2024digital}.

These effects raise feasibility questions that an unconstrained convergence proof does not contain: an assigned horizon may demand a magnitude or bandwidth the actuator cannot deliver, a fast transient may cross a safety boundary, a performance envelope may be incompatible with the horizon, the initial set, or the authority, and a continuous-time proof may rely on updates that no sampled or event-triggered implementation reproduces. Section~\ref{subsec:feasible_decay} stated the limit abstractly; Section~\ref{sec:numerical} gives its two quantitative forms for the benchmark, the authority condition \eqref{eq:feasibility_necessary} and the two-cycle residual scale \eqref{eq:sampling_cycle} of the sampled power laws, a property of that emulation rather than a general accuracy bound. Preserving a finite-, fixed-, predefined-, or prescribed-time property under constraints therefore generally requires a constraint-aware redesign and a separate certificate for the implemented closed loop \citep{polyakov2023consistent,ning2023eventPrescribed}; when such a certificate cannot be established, the statement is weakened as in \eqref{eq:claim_calibration}, and the trajectory must remain admissible before, at, and after the claimed convergence time.

The three subsections treat these issues as cross-cutting layers superimposed on a primary or augmented architecture. Optimization, safety, and performance constraints determine which closed-loop trajectories are acceptable (Section~\ref{subsubsec:safety_performance}); saturation, rate limits, and regularization determine which nominal commands can be generated (Section~\ref{subsubsec:actuation_regularization}); sampling and event-triggered updates determine when the feedback information is refreshed (Section~\ref{subsubsec:digital_implementation}).

\subsubsection{Performance-Envelope, Safety-, and Optimization-Constrained Realizations}
\label{subsubsec:safety_performance}

Optimization-, safety-, and performance-constrained realizations do not constitute a convergence-time mechanism. They either restrict the set of admissible closed-loop trajectories or select a particular input among those compatible with the required decay, so that a complete design pairs a temporal certificate with a constraint-feasibility certificate. With the admissible sets
\begin{equation}
    \mathcal X_{\mathrm{safe}}
    =
    \left\{
        x\in\mathbb R^n:
        h_r(x)\geq 0,\quad r=1,\ldots,n_h
    \right\},
    \qquad
    u\in\mathcal U,
    \label{eq:safe_input_sets}
\end{equation}
where the $h_r$ encode collision avoidance, state limits, obstacle boundaries, or operating envelopes, the temporal controller must reach its target without leaving $\mathcal X_{\mathrm{safe}}$ or $\mathcal U$. Three structurally different roles are distinguished below: instantaneous input selection under a temporal decrease constraint (quadratic programs, QPs, combining CLF and control-barrier-function, CBF, constraints), prediction-based constraint handling (model predictive control), and time-varying admissibility of the error (performance envelopes and funnels). None of them changes the decay law of Section~\ref{sec:mechanisms}; each changes what the certificate can say about it.

The CLF-CBF quadratic program combines a nominal temporal controller with Lyapunov and barrier constraints. For $\dot x=f(x)+g(x)u$,
\begin{equation}
\begin{aligned}
    (u^\star,\delta^\star)
    =
    \arg\min_{u,\delta}\quad&
        \frac{1}{2}
        \left\|u-u_{\mathrm{nom}}\right\|_{H}^{2}
        +
        p\delta^2
        \\
    \mathrm{subject\ to}\quad&
        L_fV(x)+L_gV(x)u
        \leq
        -\Psi(V,t)+\delta,
        \\
    &
        L_fh_r(x)+L_gh_r(x)u
        \geq
        -\alpha_{h,r}\!\left(h_r(x)\right),
        \quad r=1,\ldots,n_h,
        \\
    &
        u\in\mathcal U ,
\end{aligned}
\label{eq:temporal_clf_cbf_qp}
\end{equation}
where $\Psi$ is the decay law of the selected class, the barrier inequalities impose forward invariance of the safe set, and $u_{\mathrm{nom}}$ may already contain the convergence mechanism, the program acting as a minimally invasive filter \citep{ames2017cbf,garg2022fixedqp}. Relaxing only the Lyapunov constraint fixes the priority: safety is hard, the convergence rate may yield. With a soft CLF the decrease condition may become $\dot V\leq-KV+\delta^\star(x,t)$. A uniform residual-set estimate requires, for example, $\delta^\star\leq\bar\delta<\infty$, giving an ultimate level $\bar\delta/K$ for $K>0$; the quadratic penalty alone supplies no such bound. Safety additionally requires feasibility and the barrier hypotheses. These qualifications apply when interpreting soft-CLF safety and connectivity designs \citep{Zhu2026KConnectedNMSV}. The formulation exposes feasibility directly: if the input set, the barrier constraints, and the temporal constraint have empty intersection, no input preserves both safety and the requested horizon, and increasing the penalty $p$ only changes how the infeasible request is traded against the cost. Fixed-time CLF-CBF programs relate the admissible initial set, the input bounds, the spatial constraints, and the achievable horizon explicitly \citep{garg2022fixedqp}, which is the constrained form of the admissible-set description \eqref{eq:direct_clf_set} of Section~\ref{subsubsec:direct_feedback}.

Model predictive control handles constraints by prediction,
\begin{equation}
\begin{aligned}
    \boldsymbol u_k^\star
    =
    \arg\min_{\boldsymbol u_k}\quad&
        \sum_{\ell=0}^{N-1}
        \ell_x(x_{\ell|k},u_{\ell|k})
        +
        V_f(x_{N|k})
        \\
    \mathrm{subject\ to}\quad&
        x_{\ell+1|k}
        =
        F(x_{\ell|k},u_{\ell|k}),
        \qquad
        x_{\ell|k}\in\mathcal X_{\mathrm{safe}},
        \qquad
        u_{\ell|k}\in\mathcal U,
        \qquad
        x_{N|k}\in\mathcal X_f ,
\end{aligned}
\label{eq:temporal_mpc}
\end{equation}
and accommodates energy, economic, tracking, and effort objectives \citep{mayne2000mpc}. The prediction horizon $N$ is not a convergence-time guarantee: a finite-horizon optimization yields finite-, fixed-, or prescribed-time convergence only when the terminal set, terminal cost, recursive feasibility, and decrease conditions establish that property. Otherwise MPC supplies constraint handling and optimal input selection without a settling-time certificate. Distributed MPC with recursive feasibility under Markovian topology switching establishes asymptotic, not finite-time, platoon stability \citep{He2026MarkovianDMPCPlatooning}, and an iteration-convergence threshold in rolling distributed optimization is not a closed-loop stability result \citep{Gao2026MEVPPDMPC}.

Prescribed-performance control constrains the error directly through an envelope
\begin{equation}
    -\underline{\rho}_i(t)
    <
    e_i(t)
    <
    \overline{\rho}_i(t),
    \qquad
    i=1,\ldots,n_e,
    \label{eq:prescribed_performance_envelope}
\end{equation}
whose functions fix the admissible overshoot, transient decay, and terminal accuracy; an error transformation or a barrier Lyapunov function turns envelope violation into an unbounded transformed error \citep{bechlioulis2008robust,ma2022prescribedPerformance}. If the envelope reaches its terminal width at an assigned instant \citep{Stihi2026FTPPFTVSMC,Lv2026PredefinedLFCFOSMC,Luo2026HiTLUAVUSV} it supports a prescribed-time performance statement, but staying inside a shrinking envelope is equivalent to exact convergence only if the terminal width is zero and the resulting input is admissible; a nonzero terminal width makes the result practical or an envelope result (codes P and U in Table~\ref{tab:audit_full}), as the audited designs \citep{su2025second,peng2024antisaturation} of Sections~\ref{subsubsec:smc} and~\ref{subsubsec:recursive} show. Shifted or transformed envelopes remove the requirement that the initial error lie inside the envelope \citep{Xu2026AdaptivePrescribedPerformanceMAS,Zhu2026OptimizedSlidingModePsTFormation}, bounded performance-rate functions avoid terminal blow-up at the price of practical entry followed by asymptotic convergence \citep{Golestani2026PracticalPrescribedTimePPC}, and a finite-time envelope may coexist with only asymptotic state convergence \citep{Yao2026FlexibleManipulatorsPrescribedPerformance}. Funnel control imposes $\varphi(t)\|e(t)\|<1$ with a gain that grows as the error approaches the boundary and needs no reference model; input saturation can break the invariance of an aggressively shrinking funnel, and input-constrained funnel control restores feasibility by widening the boundary when the required input would leave $\mathcal U$ \citep{ilchmann2009pi,berger2024funnel}, so the relaxed funnel must not be reported as the original guarantee. When approximation errors, disturbances, sampling, or saturation prevent exact regulation, the appropriate statement is entry into an explicit accuracy region,
\begin{equation}
    e(t)\in\mathcal B_{\varepsilon}
    :=
    \left\{
        e:\|e\|\leq\varepsilon
    \right\},
    \qquad
    t\geq T_c ,
    \label{eq:practical_accuracy_region}
\end{equation}
with $\varepsilon$ related to the disturbance, the approximation error, the sampling period, and the authority \citep{cao2022practical,li2023adaptive}; a practical predefined-time result requires both $T_c$ and $\varepsilon$, since quoting the entry time alone conceals the accuracy lost to the constraint-handling mechanism.

Most of the constrained designs examined here are coded in Table~\ref{tab:audit_full}, grouped by temporal class; three boundaries are marked here. Fixed-time CLF-CBF programming applied to adaptive cruise control and multi-robot motion planning \citep{garg2022fixedqp} is the realization in which the horizon, the initial set, and the authority are related in one feasibility statement. The teleoperation design of \citet{li2023adaptive}, with simultaneous input saturation, uncertain dynamics, and output-error constraints, is explicitly practical: the error reaches an adjustable neighborhood within the selected time. The prescribed-time safety design of \citet{abel2024prescribedSafety} prescribes the time at which a safe-set boundary is reached, which is a different temporal claim from the time at which a regulation error vanishes, and the two should not be presented interchangeably; high-order finite-time neural barrier functions extend finite-time safe-set recovery to constraints of arbitrary relative degree, with exact tracking only asymptotic once safety is enforced \citep{Wang2026HFNCBFRoboticSafeTracking}. Time-scale transformation of an asymptotically safe potential-field controller gives prescribed-time spacecraft proximity tracking with collision avoidance and visibility maintenance; the ideal law is singular at $T$, its continuous truncation is practical, and arrival presumes that the potential field has no local minimum \citep{Shao2026PsTSpacecraftSafety}. Finite-time composite learning with time-varying asymmetric output constraints reports exact finite-time convergence of both the tracking and parameter-estimation errors \citep{zhang2023finiteRobot}. The remaining designs follow the pattern described above: finite-time control of saturated systems with tracking-error constraints and antisaturation dynamics \citep{sun2021finiteSaturation} carries residual terms from approximation or compensation, and the reinforcement-learning unmanned surface vehicle (USV) controller cited in Section~\ref{subsubsec:adaptive_learning} \citep{wang2022reinforcement} improves performance without certifying a convergence time; the current-constrained fixed-time converter controller of \citet{song2026fixedConverter} shows experimentally that independence from the initial condition does not remove dependence on the operating region; predefined-time designs with deferred full-state constraints \citep{chen2025predefinedMobile}, global output constraints in networked USV formation \citep{zhang2025predefinedUSV}, and command filtering with preassigned performance metrics \citep{liu2024commandFilter} must be read with their residual sets; prescribed-time tracking with a transient envelope \citep{ning2024prescribedMobile} and the USV design with an extended-state observer (ESO), envelope, and saturation audited in Section~\ref{subsubsec:observer_output} \citep{sui2025usv} act continuously through the error transformation or barrier term, whereas CBF safety filtering of a prescribed-time mobile-robot controller \citep{nasab2025safeMobile} modifies the temporal input only when the safe set would be violated.

The roles are summarized as
\begin{equation*}
\begin{aligned}
    \text{CLF-CBF/QP}
    &:\quad
    \text{temporal decrease}
    +
    \text{set invariance}
    \\
    &\qquad
    +\text{instantaneous input selection},
    \\
    \text{MPC}
    &:\quad
    \text{prediction}
    +
    \text{constraint handling}
    \\
    &\qquad
    +
    \text{finite-horizon optimization},
    \\
    \text{prescribed performance/funnel}
    &:\quad
    \text{time-varying error admissibility},
    \\
    \text{practical accuracy region}
    &:\quad
    \text{finite-horizon entry into }
    \mathcal B_{\varepsilon}.
\end{aligned}
\end{equation*}
A constraint-aware temporal result should report which constraints are hard and which relaxable, the initial set over which the optimization is feasible, the required input magnitude and update rate, whether the target is the origin, a safe set, an envelope, or a residual region, and whether the horizon is finite, uniform, selectable, or assigned.

\subsubsection{Actuator-Constrained and Regularized Realizations}
\label{subsubsec:actuation_regularization}

Actuator constraints act at the interface between the computed command $u_c$ and the applied input $u_a$. A representative model is
\begin{equation}
    u_a
    =
    \operatorname{sat}_{\mathcal U}(u_c),
    \qquad
    \|\dot u_a\|\leq r_{\max},
    \qquad
    \Delta_u:=u_a-u_c,
    \label{eq:actuator_limited_input}
\end{equation}
with amplitude set $\mathcal U$, rate limit $r_{\max}$, and saturation deficiency $\Delta_u$; dead zones, backlash, hysteresis, quantization, and faults produce related discrepancies. These effects matter for every class of Section~\ref{sec:notions} because shortening the horizon raises the transient authority required (Section~\ref{subsec:feasible_decay}). Clipping an unconstrained controller does not preserve its proof: if the analysis is carried out with $u_c$ while the plant receives $u_a=u_c+\Delta_u$, the derivative becomes
\begin{equation}
    \dot V
    \leq
    -\Psi(V,t)
    +
    \left\langle\nabla V,g(x)\Delta_u\right\rangle ,
    \label{eq:saturation_lyapunov_mismatch}
\end{equation}
an instance of \eqref{eq:general_decay_sec3} with $\Delta=\langle\nabla V,g\Delta_u\rangle$, to which Section~\ref{subsec:specification_to_design} applies. For the scalar integrator of Section~\ref{sec:numerical}, $|x(t)|\geq|x_0|-u_{\max}t$ makes global fixed-time stabilization on an unbounded initial set impossible under a fixed input bound. This conclusion is specific to the integrator and analogous bounded-speed settings; an input bound alone does not establish impossibility for every nonlinear plant. For the benchmark tuning $x_0=T_c=1$, $k_f=2$, and $\varepsilon=10^{-3}$, the saturated finite-time law needs $u_{\max}\gtrsim1.5563$ to meet the tolerance deadline. That is a limit of this calibrated law, not the minimum attainable over its entire feedback family.

Four realizations are distinguished. \emph{Bounded-by-design feedback} uses bounded homogeneous terms, nested saturation, or smooth bounded functions so that $u_c\in\mathcal U$ by construction; bounded finite-time sliding-mode feedback for saturated linear systems \citep{ma2025boundedFinite} and experimentally validated finite-time practical tracking of nonholonomic systems under input and state constraints \citep{liu2023nonholonomic} are examples, and their regional or practical qualifications must be read from the corresponding plant and admissibility assumptions. Bounded input does not, by itself, rule out global finite-time stability with an initial-condition-dependent time. \emph{Anti-windup augmentation} introduces an auxiliary state driven by the unavailable part of the command,
\begin{equation}
    \dot z
    =
    A_z z+B_z\Delta_u,
    \qquad
    u_c
    =
    u_{\mathrm{nom}}+K_z z,
    \label{eq:antisaturation_auxiliary}
\end{equation}
which modifies the law until the actuator returns to its admissible regime; the Lyapunov function must then contain both the plant error and $z$, otherwise convergence of the nominal subsystem says nothing about the implemented loop. This is the structure of the antisaturation systems audited in Section~\ref{subsubsec:recursive} \citep{peng2024antisaturation,hao2026predefined}, of predefined-time tracking of high-order systems \citep{gong2022predefinedSaturation} and actor-critic predefined-time control \citep{yang2024actorCriticPredefined} under saturation; fixed-time anti-windup sliding-mode control implemented on piezoelectric stages with saturation and hysteresis \citep{zhou2025fixedAntiwindup} adds the compensator state but certifies it separately from the tracking error, as noted below. \emph{Performance relaxation} enlarges the envelope temporarily according to a saturation signal and recovers it when authority returns \citep{ji2022saturationTolerant,ji2023saturation}; this preserves boundedness and avoids envelope violation but does not preserve the original envelope, and practical prescribed-time tracking under saturation guarantees $\|e(t)\|\le\varepsilon$ for $t\ge T_c$ rather than exact arrival \citep{lin2025practicalPrescribed}, so both $T_c$ and $\varepsilon$ must be reported. Under impulsive actuation with saturation compensation the residual radius grows explicitly with the saturation level \citep{Kim2026EventImpulsivePrescribedTimeUUV}. \emph{Regularization} addresses switching rather than amplitude: the discontinuous term $-k\operatorname{sgn}(s)$ of a sliding-mode or homogeneous law is replaced by $-k\operatorname{sat}(s/\phi)$, which removes high-frequency switching but replaces exact sliding by convergence to a $\phi$-dependent neighborhood, so that a finite- or prescribed-time proof for the discontinuous law does not transfer to its smooth implementation; this is the remedy deferred here from Section~\ref{subsubsec:smc}. Higher-order sliding modes generate a continuous applied input while retaining finite-time sliding under matching and disturbance-regularity assumptions \citep{levant1993sliding}, but sampling and saturation then create a new closed loop, and the implicit discretization of the super-twisting controller, with a conditioned version for actuator saturation, has been developed to suppress numerical chattering and windup while retaining explicit discrete-time properties \citep{seeber2025implicitSTA}.

The audited designs are listed in Table~\ref{tab:audit_full}; two mark the boundaries. The piezoelectric-stage controller of \citet{zhou2025fixedAntiwindup} marks the limit of an anti-windup certificate in hardware: its fixed-time result is practical and concerns only the compensator state, whereas the tracking error converges only asymptotically, so the temporal claim does not transfer to the regulated variable. The time-base-generator design of \citet{lin2025practicalPrescribed} is the clearest instance of a prescribed-time claim that saturation converts into a practical one with an explicit $(T_c,\varepsilon)$ pair, the actuator-level counterpart of what the continuation of the benchmark law shows under saturation in Section~\ref{subsec:numerical_constraints}.

An actuator-aware certificate states the magnitude and rate limits, the duration of possible saturation, and whether the guarantee applies to $u_c$ or to $u_a$. A derived input bound is meaningful only if it lies inside the physical actuator set; otherwise the claim is weakened to regional or practical convergence, or the horizon is restricted by a feasibility bound such as \eqref{eq:feasibility_necessary}.

\subsubsection{Sampled-Data, Event-Triggered, and Hybrid Implementations}
\label{subsubsec:digital_implementation}

Digital implementation turns a continuously evaluated law into a system with discrete measurements, held inputs, and possibly state-dependent update instants, so the implemented loop is not the continuous-time model of the proof. Under periodic sampling with a zero-order hold,
\begin{equation}
\begin{aligned}
    u_a(t)&=\kappa\!\left(x(t_k),t_k\right),
    \qquad t\in[t_k,t_{k+1}),\\
    e_s(t)&=x(t_k)-x(t),
    \qquad
    h_k=t_{k+1}-t_k ,
\end{aligned}
\label{eq:sample_hold_implementation}
\end{equation}
where $e_s$ is the intersample error, and the sampled implementation of a law satisfying $\dot V\le-\Psi(V,t)$ yields
\begin{equation}
    \dot V
    \leq
    -\Psi(V,t)
    +
    \Gamma\!\left(x,e_s,h_k\right),
    \label{eq:sampled_lyapunov_error}
\end{equation}
again an instance of \eqref{eq:general_decay_sec3}. Stability requires an explicit bound on the admissible sampling interval, not the assumption that the rate is high enough \citep{nesic2004sampledFramework}. For the classes of this review the bound is sharper than for asymptotic designs, because the terms that produce finite settling are exactly the ones a held update handles worst: Section~\ref{sec:numerical} identifies the sublinear two-cycle scale $(hk/2)^{1/(1-p)}$, the large-state contraction restriction for the full mixed-power update, and that the pure prescribed-time gain loses contraction on the last $\eta h/2$ of its horizon unless it is capped below $2/h$. Three approaches must be kept apart: emulation of a continuous-time law (what Section~\ref{sec:numerical} tests), direct design on the sampled-data model, and consistent or implicit discretization that preserves the continuous-time class \citep{polyakov2019consistent,jimenez2020consistent,polyakov2023consistent,seeber2025implicitSTA}.

Sampling does not exclude exact finite-time convergence, but such a result requires a controller designed for the sampled-data model: finite-time sampled-data output-feedback stabilization has been established for feedforward nonholonomic systems by building observer and stabilizer around the sampled measurements \citep{mao2023finiteSampled}, whereas approximation errors, input delays, and held measurements more commonly yield practical results, as in neural sampled-data designs whose residual set and settling bound depend on the design and sampling parameters \citep{mao2025fixedSampled}. Event-triggered implementation replaces periodic updates with
\begin{equation}
    t_{k+1}
    =
    \inf
    \left\{
        t>t_k:
        \mathcal G\!\left(x(t),e_s(t),\eta_e(t),t\right)\geq0
    \right\},
    \label{eq:event_trigger_general}
\end{equation}
with an optional dynamic threshold state $\eta_e$; static triggers use an algebraic threshold, dynamic triggers an internal state, and self-triggered implementations compute $t_{k+1}$ at $t_k$ from a model-based prediction \citep{heemels2012eventTriggered}. The event-induced error must enter the temporal analysis through a budget of the form
\begin{equation}
    \Gamma_e(e_s,x,t)
    \leq
    \sigma\Psi(V,t)+\varepsilon_e(t),
    \qquad 0<\sigma<1,
    \label{eq:event_error_budget}
\end{equation}
which leaves the fraction $1-\sigma$ of the decay for convergence and, when $\varepsilon_e\neq0$, yields practical convergence. If the threshold shrinks too fast as $t\to T_c$ the events accumulate near the horizon, so preserving the convergence time is not enough: a strictly positive minimum inter-event time must be established. Periodic event-triggered control evaluates the trigger only on a grid $t_k\in\{0,h,2h,\ldots\}$, removing continuous monitoring and bounding the inter-event time below by $h$, but the admissible $h$ must still come from the closed-loop analysis. Sampled-data and event-triggered loops are naturally hybrid,
\begin{equation}
    \begin{cases}
        \dot \xi=F(\xi,t),
        & \xi\in\mathcal C,\\[1mm]
        \xi^+=G(\xi,t),
        & \xi\in\mathcal D,
    \end{cases}
    \label{eq:hybrid_digital_model}
\end{equation}
with $\xi$ collecting plant state, held measurements, controller states, timers, and trigger variables, and hybrid-system analysis establishes stability, completeness of solutions, inter-event bounds, and robustness in one model \citep{postoyan2015eventFramework}. The distinction between physical time and jump count must remain explicit: convergence after finitely many jumps is not a finite physical settling time, convergence by a prescribed physical time is not implementable if it requires infinitely many jumps before that time, and the regulated state must remain invariant after convergence under the held input and the subsequent jumps. Under denial-of-service attacks the wall-clock bound depends on a finite-prefix bound for accumulated attack-free time, including any additive outage allowance; a factor $1/(1-\rho)$ alone is justified only under the corresponding zero-offset condition when decay is suspended during attacks \citep{Yang2026FixedTimeMicrogridDoS}, or the prescribed-time clock advances only in attack-free time \citep{Zhang2026DCMicrogridPT}; exact prescribed-time consensus under DoS and delay rests on piecewise or delay-dependent scaling inequalities \citep{Zhang2026PrescribedTimeConsensusDelayDoS,Zhu2026PrescribedTimeDelayDoS} (Table~\ref{tab:temporal_catalogue}).

Fixed-time event- and self-triggered consensus \citep{liu2020fixedEvent} illustrates the dual requirement of an initial-condition-independent bound and a triggering law without accumulation. Predefined-time event-triggered tracking for strict-feedback systems with full-state constraints \citep{wu2024predefinedEvent} reaches a prescribed neighborhood within the selected time and is therefore practical. Prescribed-time event-triggered control couples the time-varying gain to the threshold: \citet{ning2023eventPrescribed} regulates uncertain nonlinear systems within a selected interval while excluding Zeno behavior, and the adaptive extension \citep{ning2023adaptiveEventPrescribed} characterizes the relation among trigger parameters, event intervals, and prescribed-time performance. Exact event-triggered prescribed-time stabilization with unknown control direction uses a threshold term that vanishes at $t_f$, so that the state and the held input are zero afterwards \citep{Li2026EventTriggeredUnknownCoefficientsPsT}, and dynamic self-triggering ties the execution interval to the scaling function and the estimated actuator effectiveness while proving a positive minimum interval \citep{Zhang2026DynamicSelfTriggeredBipartitePsT}. These results show that updates can be reduced without abandoning the horizon, but only when the event-induced error is inside the certificate. The command-filtered design \citep{zhang2025command} and the helicopter controller \citep{song2026rl} audited in Sections~\ref{subsubsec:recursive} and~\ref{subsubsec:adaptive_learning} are the event-triggered members of the audited set with hardware and numerical validation, respectively.

The principal digital realizations are periodic sampling ($t_{k+1}=t_k+h_k$), event triggering ($t_{k+1}=\inf\{t>t_k:\mathcal G(t)\ge0\}$), self triggering ($t_{k+1}=t_k+\tau(x(t_k))$), and hybrid implementation (flows and jumps). A credible temporal certificate states the allowable sampling period and jitter, the behavior between samples, the minimum inter-event time, whether continuous trigger monitoring is required, whether convergence is exact or practical, and whether the bound is expressed in physical time, discrete steps, or hybrid time.

\subsection{Architectural Synthesis and Design Guidance}
\label{subsec:design_guidance}

\paragraph{Evaluation criteria.}
\label{subsec:comparison_criteria}
The studies examined in Sections~\ref{subsec:primary_architectures}-\ref{subsec:implementation_architectures} are compared along three dimensions, each posed as a question about the certificate rather than about the reported class.
\begin{description}
\item[Established result.] For which variable or set, at what accuracy, and with which temporal guarantee: an initial-condition-dependent settling time; a uniform, assignable, or tight bound; a terminal limit; or exact first arrival. Exact convergence is distinguished from entry into a stated residual set, together with retention after entry.
\item[Conditions of validity.] Under which initial set (local, semiglobal, or global), uncertainty and noise assumptions, and auxiliary dynamics: which observer, adaptation, filter, learning, or dynamic-gain variables enter the certificate, and whether they are proved bounded or convergent. A design parameter $T_c$ alone does not make an autonomous closed loop time dependent.
\item[Implementation and evidence.] Whether input amplitude, rate, and internal-signal admissibility, sampling, and the continuation at and beyond $T_c$ or $T_p$ are covered, a bounded physical input does not imply bounded scaling variables, nor does an unbounded scaling gain imply an unbounded input, and at which level the claim is supported: analytical certificate, numerical illustration, hardware-in-the-loop validation, or physical experiment.
\end{description}
Each study is coded from the strongest property its theorems prove for the certified variable, which is the regulated variable unless stated otherwise, not from its title or conclusions; when two variables are certified with different guarantees, both codes are given. Numerical entry into a tolerance region is not identified with exact settling, and fast recovery or stable decisions reported without a theorem, even under real-time validation, are observations rather than certificates \citep{Su2026OffshoreWindDCC,Gao2026MemristiveDecisionNavigation}. The same distinctions are used in the numerical comparison of Section~\ref{sec:numerical} and motivate Section~\ref{sec:open-problems}.

\paragraph{Audit.}

Table~\ref{tab:audit_full} in Appendix~\ref{app:audit} records, for the design studies examined in Sections~\ref{subsec:primary_architectures}-\ref{subsec:implementation_architectures} and for additional representative designs, the architecture that generates its physical input together with its augmentations, the temporal class reported by the source, the certified variable, the exact or practical nature of the guarantee, the auxiliary dynamics that enter the certificate, and the type of validation reported; Table~\ref{tab:audit} reproduces the entries that mark the boundaries discussed in those sections. The tables record a compact subset of these criteria, temporal class, certified variable, guarantee, auxiliary dynamics, and validation, while the discussion of Sections~\ref{subsec:primary_architectures}-\ref{subsec:implementation_architectures} addresses domain, uncertainty, admissibility, and continuation. Source terminology and mechanism-based interpretation are distinct: a work labeled predefined-time may employ the time-varying mechanism discussed under prescribed-time control. The source-class entry is therefore read together with the certified variable and guarantee, rather than used as an independent certification. The audited studies form a representative comparative set, not a population estimate or a systematic-review sample, and each design is placed in one principal group although its mechanisms and augmentations may overlap. Within this set, hardware and hardware-in-the-loop validation cluster in fast-sampled electromechanical platforms, electrical drives, power converters, magnetic levitation, and precision positioning, and are reported less often for distributed and for constraint- and implementation-aware designs. This is a property of the examined set, not of the architectures; it matters because communication effects and constraint handling act directly on the auxiliary dynamics that the corresponding certificates must contain.

\noindent Boundary cases from the architectural audit (complete audit: Table~\ref{tab:audit_full} in Appendix~\ref{app:audit}), coded by the criteria of Section~\ref{subsec:comparison_criteria}. Selected entries from each group mark how far the certificate of each realization extends. \emph{Source class}: FT finite-, FxT fixed-, PdT predefined-, PT prescribed-time. \emph{G} (guarantee): E exact for the certified variable; E\textsubscript{r} exact for the regulated variable, auxiliary errors only bounded or residual; P practical (entry into a residual set); P$^\ast$ practical with a qualified horizon (see text); U uniform ultimate boundedness or envelope result; E$^\dagger$ exact for a safety target. \emph{Aux.}: auxiliary dynamics in the certificate (DO disturbance observer, Obs observer, Ad adaptation, Cr actor-critic, AW antisaturation state, Trg trigger, Sw switching, PPF performance transformation); the remaining codes are defined in Appendix~\ref{app:audit}. \emph{V}: N numerical, HIL hardware-in-the-loop, Hw physical experiment. Symbols: $e$ tracking error, $x$ plant state, $s$ sliding variable, $\tilde x$ estimation error.

\setlength{\tabcolsep}{2.4pt}\renewcommand{\arraystretch}{1.10}\fontsize{9}{11}\selectfont
\begin{longtable}{@{}>{\raggedright\arraybackslash}p{2.25cm} >{\raggedright\arraybackslash}p{4.7cm} l >{\raggedright\arraybackslash}p{2.6cm} l >{\raggedright\arraybackslash}p{2.05cm} l@{}}
\caption{Boundary cases from the architectural audit.}\label{tab:audit}\\
\toprule \textbf{Study} & \textbf{Realization (primary + augmentation; platform)} & \textbf{Source class} & \textbf{Certified variable} & \textbf{G} & \textbf{Aux.} & \textbf{V}\\ \midrule\endfirsthead
\multicolumn{7}{@{}l}{\textit{Table~\ref{tab:audit} (continued)}}\\ \toprule \textbf{Study} & \textbf{Realization (primary + augmentation; platform)} & \textbf{Source class} & \textbf{Certified variable} & \textbf{G} & \textbf{Aux.} & \textbf{V}\\ \midrule\endhead
\midrule\multicolumn{7}{r@{}}{\textit{continued on next page}}\\ \endfoot \bottomrule\endlastfoot
\multicolumn{7}{@{}l}{\textbf{Direct state feedback (Section \ref{subsubsec:direct_feedback})}}\\
\citealp{nezhad2026finite} & Robust voltage-based SF; flexible-joint manipulator & FT & $e$ & E & -- & HIL\\
\midrule
\multicolumn{7}{@{}l}{\textbf{Sliding-mode realizations (Section \ref{subsubsec:smc})}}\\
\citealp{shi2024prescribed} & Nonsingular TSM, bounded TV gains; PMSM & PT & reaching + sliding phases & E & -- & Hw\\
\citealp{yin2025nonsingular} & Nonsingular fast SMC + DO; construction robot & PdT & $x$ ($\tilde x$ uncertified) & E\textsubscript{r} & DO & N\\
\citealp{ding2026switching} & Switching adaptive second-order SMC & PT & $s$, $x$ & P$^\ast$ & Ad, Sw & N\\
\midrule
\multicolumn{7}{@{}l}{\textbf{Backstepping realizations (Section \ref{subsubsec:recursive})}}\\
\citealp{zeng2025backstepping} & Composite backstepping; two-time-scale systems & PT & slow + fast states, $u$ & E & TV scaling & N\\
\citealp{peng2024antisaturation} & Antisaturation backstepping + BLF + PPF; slung load & FxT & variables in envelope & U & AW, BLF, PPF & N\\
\midrule
\multicolumn{7}{@{}l}{\textbf{Observer-based and output feedback (Section \ref{subsubsec:observer_output})}}\\
\citealp{you2025visual} & DO-based control, visual feedback; bolt tightening & FT & $e$ & E\textsubscript{r} & DO & Hw\\
\citealp{tan2026decentralized} & Decentralized OF, local outputs; interconnected sys. & PT & $x$, $\tilde x$, $u$ & E & Obs & N\\
\midrule
\multicolumn{7}{@{}l}{\textbf{Adaptive and learning-based (Section \ref{subsubsec:adaptive_learning})}}\\
\citealp{wang2023maglev} & Adaptive position control; magnetic levitation & FxT & position ($\tilde d$ residual) & E\textsubscript{r} & DO & Hw\\
\citealp{song2026rl} & RL optimized + PdT DO + DSC + ETC; 2-DOF helicopter & PdT & composite errors & P & Cr, DO, Trg & N\\
\midrule
\multicolumn{7}{@{}l}{\textbf{Distributed and cooperative (Section \ref{subsubsec:distributed})}}\\
\citealp{chen2024prescribed} & Coop.\ output regulation + PT exosystem observer & PT & outputs, $\tilde x$ & E & Obs & N\\
\citealp{Zhou2026PracticalPsTRigidFormation} & Frozen time-varying scaling + two-power law + distance PPF; rigid formation & PT & distance errors & P & PPF & Hw\\
\midrule
\multicolumn{7}{@{}l}{\textbf{Constraint- and implementation-aware (Section \ref{subsec:implementation_architectures})}}\\
\citealp{abel2024prescribedSafety} & Safety design; strict-feedback systems & PT & distance to safe set & E$^\dagger$ & -- & N\\
\citealp{lin2025practicalPrescribed} & Time-base generator, saturation; nonaffine & PT & output error & P & -- & N\\
\citealp{ning2023eventPrescribed} & Event-triggered control; uncertain nonlinear & PT & $x$ & E & Trg & N\\
\end{longtable}\normalsize\setlength{\tabcolsep}{6pt}\renewcommand{\arraystretch}{1.0}
\normalsize
\setlength{\tabcolsep}{6pt}
\renewcommand{\arraystretch}{1.0}


\paragraph{Conditional design guidance.}
The audit supports a conditional rather than prescriptive guide, given in Table~\ref{tab:architectural_design_guidance}. It is a starting point, not a ranking: the appropriate realization is the one that preserves the required temporal guarantee for the stated variable under the available measurements, uncertainty, constraints, and implementation resources.

Three decisions precede the choice of a realization. The first is the temporal requirement itself: an initial-condition-dependent settling time, a uniform bound, a bound selected through the parameters, or a horizon that enters the law directly, in the sense of Section~\ref{sec:notions}; this choice should be made before the observer, sliding-mode, backstepping, learning, or distributed realization is selected. The second is the target of the guarantee: the reported settling time must correspond to the variable in the problem statement, not to an intermediate signal, which is the recurring question of Sections~\ref{subsec:primary_architectures} and~\ref{subsec:augmented_architectures}. The third is feasibility: a shorter horizon requires more authority, faster updates, or larger transient gains, and when the requested horizon cannot be met over the stated initial set, increasing a formal gain parameter does not resolve the conflict. The admissible claim is then reformulated as
\begin{equation}
\begin{aligned}
    \text{exact}
    &\rightarrow
    \text{practical},\\
    \text{global}
    &\rightarrow
    \text{regional},\\
    \text{arbitrarily assignable horizon}
    &\rightarrow
    \text{horizon satisfying }T_c\geq T_{\min},\\
    \text{nominal guarantee}
    &\rightarrow
    \text{robust guarantee with an explicit margin},
\end{aligned}
\label{eq:claim_calibration}
\end{equation}
with $T_{\min}$ as in \eqref{eq:min_feasible_time}. The theorem that certifies an implementable design therefore concerns the complete augmented system, reported with the criteria of Section~\ref{subsec:comparison_criteria}, and when preservation of the nominal guarantee cannot be proved, the guarantee is weakened explicitly rather than inferred from the ideal continuous-time model.

\noindent Conditional guidance for selecting a convergence-time architecture. The required certificate follows the criteria of Section~\ref{subsec:comparison_criteria} and the architecture-specific questions of Sections~\ref{subsec:primary_architectures}-\ref{subsec:implementation_architectures}.

{
\footnotesize
\setlength{\tabcolsep}{4pt}
\renewcommand{\arraystretch}{1.15}
\begin{xltabular}{\textwidth}{@{}>{\raggedright\arraybackslash}p{0.22\textwidth} >{\raggedright\arraybackslash}p{0.28\textwidth} >{\raggedright\arraybackslash}X@{}}
\caption{Design choices and certificate requirements.}\label{tab:architectural_design_guidance}\\
\toprule
\textbf{Design condition} & \textbf{Natural starting point} & \textbf{Required certificate}\\
\midrule
\endfirsthead
\multicolumn{3}{@{}l}{\footnotesize\textit{Table~\ref{tab:architectural_design_guidance} (continued)}}\\
\toprule
\textbf{Design condition} & \textbf{Natural starting point} & \textbf{Required certificate}\\
\midrule
\endhead
\midrule\multicolumn{3}{r@{}}{\footnotesize\textit{continued on next page}}\\
\endfoot
\bottomrule\endlastfoot

Full state available, model sufficiently known & Direct state feedback (Section~\ref{subsubsec:direct_feedback}) & Temporal Lyapunov inequality for the plant state or tracking error; admissibility $u_c\in\mathcal U$ over the operating domain\\
Matched uncertainty, strong robustness requirement & Sliding-mode or higher-order sliding-mode realization (Section~\ref{subsubsec:smc}) & Reaching bound, control of $|e(t_r)|$, manifold invariance, reduced-order convergence, auxiliary controller states, behavior under regularization and sampling\\
Strict-feedback or recursive structure & Backstepping, dynamic-surface, or command-filtered realization (Section~\ref{subsubsec:recursive}) & Target inequality for the composite $V_n$ (stagewise conversion stated), filter and differentiator errors in the certificate; additive time budgets only for genuinely sequential subsystems\\
Only output measurements & Observer-based or dynamic output feedback (Section~\ref{subsubsec:observer_output}) & Property proved for the estimation error (convergence, boundedness, or set entry) and its sufficiency for the regulated variable; peaking and noise under a short estimation horizon\\
Unknown nonlinearities or parameters & Adaptive, neural, fuzzy, or learning-assisted realization (Section~\ref{subsubsec:adaptive_learning}) & Regulated variable named; excitation condition; approximation residual in the certificate; exact versus practical stated; learned performance not substituted for a temporal certificate\\
Local neighbor information only & Distributed or cooperative realization (Section~\ref{subsubsec:distributed}) & Graph assumption; dependence of bound and gains on topology and size; distributed versus fully distributed; communication assumptions separated\\
Hard safety, state, or performance constraints & CLF-CBF/QP, MPC, barrier, prescribed-performance, or funnel realization (Section~\ref{subsubsec:safety_performance}) & Constraint feasibility over the certified initial set; compatibility of the horizon with authority and safe-set geometry; terminal envelope width\\
Input saturation, rate limits, actuator nonlinearities & Bounded-by-design feedback or anti-windup augmentation (Section~\ref{subsubsec:actuation_regularization}) & Bound for the applied input $u_a$ inside the actuator set; saturation-induced dynamics and auxiliary state in the certificate; feasibility condition such as \eqref{eq:feasibility_necessary}\\
Digital or networked implementation & Sampled-data, event-triggered, self-triggered, or hybrid realization (Section~\ref{subsubsec:digital_implementation}) & Maximum admissible sampling interval and jitter; intersample behavior; minimum inter-event time; physical versus hybrid time\\
\end{xltabular}
}

\section{Illustrative Numerical Comparisons}
\label{sec:numerical}

This section uses a common scalar benchmark to show how the distinctions of Sections~\ref{sec:notions}-\ref{sec:architectures} appear in measurable closed-loop quantities, under the criteria of Section~\ref{subsec:comparison_criteria}: how the certified guarantees appear numerically, how the observed convergence time depends on the initial condition, what control effort comparable temporal performance requires, and which guarantees survive actuator saturation and sampling. The comparison concerns explicit laws and calibrations; it does not rank the convergence-time classes or identify a universally superior controller, and every outcome below depends on stated choices of calibration, initial set, accuracy, authority, and sampling rate.

The plant is a scalar integrator, so every controller coincides with the decay law it realizes and the comparison is confined to the mechanism layer of Fig.~\ref{fig:three_layers}; reaching transients and augmented states (Sections~\ref{subsubsec:smc} and~\ref{subsec:augmented_architectures}) are not illustrated. All nominal quantities admit closed forms or one-dimensional quadratures, which separates certificate conservatism from numerical tolerance, and the fixed- and predefined-time laws belong to one two-power family (Section~\ref{subsec:predefined_time_stability}). Section~\ref{subsec:benchmark_protocol} fixes the protocol, Section~\ref{subsec:numerical_time} reports the nominal results, and Section~\ref{subsec:numerical_constraints} treats effort, saturation, and sampling.

\subsection{Benchmark System and Common Evaluation Conditions}
\label{subsec:benchmark_protocol}

\paragraph{Benchmark system.}
The plant is the scalar integrator
\begin{equation}
    \dot{x}(t)=u_a(t)+d(t),
    \qquad
    x(0)=x_0,
    \label{eq:numerical_benchmark}
\end{equation}
where $u_a$ is the input actually applied to the plant and $d$ is a disturbance. In the nominal experiments $d\equiv0$ and $u_a=u_c$, where $u_c$ is the command generated by the controller. In the constrained experiments the command is sampled with period $h$, held constant between sampling instants $t_k=kh$, and saturated at the actuator,
\begin{equation}
    u_a(t)
    =
    \operatorname{sat}_{u_{\max}}
    \bigl(u_c(t_k)\bigr),
    \qquad
    t\in[t_k,t_{k+1}),
    \label{eq:numerical_implemented_input}
\end{equation}
with $\operatorname{sat}_{u_{\max}}(v)=\min\{|v|,u_{\max}\}\operatorname{sgn}(v)$; $u_{\max}=\infty$ and $h\to0$ recover the unsaturated and continuous-time cases. The plant is integrated exactly between samples, so the constrained experiments are emulation experiments (Section~\ref{subsubsec:digital_implementation}); consistent discrete-time redesign is discussed in Section~\ref{subsec:open_digital}.

\paragraph{Representative control laws.}
One canonical law is used for each class, together with a linear law as an asymptotic baseline, with $\lceil x\rceil^{\rho}=|x|^{\rho}\operatorname{sgn}(x)$:
\begin{equation}
\begin{aligned}
    u_{\mathrm{AS}}
    &=
    -k_a x,
    &&k_a>0,
    \\
    u_{\mathrm{FT}}
    &=
    -k_f\lceil x\rceil^{\alpha},
    &&0<\alpha<1,
    \\
    u_{\mathrm{FxT}}
    &=
    -k_1\lceil x\rceil^{p}
    -k_2\lceil x\rceil^{q},
    &&0<p<1<q,
    \\
    u_{\mathrm{PdT}}
    &=
    -\frac{\gamma_{p,q}}{T_c}
    \left(
        \lceil x\rceil^{p}+\lceil x\rceil^{q}
    \right),
    \\
    u_{\mathrm{PT}}
    &=
    -\frac{\eta}{T_c-t}\,x,
    &&0\leq t<T_c,
\end{aligned}
\label{eq:benchmark_control_families}
\end{equation}
where the predefined-time normalization is
\begin{equation}
    \gamma_{p,q}
    :=
    \int_{0}^{\infty}
    \frac{ds}{s^{p}+s^{q}}
    =
    \frac{\pi}{q-p}
    \csc\left(
        \frac{\pi(1-p)}{q-p}
    \right).
    \label{eq:predefined_normalization}
\end{equation}
With $V=|x|$, the baseline produces the exponential decay $\dot V=-k_aV$, and the four class laws produce exactly the decay laws of Section~\ref{sec:mechanisms}: \eqref{eq:sec3_ft_decay} for $u_{\mathrm{FT}}$, \eqref{eq:sec33_dual_power} for $u_{\mathrm{FxT}}$, \eqref{eq:sec34_pdt_general} with $G_c=\gamma_{p,q}$ and $\Phi(v)=v^p+v^q$ for $u_{\mathrm{PdT}}$, and \eqref{eq:sec34_pt_lyap} with $\mu(t)=1/(T_c-t)$ for $u_{\mathrm{PT}}$. The corresponding certified bounds are
\begin{equation}
\begin{aligned}
    T^{\mathrm{cert}}_{\mathrm{FT}}(x_0)
    &=
    \frac{|x_0|^{1-\alpha}}{k_f(1-\alpha)},
    \qquad
    T^{\mathrm{cert}}_{\mathrm{FxT}}
    =
    \frac{1}{k_1(1-p)}+\frac{1}{k_2(q-1)},
    \\
    T^{\mathrm{cert}}_{\mathrm{PdT}}
    &=
    T_c,
    \qquad\qquad\qquad\;\;
    T^{\mathrm{cert}}_{\mathrm{PT}}
    =
    T_c,
\end{aligned}
\label{eq:benchmark_certified_bounds}
\end{equation}
from \eqref{eq17}, \eqref{eq20}, \eqref{eq:pdt_normalized_bound}, and \eqref{eq:pt_terminal}. The asymptotic law has no finite settling time; the only finite time it can certify is entry into the accuracy band, $T^{\mathrm{cert}}_{\mathrm{AS}}(x_0)=\ln(|x_0|/\varepsilon)/k_a$ for $|x_0|>\varepsilon$, which depends on $\varepsilon$ as well as on $x_0$ and is used as its deadline below. The fixed-time and predefined-time laws are not different dynamical families: $u_{\mathrm{PdT}}$ is the two-power law with $k_1=k_2=\gamma_{p,q}/T_c$, so their comparison concerns calibration and bound tightness rather than feedback structure. The prescribed-time law is defined only on $[0,T_c)$; the benchmark continues it with the linear law $u=-k_h x$ for $t\geq T_c$, which holds the origin nominally because $x(T_c)=0$, and records separately how the continuation behaves when $x(T_c)\neq0$ under constraints. A bounded-gain variant, $\min\{\eta/(T_c-t),\kappa_{\max}\}$, is included because a zero-order-hold implementation of the pure law contracts at sample $k$ only if $h\eta/(T_c-t_k)<2$, a condition that fails on the last $\eta h/2$ of the horizon; the cap $\kappa_{\max}<2/h$ removes this failure at the price of practical rather than exact terminal convergence (Section~\ref{subsubsec:time_varying_pt}).

\paragraph{Nominal closed forms.}
For $d\equiv0$, $u_{\max}=\infty$, and $h\to0$, the nominal trajectories and settling times are
\begin{equation}
\begin{aligned}
    \mathrm{FT}:\;&
    |x(t)|=\bigl(|x_0|^{1-\alpha}-k_f(1-\alpha)t\bigr)_+^{1/(1-\alpha)},
    &&
    T_{\mathrm{FT}}(x_0)=\frac{|x_0|^{1-\alpha}}{k_f(1-\alpha)},
    \\
    \mathrm{FxT}:\;&
    T_{\mathrm{FxT}}(x_0)=\int_0^{|x_0|}\frac{ds}{k_1s^{p}+k_2s^{q}},
    &&
    \sup_{x_0}T_{\mathrm{FxT}}(x_0)=\int_0^{\infty}\frac{ds}{k_1s^{p}+k_2s^{q}},
    \\
    \mathrm{PdT}:\;&
    T_{\mathrm{PdT}}(x_0)=\frac{T_c}{\gamma_{p,q}}\int_0^{|x_0|}\frac{ds}{s^{p}+s^{q}},
    &&
    \sup_{x_0}T_{\mathrm{PdT}}(x_0)=T_c,
    \\
    \mathrm{PT}:\;&
    x(t)=x_0\Bigl(1-\frac{t}{T_c}\Bigr)^{\eta},
    &&
    T_{\mathrm{PT}}(x_0)=T_c\;\;(x_0\neq0).
\end{aligned}
\label{eq:benchmark_nominal_solutions}
\end{equation}
The asymptotic law gives $x(t)=x_0e^{-k_at}$, so that $T^{\mathrm{exact}}_{\mathrm{AS}}=\infty$ for $x_0\neq0$ and its band-entry time grows as $\ln|x_0|$. The prescribed-time input is $u_{\mathrm{PT}}(t)=-\eta x_0(1-t/T_c)^{\eta-1}/T_c$: bounded and vanishing at $T_c$ for $\eta>1$, constant for $\eta=1$, and unbounded for $\eta<1$, with $\int_0^{T_c}u_{\mathrm{PT}}^2\,dt=\eta^2x_0^2/((2\eta-1)T_c)$ for $\eta>1/2$. The exponent $\eta$ therefore decides admissibility, a direct illustration of \eqref{eq:sec34_false_implication}: the gain diverges while the input does not. Three facts organize the nominal experiments. The finite-time certificate is tight. The fixed-time certificate is not: for $k_1=k_2=k$ and $(p,q)=(1/2,3/2)$ the actual supremum is $\pi/k$ whereas the certificate is $4/k$, a ratio of $4/\pi\approx1.27$ that measures the conservatism of \eqref{eq20} relative to the tight normalization \citep{aldana2019enhancing}. The predefined-time supremum $T_c$ is approached only as $|x_0|\to\infty$, so on any bounded initial set the observed maximum lies strictly below $T_c$.

\paragraph{Calibration protocols.}
A law tuned for a short observed time may be granted more authority than another, so two protocols are used, both referenced to a common initial condition $x_{\mathrm{ref}}$.

\emph{Certificate-matched calibration.} The certified bound of each law at $x_{\mathrm{ref}}$ is set equal to the common horizon $T_c$:
\begin{equation}
    k_f=\frac{|x_{\mathrm{ref}}|^{1-\alpha}}{(1-\alpha)T_c},
    \qquad
    k_1=k_2=\frac{1}{T_c}\Bigl(\frac{1}{1-p}+\frac{1}{q-1}\Bigr),
    \qquad
    u_{\mathrm{PdT}},\,u_{\mathrm{PT}}\ \text{as in \eqref{eq:benchmark_control_families}}.
    \label{eq:certificate_matched}
\end{equation}
For the asymptotic baseline the same protocol sets its $\varepsilon$-entry time at $x_{\mathrm{ref}}$ equal to $T_c$, $k_a=\ln(|x_{\mathrm{ref}}|/\varepsilon)/T_c$. A common certificate does not equalize the actual settling-time functions, their worst cases, or their tolerance effects, and the observed times are deliberately not equalized: matching a tolerance-dependent observable would hide the certificate-observation gap that this section measures.

\emph{Effort-matched calibration.} The peak nominal command at $x_{\mathrm{ref}}$ is set equal to a common bound $\bar u$. Since all five laws attain their peak at $t=0$ along the nominal trajectory (for $\eta\geq1$), $k_a=\bar u/|x_{\mathrm{ref}}|$ for the baseline and
\begin{equation}
    k_f=\frac{\bar u}{|x_{\mathrm{ref}}|^{\alpha}},
    \qquad
    k_1=k_2=\frac{\bar u}{|x_{\mathrm{ref}}|^{p}+|x_{\mathrm{ref}}|^{q}},
    \qquad
    T_c^{\mathrm{PdT}}=\frac{\gamma_{p,q}\bigl(|x_{\mathrm{ref}}|^{p}+|x_{\mathrm{ref}}|^{q}\bigr)}{\bar u},
    \qquad
    \eta=\frac{\bar u\,T_c}{|x_{\mathrm{ref}}|}.
    \label{eq:effort_matched}
\end{equation}
Two consequences follow. Under effort matching the fixed-time and predefined-time laws coincide, the same-family statement in its sharpest form, and the prescribed-time law is admissible only if $\bar u\geq|x_{\mathrm{ref}}|/T_c$, the plant condition \eqref{eq:feasibility_necessary} appearing at the calibration stage. Certificate matching exposes the differences in control effort; effort matching exposes the temporal performance obtainable under equal authority. Both are reported.

\paragraph{Performance indices.}
A numerical solver cannot detect exact arrival at zero, so the observed settling time is defined through an accuracy threshold $\varepsilon>0$ and a persistence window $\Delta_{\mathrm{hold}}>0$,
\begin{equation}
    T_{\varepsilon}
    :=
    \inf
    \left\{
        t\geq0:
        |x(\tau)|\leq\varepsilon
        \quad
        \forall \tau\in[t,t+\Delta_{\mathrm{hold}}]
    \right\},
    \label{eq:observed_epsilon_settling_time}
\end{equation}
where the window prevents transient threshold crossings from being counted as convergence. $T_\varepsilon$ is not a verification of exact finite-time convergence: exact convergence follows from the analytical certificate, and the simulation illustrates only entry into and persistence within an accuracy region. Since $T^{\mathrm{exact}}(x_0)$ is available from \eqref{eq:benchmark_nominal_solutions}, the ratio of certificate to observation decomposes as
\begin{equation}
\begin{aligned}
\mathcal C_T(x_0)&:=\frac{T^{\mathrm{cert}}}{T_\varepsilon(x_0)}\\
&=\underbrace{\frac{T^{\mathrm{cert}}}{T_{\sup}}}_{\text{bound conservatism}}
\underbrace{\frac{T_{\sup}}{T^{\mathrm{exact}}(x_0)}}_{\text{initial-condition effect}}
\underbrace{\frac{T^{\mathrm{exact}}(x_0)}{T_\varepsilon(x_0)}}_{\text{tolerance effect}},
\qquad T_{\sup}:=\sup_{x\in\mathcal X_0}T^{\mathrm{exact}}(x).
\end{aligned}
\label{eq:settling_time_conservatism_index}
\end{equation}
Here $T^{\mathrm{cert}}$ is a uniform certificate on the same initial set $\mathcal X_0$, and the factorization is used only when $0<T_\varepsilon(x_0)\leq T^{\mathrm{exact}}(x_0)\leq T_{\sup}<\infty$. The first factor measures the tightness of a uniform certificate, the second where $x_0$ lies relative to the worst case, and the third the effect of the accuracy threshold. For the finite-time law on an unbounded initial set $T_{\sup}=\infty$ and the factorization is inapplicable; its pointwise ratio is instead $[T^{\mathrm{cert}}(x_0)/T^{\mathrm{exact}}(x_0)][T^{\mathrm{exact}}(x_0)/T_\varepsilon(x_0)]$, with first factor one for the benchmark. Zero initial conditions and runs that never enter the band are excluded. The control-effort and terminal indices use the applied input $u_a$:

\begin{equation}
\begin{aligned}
    J_{\mathrm{peak}}
    &:=\max_{t\in[0,T_{\mathrm{sim}}]}|u_a(t)|,
    &
    J_1
    &:=\int_0^{T_{\mathrm{sim}}}|u_a(t)|\,dt,
    &
    J_2
    &:=\int_0^{T_{\mathrm{sim}}}u_a^2(t)\,dt,
    \\
    J_{\mathrm{TV}}
    &:=\operatorname{TV}\!\left(u_a;[0,T_{\mathrm{sim}}]\right),
    &
    E_T
    &:=|x(T^{\mathrm{cert}})|,
    &
    E_{\mathrm{post}}
    &:=\max_{t\in[T^{\mathrm{cert}},\,T^{\mathrm{cert}}+\Delta_T]}|x(t)|,
\end{aligned}
\label{eq:numerical_performance_indices}
\end{equation}
where $\operatorname{TV}$ is the total variation, which records chattering under sampling, $E_T$ is the terminal miss at the certified time, and $E_{\mathrm{post}}$ records whether the origin is retained after it (the terminal-continuation criterion of Section~\ref{subsec:comparison_criteria}). Under saturation and sampling two further indices are recorded,
\begin{equation}
\begin{aligned}
    R_{\mathrm{sat}}
    &:=
    \frac{1}{T_{\mathrm{sim}}}
    \operatorname{meas}
    \left\{
        t:|u_c(t)|>u_{\max}
    \right\},
    \qquad
    \Delta T_{\mathrm{sat}}
    :=
    T_{\varepsilon}^{\mathrm{sat}}
    -
    T_{\varepsilon}^{\mathrm{nom}},
    \\
    E_{\mathrm{inter}}
    &:=
    \max_k
    \max_{t\in[t_k,t_{k+1})}
    |x(t)-x(t_k)|,
\end{aligned}
\label{eq:saturation_metrics}
\end{equation}
where $R_{\mathrm{sat}}$ is the fraction of time in saturation, $\Delta T_{\mathrm{sat}}$ the delay it causes, and $E_{\mathrm{inter}}$ the intersample deviation; under sampling, $T_\varepsilon$ and $E_{\mathrm{post}}$ are evaluated at the sampling instants. Because the held input makes the intersample trajectory affine, $|x(t)|\le\max\{|x(t_k)|,|x(t_{k+1})|\}$ on each interval, so a sample-based check cannot miss a band violation; the reported $T_\varepsilon$ is the first sampling instant of persistent entry and exceeds the exact entry time by less than one period $h$. The indices $J_1$ and $E_{\mathrm{inter}}$ are computed for completeness but not tabulated.

\paragraph{Feasibility classes.}
Every constrained run is assigned to one of four classes, with priority D, L, P, N so that each run receives a single label: \emph{deadline satisfied} (D), $T_\varepsilon\leq T^{\mathrm{cert}}$ and $E_{\mathrm{post}}\leq\varepsilon$; \emph{late convergence} (L), $T_\varepsilon<\infty$ on the simulation horizon but $T_\varepsilon>T^{\mathrm{cert}}$; \emph{practical convergence} (P), the terminal amplitude over the last two persistence windows is at most $\rho|x_0|$ (with $\rho=0.1$) but the accuracy region is either never entered for a full window or not retained after $T^{\mathrm{cert}}$; and \emph{not converged} (N), comprising divergence, sustained oscillation, and failure to converge within the horizon. For the asymptotic law, whose $T^{\mathrm{cert}}$ is the instant at which $|x|=\varepsilon$, the timing tolerance is one grid or sampling step. The classes are observational, defined through $\varepsilon$, $\Delta_{\mathrm{hold}}$, and $T_{\mathrm{sim}}$; they do not restate the analytical certificates. Two analytical boundaries are overlaid on the maps. Since $|x(t)|\geq|x_0|-u_{\max}t$ under \eqref{eq:numerical_implemented_input} with $d\equiv0$, no controller can bring the state to zero by $T_c$ unless
\begin{equation}
    u_{\max}T_c\geq|x_0|,
    \qquad\text{and into the accuracy band unless}\qquad
    u_{\max}T_c\geq(|x_0|-\varepsilon)_+,
    \label{eq:feasibility_necessary}
\end{equation}
the first being the plant's minimum-time limit of Section~\ref{subsec:feasible_decay} for exact arrival and the second its practical counterpart for $T_\varepsilon$; both are limits of the plant, as opposed to the feasibility limit of a controller family. For the sampled scalar law $x_{k+1}=x_k-hk\lceil x_k\rceil^p$, $0<p<1$, a symmetric two-cycle has amplitude
\begin{equation}
a_h=\left(\frac{hk}{2}\right)^{1/(1-p)}.
\label{eq:sampling_cycle}
\end{equation}
This gives a characteristic residual scale for this emulation, not a universal accuracy floor: exceptional grid-dependent trajectories can hit zero exactly, and other discretizations have different properties. The boundary test $hk\varepsilon^{p-1}\leq2$ checks the update from $x_k=\varepsilon$; for the benchmark exponent $p=1/2$ it also ensures invariance of $[-\varepsilon,\varepsilon]$. For the two-power law, one-step contraction at a nonzero state requires
\begin{equation}
h\left(k_1|x_k|^{p-1}+k_2|x_k|^{q-1}\right)<2.
\label{eq:full_sampled_contraction}
\end{equation}
The sublinear term limits small-error accuracy, while the superlinear term can destroy contraction at large initial states. A line computed from the sublinear term alone is therefore only a reference scale for the two-power laws, and neither reference line is a universal feasibility boundary for all five controllers. Both restrictions belong to the explicit held update; consistent or implicit discretizations designed to preserve finite-, fixed-, or predefined-time convergence avoid them \citep{polyakov2019consistent,jimenez2020consistent,brogliato2020implicit}, but they are discrete-time redesigns rather than the emulation tested here.

\paragraph{Scenario matrix.}
Table~\ref{tab:scenario_matrix} defines the experiments, and Table~\ref{tab:benchmark_laws} lists the laws, parameters, and certified guarantees used in them. The nominal, initial-condition, and effort-matched experiments are analytical and are checked against simulation; the remaining five are simulated. Measurement noise and a full disturbance sweep are not reported; the single bounded-disturbance case illustrates the residual-set qualification of Section~\ref{subsec:specification_to_design}, and the bounded-gain case compares the pure and capped prescribed-time laws under sampling.

\noindent Scenario matrix of the numerical comparison. All experiments except the effort-matched one use certificate-matched tuning; the effort-matched experiment gives the explicit comparison under equal peak input. The laws are defined in \eqref{eq:benchmark_control_families}.

{
\footnotesize
\setlength{\tabcolsep}{4pt}
\renewcommand{\arraystretch}{1.15}
\begin{xltabular}{\textwidth}{@{}l >{\raggedright\arraybackslash}p{0.18\textwidth} >{\raggedright\arraybackslash}p{0.16\textwidth} >{\raggedright\arraybackslash}p{0.14\textwidth} >{\raggedright\arraybackslash}X@{}}
\caption{Numerical comparison scenarios.}\label{tab:scenario_matrix}\\
\toprule
\textbf{Scenario} & \textbf{Initial set} & \textbf{Saturation} & \textbf{Sampling} & \textbf{Purpose}\\
\midrule
\endfirsthead
\multicolumn{5}{@{}l}{\footnotesize\textit{Table~\ref{tab:scenario_matrix} (continued)}}\\
\toprule
\textbf{Scenario} & \textbf{Initial set} & \textbf{Saturation} & \textbf{Sampling} & \textbf{Purpose}\\
\midrule
\endhead
\midrule\multicolumn{5}{r@{}}{\footnotesize\textit{continued on next page}}\\
\endfoot
\bottomrule\endlastfoot

Nominal & $x_0=x_{\mathrm{ref}}$ & none & continuous & Convergence geometry of the five laws\\
$x_0$ sweep & $|x_0|\in[10^{-2},10^{3}]$ & none & continuous & Dependence of $T^{\mathrm{exact}}$, $T_\varepsilon$ on $x_0$; certificate tightness\\
Effort, saturation & $x_0=x_{\mathrm{ref}}$ & $u_{\max}$ sweep & continuous & Control effort; authority feasibility \eqref{eq:feasibility_necessary}\\
Sampling & $x_0\in\{x_{\mathrm{ref}},\,100x_{\mathrm{ref}}\}$ & none & $h$ sweep & Emulation sensitivity; two-cycle residual; full one-step bound; chattering ($J_{\mathrm{TV}}$)\\
Joint map & $x_0=x_{\mathrm{ref}}$ & $u_{\max}$ sweep & $h$ sweep & Joint feasibility map with the analytical boundaries\\
Disturbance & $x_0=x_{\mathrm{ref}}$ & none & continuous & Bounded disturbance $d(t)=0.05\sin5t$; residual-set qualification\\
Effort-matched & $x_0=x_{\mathrm{ref}}$, $|x_0|\in[10^{-2},10^{3}]$ & none & continuous & Effort-matched calibration \eqref{eq:effort_matched}, $\bar u=4$\\
Bounded-gain PT & $x_0\in\{x_{\mathrm{ref}},\,100x_{\mathrm{ref}}\}$ & none, $u_{\max}$ sweep & $h$ sweep & Pure versus bounded-gain prescribed-time law, $\kappa_{\max}=0.8/h$\\
\end{xltabular}
}

\noindent Benchmark laws, parameters, and certified guarantees. Numerical values are those of the reported experiments. Common: $T_c=1$, $x_{\mathrm{ref}}=1$, $\varepsilon=10^{-3}$, $\Delta_{\mathrm{hold}}=0.1$, $\Delta_T=0.5$, base $T_{\mathrm{sim}}=2$, $k_h=\eta/T_c$; extended windows are specified below.

{
\footnotesize
\setlength{\tabcolsep}{4pt}
\renewcommand{\arraystretch}{1.15}
\begin{xltabular}{\textwidth}{@{}l >{\raggedright\arraybackslash}p{0.24\textwidth} >{\raggedright\arraybackslash}X >{\raggedright\arraybackslash}p{0.22\textwidth} l@{}}
\caption{Benchmark laws and certified guarantees.}\label{tab:benchmark_laws}\\
\toprule
\textbf{Law} & \textbf{Mechanism (Section~\ref{sec:mechanisms})} & \textbf{Parameters} & \textbf{Certified guarantee} & \textbf{Tight?}\\
\midrule
\endfirsthead
\multicolumn{5}{@{}l}{\footnotesize\textit{Table~\ref{tab:benchmark_laws} (continued)}}\\
\toprule
\textbf{Law} & \textbf{Mechanism (Section~\ref{sec:mechanisms})} & \textbf{Parameters} & \textbf{Certified guarantee} & \textbf{Tight?}\\
\midrule
\endhead
\midrule\multicolumn{5}{r@{}}{\footnotesize\textit{continued on next page}}\\
\endfoot
\bottomrule\endlastfoot

$u_{\mathrm{AS}}$ & Exponential decay (baseline) & $k_a=\ln(|x_{\mathrm{ref}}|/\varepsilon)/T_c$ or $\bar u/|x_{\mathrm{ref}}|$ & Asymptotic only; $\varepsilon$-entry at $T^{\mathrm{cert}}_{\mathrm{AS}}(x_0)$ & Yes (band entry)\\
$u_{\mathrm{FT}}$ & Sublinear decay \eqref{eq:sec3_ft_decay} & $\alpha=0.5$; $k_f$ by \eqref{eq:certificate_matched}/\eqref{eq:effort_matched} & Exact, $T\le T^{\mathrm{cert}}_{\mathrm{FT}}(x_0)$ & Yes\\
$u_{\mathrm{FxT}}$ & Two-power decay \eqref{eq:sec33_dual_power} & $(p,q)=(0.5,1.5)$; $k_1=k_2$ & Exact, $T\le T^{\mathrm{cert}}_{\mathrm{FxT}}$ uniform & No ($4/\pi$)\\
$u_{\mathrm{PdT}}$ & Normalized two-power \eqref{eq:sec34_pdt_general} & $(p,q)=(0.5,1.5)$; $\gamma_{p,q}=\pi$ & Exact, $T\le T_c$ uniform & In the supremum\\
$u_{\mathrm{PT}}$ & Time-varying gain \eqref{eq:sec34_pt_lyap} & $\eta=3$; continuation $u=-k_hx$, $t\ge T_c$ & Exact at $T_c$, $x_0$-independent & Exact arrival\\

\end{xltabular}
}

\paragraph{Reproducibility.}
All continuous-time runs use MATLAB \texttt{ode45} with relative tolerance $10^{-8}$, absolute tolerance $10^{-10}$, and maximum step $10^{-3}T_c$, reported on a uniform grid of step $10^{-3}T_c$; the prescribed-time law is integrated on $[0,T_c-10^{-6}T_c]$ and continued from $T_c$, the neglected interval contributing at most $10^{-6}T_cu_{\max}$ to the state in saturated disturbance-free runs (in unsaturated runs the error is bounded through the actual input or the nominal closed form). For nominal continuous-time checks with $d\equiv0$, a zero-state continuation below $10^{-9}\max\{1,|x_0|\}$ represents the analytically established invariant equilibrium; such clamping is never applied to sampled, disturbed, or regularized runs, whose intersample dynamics determine retention and residual oscillations. Exact settling times are computed from \eqref{eq:benchmark_nominal_solutions}, by adaptive quadrature where an integral is involved, not inferred from the solver, and $T_\varepsilon$ is computed from the reported trajectory. In sampled-data runs the command is evaluated at the sampling instants and held, the plant is integrated exactly over each interval, and the affine intersample trajectory is used for all indices; saturation acts on the applied input. The base window $T_{\mathrm{sim}}=2$ is extended for larger initial states to $\max\{2,1.5\max_\ell T^{\mathrm{exact}}_\ell(x_0),\max_\ell T^{\mathrm{cert}}_\ell(x_0)+\Delta_T\}$, where $\ell$ indexes the compared laws and the $\varepsilon$-entry time replaces $T^{\mathrm{exact}}$ for the asymptotic law; the finite-time law at $x_0=100$, for example, settles at $10$. Sweeps use logarithmic grids of $13$ points for $x_0$ and $9$ for $u_{\max}$ and $h$. The cap $\kappa_{\max}=0.8/h$ avoids $1/h$, the deadbeat gain of the held integrator, which would produce a plant-specific exact zero; sampling periods are non-commensurate with $T_c$, since for integer $T_c/h$ and $\eta$ the held trajectory reaches zero one step after $t_k=T_c-\eta h$ as an artifact of the grid; and $\eta=3$ avoids coincidence of the prescribed- and finite-time trajectories at $x_{\mathrm{ref}}$, which occurs for $\eta=1/(1-\alpha)$.

\subsection{Temporal Guarantees and Observed Convergence}
\label{subsec:numerical_time}

\begin{figure*}[pos=!t]
\centering
\reviewfigure{0.85\textwidth}{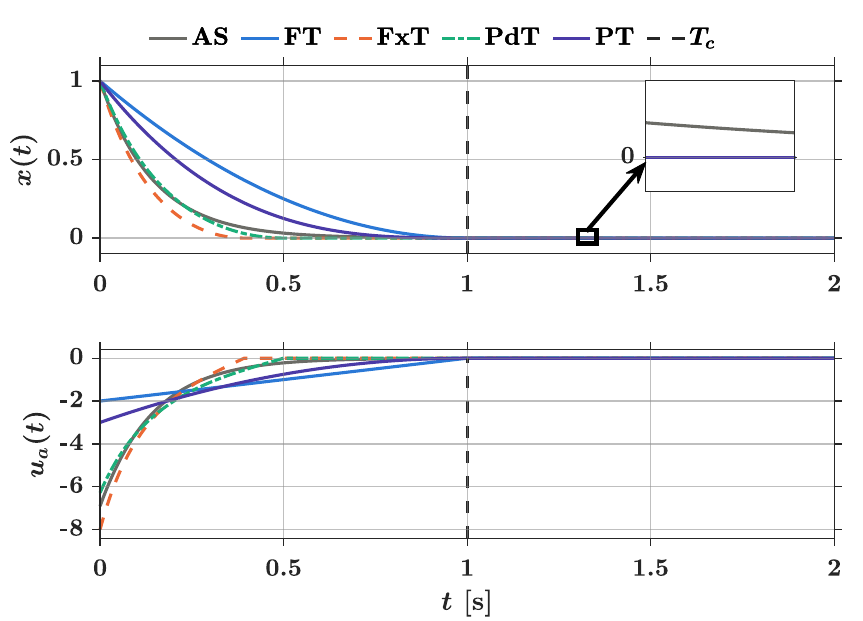}
\caption{Nominal responses of the five laws from $x_0=x_{\mathrm{ref}}$ under certificate-matched calibration.}
\label{fig:S0_trajectories}
\end{figure*}

\paragraph{Convergence geometry.}
Figure~\ref{fig:S0_trajectories} shows the nominal responses from $x_{\mathrm{ref}}$ under certificate matching, with gains $k_a=\ln10^{3}=6.91$, $k_f=2$, $k_1=k_2=4$, $\gamma_{p,q}/T_c=\pi$, and $\eta=3$. The finite- and prescribed-time laws arrive at $T_c=1$ exactly, as their closed forms $|x|=(1-t)^2$ and $x=(1-t)^3$ require, whereas the two-power laws arrive at $0.393$ (FxT) and $0.500$ (PdT), well before the common certificate. The tolerance factor $T^{\mathrm{exact}}/T_\varepsilon$ lies between $1.03$ and $1.11$ (Table~\ref{tab:certificate_vs_observed}); it is largest for the prescribed-time law, whose trajectory crosses $\varepsilon$ at $t=1-\varepsilon^{1/3}\approx0.90$. Arriving early costs effort: the two laws calibrated to arrive at $T_c$ are the cheapest ($J_{\mathrm{peak}}=2$ and $3$), the two-power laws and the asymptotic baseline the most expensive ($8$, $6.28$, and $6.91$). The asymptotic baseline enters the band one reporting step after its $\varepsilon$-entry certificate and needs $J_{\mathrm{peak}}=6.91$, as much as the two-power laws, which reach the origin itself in half the time. The prescribed-time input is bounded and vanishes at $T_c$, the concrete instance of \eqref{eq:sec34_false_implication}. The closed-form and \texttt{ode45} trajectories agree to $5\times10^{-11}$, and \eqref{eq:predefined_normalization} agrees with adaptive quadrature to six digits, which validates the analytical route used for the nominal experiments.

\begin{figure*}[pos=!t]
\centering
\begin{subfigure}[t]{0.49\textwidth}
\centering
\reviewfigure{\linewidth}{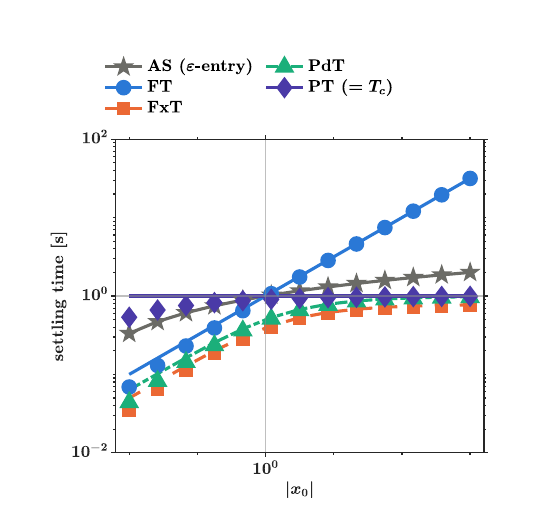}
\caption{Certificate-matched calibration.}
\label{fig:S1_x0_sweep}
\end{subfigure}\hfill
\begin{subfigure}[t]{0.49\textwidth}
\centering
\reviewfigure{\linewidth}{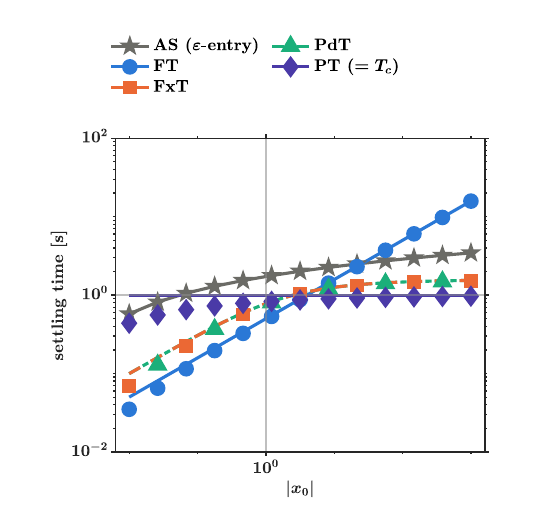}
\caption{Effort-matched calibration, $\bar u=4$.}
\label{fig:S6_effort_matched}
\end{subfigure}
\caption{Settling time versus initial condition under the two calibrations (nominal).}
\label{fig:x0_sweeps}
\end{figure*}

\paragraph{Dependence on the initial condition.}
Figure~\ref{fig:S1_x0_sweep} and Table~\ref{tab:certificate_vs_observed} report the sweep $|x_0|\in[10^{-2},10^{3}]$; lines are exact settling times from \eqref{eq:benchmark_nominal_solutions}, markers the observed $T_\varepsilon$, and their gap the tolerance effect of \eqref{eq:settling_time_conservatism_index}, largest for small $|x_0|$. Four shapes appear. The finite-time settling time grows without bound, as $2|x_0|^{1/2}/k_f$, and its certificate is tight at every $x_0$. The fixed-time settling time remains bounded and converges to $\pi/k=0.785$, under its uniform certificate $1.000$: the certificate is calibrated to the common bound but conservative by $1.273$. The predefined-time settling time approaches $T_c$ from below without reaching it on any bounded set: the normalization fixes the supremum, not the time at a finite $x_0$. The prescribed-time law arrives at $T_c$ for every $x_0$, so the variation of its $T_\varepsilon$ is the tolerance effect alone. For the autonomous laws the tolerance offset at large $|x_0|$ is nearly constant, so the markers track the exact curves and the shape of the dependence is read off directly. The asymptotic band-entry time grows logarithmically and, unlike the others, depends on $\varepsilon$ itself. These calibrations illustrate settling-time functions; they do not add defining requirements to the classes of Section~\ref{sec:notions}.

\noindent Certified, exact, and observed settling times (certificate-matched calibration, $x_{\mathrm{ref}}=1$, $\varepsilon=10^{-3}$). ``Exact'' is the analytical claim; $T_\varepsilon$ is a numerical observation and does not verify exact convergence. $\max_{\mathcal X_0}$ denotes the maximum over the swept initial set $\mathcal X_0=\{x_0:|x_0|\in[10^{-2},10^{3}]\}$ (with the extended horizon of the reproducibility paragraph); $\sup$ is the global supremum over $x_0\in\mathbb R$. The FT and AS entries in the last column are inapplicable: their global suprema are infinite, although the FT pointwise certificate is tight. For AS, $T^{\mathrm{cert}}$ is the $\varepsilon$-entry time ($1.000$ at $x_{\mathrm{ref}}$) and $T^{\mathrm{exact}}=\infty$.

{
\scriptsize
\setlength{\tabcolsep}{3pt}
\renewcommand{\arraystretch}{1.15}
\begin{longtable}{@{}l >{\raggedright\arraybackslash}p{0.17\textwidth} *{7}{>{\centering\arraybackslash}p{0.091\textwidth}}@{}}
\caption{Certified, exact, and observed settling times.}\label{tab:certificate_vs_observed}\\
\toprule
\textbf{Law} & \textbf{Claim} & $T^{\mathrm{cert}}$ & \tabhead{$T^{\rm exact}$\\$(x_{\rm ref})$} & \tabhead{$T_\varepsilon$\\$(x_{\rm ref})$} & \tabhead{$\max_{\mathcal X_0}$\\$T^{\rm exact}$} & \tabhead{$\max_{\mathcal X_0}$\\$T_\varepsilon$} & \tabhead{$\sup$\\$T^{\rm exact}$} & \tabhead{$T^{\rm cert}/$\\$\sup T^{\rm exact}$}\\
\midrule
\endfirsthead
\multicolumn{9}{@{}l}{\footnotesize\textit{Table~\ref{tab:certificate_vs_observed} (continued)}}\\
\toprule
\textbf{Law} & \textbf{Claim} & $T^{\mathrm{cert}}$ & \tabhead{$T^{\rm exact}$\\$(x_{\rm ref})$} & \tabhead{$T_\varepsilon$\\$(x_{\rm ref})$} & \tabhead{$\max_{\mathcal X_0}$\\$T^{\rm exact}$} & \tabhead{$\max_{\mathcal X_0}$\\$T_\varepsilon$} & \tabhead{$\sup$\\$T^{\rm exact}$} & \tabhead{$T^{\rm cert}/$\\$\sup T^{\rm exact}$}\\
\midrule
\endhead
\midrule\multicolumn{9}{r@{}}{\footnotesize\textit{continued on next page}}\\
\endfoot
\bottomrule\endlastfoot

AS  & asymptotic, $\varepsilon$-entry & $T(x_0)$ & $\infty$ & 1.001 & $\infty$ & 2.001 & $\infty$ & ---\\
FT  & exact, $x_0$-dependent    & $T(x_0)$ & 1.000 & 0.969 & 31.62 & 31.59 & $\infty$ & ---\\
FxT & exact, uniform bound      & 1.000 & 0.393 & 0.377 & 0.770 & 0.754 & 0.785 & 1.273\\
PdT & exact, selectable bound   & 1.000 & 0.500 & 0.480 & 0.980 & 0.960 & 1.000 & 1 (supremum)\\
PT  & exact arrival at $T_c$    & 1.000 & 1.000 & 0.901 & 1.000 & 0.990 & 1.000 & 1\\
\end{longtable}
}

\paragraph{Effort-matched calibration.}
Under \eqref{eq:effort_matched} with $\bar u=4$, all five laws command the same peak at $x_{\mathrm{ref}}$ (Table~\ref{tab:effort_matched}). The fixed- and predefined-time laws become literally identical, so their curves coincide in Fig.~\ref{fig:S6_effort_matched}, and the table shows two certificates attached to one law: $2.000$ from \eqref{eq20} and $1.571$ from the normalization. At $x_{\mathrm{ref}}$ the finite-time law is now the fastest ($0.500$) and the asymptotic law the slowest ($1.727$), while the prescribed-time law still arrives at $T_c$, the larger $\eta$ having only smoothed its input. The ordering reverses for large initial conditions: at $|x_0|=10^{3}$ the finite-time settling time reaches $15.8$, whereas the two-power law saturates near $\pi/2$ and the prescribed-time law stays at $1$, at the price of peak inputs of $6.3\times10^{4}$ and $4\times10^{3}$ against $126$ for the finite-time law. Certificate matching shows what each guarantee costs at the design point; effort matching shows what it delivers away from it.

\noindent Effort-matched calibration at $x_{\mathrm{ref}}=1$ with $\bar u=4$. For AS, $T^{\mathrm{cert}}$ is the $\varepsilon$-entry time. The FxT and PdT rows are one and the same control law, $u=-2(\lceil x\rceil^{1/2}+\lceil x\rceil^{3/2})$, carrying two different certificates.

{
\footnotesize
\setlength{\tabcolsep}{4pt}
\renewcommand{\arraystretch}{1.15}
\begin{longtable}{@{}l >{\raggedright\arraybackslash}p{0.21\textwidth} *{6}{>{\centering\arraybackslash}p{0.104\textwidth}}@{}}
\caption{Performance under effort-matched calibration.}\label{tab:effort_matched}\\
\toprule
\textbf{Law} & \textbf{Gains} & $T^{\mathrm{cert}}$ & \tabhead{$T^{\rm exact}$\\$(x_{\rm ref})$} & \tabhead{$T_\varepsilon$\\$(x_{\rm ref})$} & \tabhead{$\sup$\\$T^{\rm exact}$} & $J_{\mathrm{peak}}$ & $J_2$\\
\midrule
\endfirsthead
\multicolumn{8}{@{}l}{\footnotesize\textit{Table~\ref{tab:effort_matched} (continued)}}\\
\toprule
\textbf{Law} & \textbf{Gains} & $T^{\mathrm{cert}}$ & \tabhead{$T^{\rm exact}$\\$(x_{\rm ref})$} & \tabhead{$T_\varepsilon$\\$(x_{\rm ref})$} & \tabhead{$\sup$\\$T^{\rm exact}$} & $J_{\mathrm{peak}}$ & $J_2$\\
\midrule
\endhead
\midrule\multicolumn{8}{r@{}}{\footnotesize\textit{continued on next page}}\\
\endfoot
\bottomrule\endlastfoot

AS  & $k_a=4$              & 1.727 & $\infty$ & 1.727 & $\infty$ & 4.00 & 2.00\\
FT  & $k_f=4$              & 0.500 & 0.500 & 0.485 & unbounded & 4.00 & 2.67\\
FxT & $k_1=k_2=2$          & 2.000 & 0.785 & 0.754 & 1.571 & 4.00 & 2.13\\
PdT & $T_c^{\mathrm{PdT}}=1.571$ & 1.571 & 0.785 & 0.754 & 1.571 & 4.00 & 2.13\\
PT  & $\eta=4$             & 1.000 & 1.000 & 0.823 & 1.000 & 4.00 & 2.29\\
\end{longtable}
}

\paragraph{Reading of the nominal experiments.}
Two observations carry over to the constrained experiments. First, the tight finite-time certificate, the conservative fixed-time certificate, the predefined-time supremum, and exact prescribed-time arrival are properties of the selected laws, not definitions of the classes: fixed-time bounds can be calibrated, predefined-time stability does not require equality of the supremum, and prescribed-time convergence need not mean first arrival at the deadline. Second, \eqref{eq:settling_time_conservatism_index} separates certificate conservatism from initial-condition and tolerance effects, whereas the asymptotic baseline can be matched only to an $\varepsilon$-dependent entry time that grows without bound in $x_0$.

\subsection{Control Effort and Sensitivity to Implementation Constraints}
\label{subsec:numerical_constraints}

\begin{figure*}[pos=!t]
\centering
\begin{subfigure}[t]{0.48\textwidth}
\centering
\reviewfigure{\linewidth}{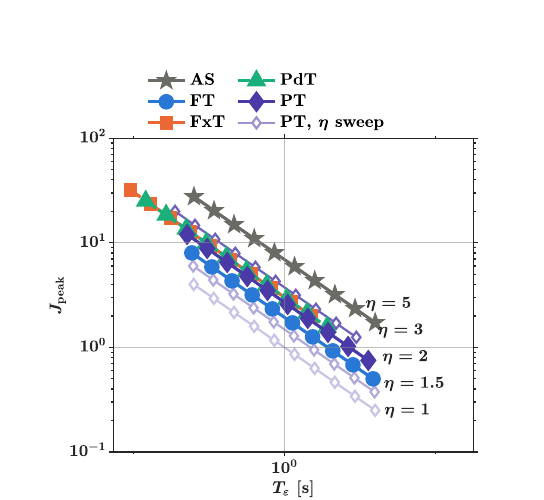}
\caption{Peak input versus observed settling time over a sweep of $T_c$.}
\label{fig:S2a_pareto}
\end{subfigure}\hfill
\begin{subfigure}[t]{0.48\textwidth}
\centering
\reviewfigure{\linewidth}{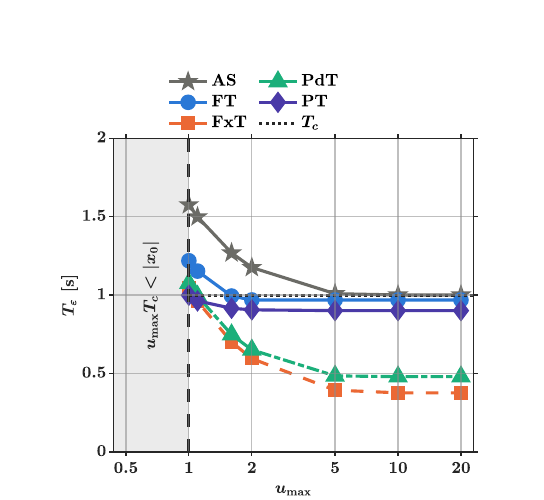}
\caption{Observed settling time under actuator saturation.}
\label{fig:S2b_saturation}
\end{subfigure}
\caption{Control effort and control-authority feasibility at $x_0=x_{\mathrm{ref}}$.}
\label{fig:S2_effort}
\end{figure*}

\paragraph{Control effort.}
Figure~\ref{fig:S2a_pareto} shows the peak input against $T_\varepsilon$ for $T_c\in[0.25,4]$ under certificate matching. All curves are hyperbolae, $J_{\mathrm{peak}}T_\varepsilon\approx\text{const}$, with constants $1.94$ (FT), $2.70$ (PT, $\eta=3$), $3.02$ for FxT and PdT, which lie on one curve, and $6.91$ for the asymptotic law, whose constant $\ln(|x_0|/\varepsilon)$ grows as the accuracy is tightened. That the fixed- and predefined-time laws share one curve is the same-family statement in effort terms: certificate matching merely places them at different points on it. Among the prescribed-time curves, $\eta=1$ gives the constant input $u=-x_0/T_c$ with $J_{\mathrm{peak}}T_c=|x_0|$, the plant's minimum-time limit of Section~\ref{subsec:feasible_decay}; larger $\eta$ buys a smoother input at the price of a peak $\eta|x_0|/T_c$.

\paragraph{Control authority.}
Figure~\ref{fig:S2b_saturation} and Table~\ref{tab:constrained_indices} report the saturation sweep at $x_{\mathrm{ref}}$; three findings carry the message. First, below the plant limit $u_{\max}T_c=|x_0|$ of \eqref{eq:feasibility_necessary} no law can meet the deadline: at $u_{\max}=0.5$ all five are saturated for most of the interval and none converges within the horizon. At the boundary $u_{\max}=1$ the four autonomous laws converge late, whereas the prescribed-time law meets the deadline only because its saturated trajectory $x=1-t$ is the minimum-time trajectory, a zero-margin boundary case, not a robustness property. Second, the fixed calibrations need different surpluses of authority above that limit. The two-power laws meet the deadline from $u_{\max}=1.1$; at $u_{\max}=2$ they still spend about a fifth of the interval in saturation and lose $0.17$-$0.22$ relative to their nominal times, while the finite-time law, whose nominal peak is exactly $2$, is unaffected. The asymptotic law, whose calibration places band entry exactly at $T_c$, is late for every $u_{\max}\leq5$ and meets its deadline only once it no longer saturates ($u_{\max}\geq10$). After a saturated phase the prescribed-time gain re-plans the remaining horizon and arrives at $T_c$ for every $u_{\max}\geq1$. Third, the finite-time threshold belongs to the calibration. With $k_f=2$ the trajectory saturates while $|x|>u_{\max}^2/4$, giving $T_\varepsilon=1/u_{\max}+u_{\max}/4-\sqrt\varepsilon$ for $2\sqrt\varepsilon<u_{\max}<2$ and the threshold $u_{\max}\approx1.55631$ for $x_0=T_c=1$ and $\varepsilon=10^{-3}$, bracketed by the observed $T_\varepsilon=1.153$ at $u_{\max}=1.1$ and $0.994$ at $1.6$; it is not a controller-family optimum.

Retuning changes that threshold. For $u=-\operatorname{sat}_{u_{\max}}(k\lceil x\rceil^{1/2})$, $x_0>\varepsilon$, let $a=(u_{\max}/k)^2$. If $\varepsilon<a<x_0$, direct integration gives
\begin{equation}
T_\varepsilon=\frac{x_0}{u_{\max}}+\frac{u_{\max}}{k^2}-\frac{2\sqrt\varepsilon}{k}.
\label{eq:retuned_saturation_time}
\end{equation}
If $a\leq\varepsilon$, the trajectory remains saturated until it enters the band and $T_\varepsilon=(x_0-\varepsilon)/u_{\max}$. For example, $k=20$, $u_{\max}=x_0=T_c=1$, and $\varepsilon=10^{-3}$ give $T_\varepsilon\approx0.999338<1$, although the $k=2$ calibration misses the deadline. The benchmark therefore establishes feasibility limits of specified tunings; a family minimum requires optimization over all admissible parameters on a stated initial set and accuracy criterion.

\begin{figure*}[pos=!t]
\centering
\reviewfigure{0.9\textwidth}{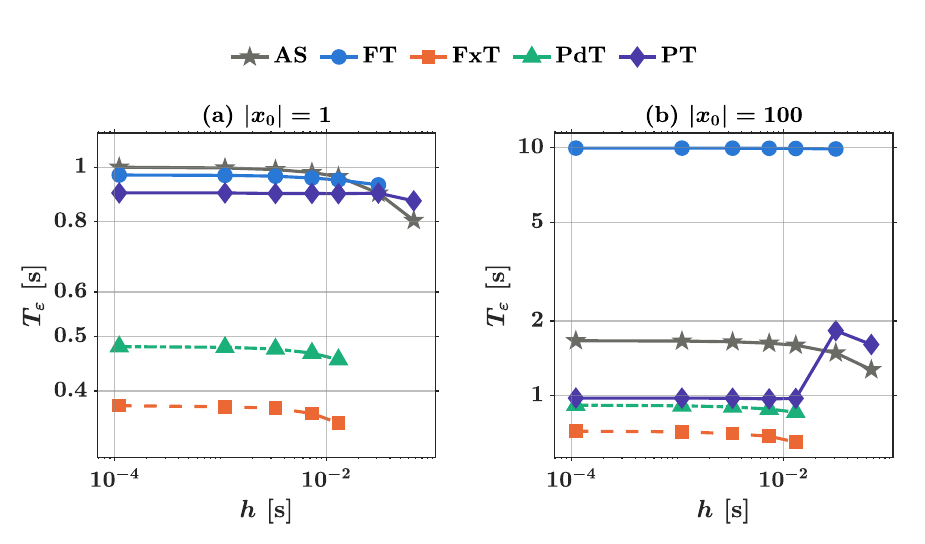}
\caption{Observed settling time under zero-order-hold sampling for $x_0=x_{\mathrm{ref}}$ and $100x_{\mathrm{ref}}$.}
\label{fig:S3_sampling}
\end{figure*}

\paragraph{Digital implementation.}
Figure~\ref{fig:S3_sampling} and the sampling rows of Table~\ref{tab:constrained_indices} report the emulation sweep $h\in[1.1\times10^{-4},6.7\times10^{-2}]$ for $x_0=x_{\mathrm{ref}}$ and $100\,x_{\mathrm{ref}}$, without saturation; missing markers denote runs that did not reach the accuracy region. The autonomous power laws fail in the two ways predicted in Section~\ref{subsec:benchmark_protocol}. Near the origin their terminal amplitudes match the two-cycle scale \eqref{eq:sampling_cycle} (at $h=3.1\times10^{-2}$, observed $9.61\times10^{-4}$, $3.87\times10^{-3}$, and $2.38\times10^{-3}$ for FT, FxT, and PdT against reference values $9.6\times10^{-4}$, $3.8\times10^{-3}$, and $2.4\times10^{-3}$), so the two-power laws become practical before the finite-time law, and at $h=6.7\times10^{-2}$ all three are practical. Far from the origin the superlinear term destroys contraction: at $x_0=100$, \eqref{eq:full_sampled_contraction} requires $h<4.9505\times10^{-2}$ (FxT) and $h<6.3032\times10^{-2}$ (PdT), and both laws diverge at $h=6.7\times10^{-2}$; when they converge, their input variation $J_{\mathrm{TV}}$ is several times (at $x_0=1$) to two orders of magnitude (at $x_0=100$) larger than that of the finite-time law. The pure prescribed-time law meets the deadline at every $h$ for $x_0=1$, because the state is already of order $h^{\eta}$ when contraction fails near $T_c$; for $x_0=100$ and $h\geq3.1\times10^{-2}$ the terminal overshoot is no longer negligible and convergence is late, through the continuation. The asymptotic baseline is insensitive to sampling over the sweep: its held update $x_{k+1}=(1-hk_a)x_k$ contracts at every state for $h<2/k_a=0.29$, and it meets its $\varepsilon$-entry deadline at every $h$ for both initial conditions, entering the band earlier as $h$ grows because $1-hk_a<e^{-hk_a}$.

\paragraph{Bounded-gain prescribed-time variant.}
Capping the gain at $\kappa_{\max}=0.8/h$ freezes it from $t=T_c-\eta h/0.8$, after which the held update contracts by a factor $0.2$ per step. This removes the large-period failures of the pure law: at $x_{\mathrm{ref}}$ and $h=0.107$ the post-certificate miss drops from $5.0\times10^{-3}$ (P) to $3.0\times10^{-5}$ (D), and at $x_0=100$ the capped law meets the deadline at $h=3.1\times10^{-2}$, where the pure law is late. In the authority-period map the cap turns the row $h=0.107$ with $u_{\max}\geq1.8$ into deadline cells, while the authority-limited columns $u_{\max}\leq1$ remain late or practical. The price, stated in Section~\ref{subsubsec:time_varying_pt}, is exponential rather than exact terminal convergence from the amplitude $|x_0|(\eta h/0.8)^{\eta}$ at the freezing instant.

\begin{figure*}[pos=!t]
\centering
\reviewfigure{\textwidth}{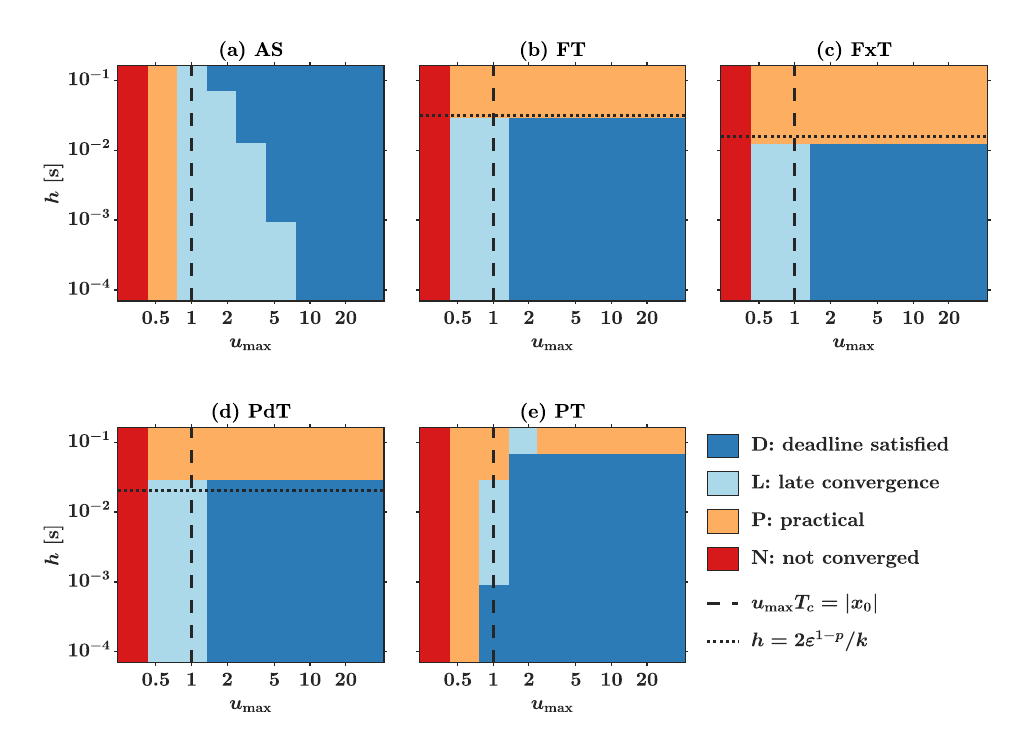}
\caption{Feasibility classes over control authority $u_{\max}$ and sampling period $h$ at $x_0=x_{\mathrm{ref}}$ (pure prescribed-time law in panel (e)). Green, yellow, orange, and red: D, L, P, and N; white dashed line: plant limit \eqref{eq:feasibility_necessary}; dotted line: two-cycle reference scale $h=2\varepsilon^{1-p}/k$ of the autonomous power laws.}
\label{fig:S4_feasibility_map}
\end{figure*}

\paragraph{Joint feasibility.}
Figure~\ref{fig:S4_feasibility_map} assembles both constraints. Green, yellow, orange, and red mark D, L, P, and N; the white dashed line is the plant limit \eqref{eq:feasibility_necessary}, which is sufficient for the selected prescribed-time law in continuous time under the stated assumptions, and the dotted line is the sublinear two-cycle reference scale $h=2\varepsilon^{1-p}/k$, which applies to the autonomous power laws. To the left of the dashed line no law meets the deadline; the next column is late for the autonomous power laws and practical for the asymptotic and prescribed-time laws, whose exponential tails do not reach $\varepsilon$ with the required persistence before $T_{\mathrm{sim}}$. To its right the autonomous power laws satisfy the deadline for all $u_{\max}\geq1.8$ below their reference scales ($h\leq3.2\times10^{-2}$, $1.6\times10^{-2}$, and $2.0\times10^{-2}$ for FT, FxT, and PdT), independently of $u_{\max}$; the pure prescribed-time law meets it for every $u_{\max}\geq1.8$ at $h\leq4.5\times10^{-2}$ on the reported grid, but in the zero-margin column $u_{\max}=1$ only for $h\leq6.0\times10^{-4}$; at $h=0.107$ it becomes late ($u_{\max}=1.8$) or practical ($u_{\max}\geq3.2$), because the last-sample overshoot is not absorbed by the continuation within $\Delta_T$. The asymptotic map has the opposite shape: it needs the most authority, and its threshold falls stepwise from $u_{\max}\geq10$ at the finest periods to $u_{\max}\geq1.8$ at $h=0.11$ because the held linear update decays faster than its continuous-time counterpart. A horizon is thus realizable only where authority and implementation rate are both compatible with it, the mechanism-level form of \eqref{eq:min_feasible_time}; within that region the class laws differ only near its edges, so the temporal class decides what happens when a constraint binds, not whether the deadline is met when none does. The fixed-tuning thresholds lie between the map columns $u_{\max}=1$ and $1.8$ and are located in Table~\ref{tab:constrained_indices}.

\FloatBarrier
\noindent Nominal and constrained performance indices at $x_0=x_{\mathrm{ref}}$ (certificate-matched). Saturation rows: continuous time; sampling rows: zero-order hold, no saturation; for the sampling rows the last column is $J_{\mathrm{TV}}$ and $\Delta T_{\mathrm{sat}}$, $R_{\mathrm{sat}}$ do not apply. All effort indices use the applied input $u_a$. Class: feasibility class of Section~\ref{subsec:benchmark_protocol} (D deadline satisfied, L late convergence, P practical, N not converged within $T_{\mathrm{sim}}$). For AS, $T^{\mathrm{cert}}$ is the $\varepsilon$-entry time, resolved to one reporting step. The classes are observational labels for the finite window $T_{\mathrm{sim}}$, not analytical deadline certificates; D in particular records an observed run, not a proved deadline.

{

\footnotesize
\setlength{\tabcolsep}{4pt}
\renewcommand{\arraystretch}{1.12}
\begin{longtable}{@{}l l c c c c c c c c@{}}
\caption{Nominal and constrained performance indices.}\label{tab:constrained_indices}\\
\toprule
\textbf{Case} & \textbf{Law} & $T_\varepsilon$ & $\Delta T_{\mathrm{sat}}$ & $R_{\mathrm{sat}}$ & $E_T$ & $E_{\mathrm{post}}$ & $J_{\mathrm{peak}}$ & $J_2$ / $J_{\mathrm{TV}}$ & \textbf{Class}\\
\midrule
\endfirsthead
\multicolumn{10}{@{}l}{\footnotesize\textit{Table~\ref{tab:constrained_indices} (continued)}}\\
\toprule
\textbf{Case} & \textbf{Law} & $T_\varepsilon$ & $\Delta T_{\mathrm{sat}}$ & $R_{\mathrm{sat}}$ & $E_T$ & $E_{\mathrm{post}}$ & $J_{\mathrm{peak}}$ & $J_2$ / $J_{\mathrm{TV}}$ & \textbf{Class}\\
\midrule
\endhead
\midrule\multicolumn{10}{r@{}}{\footnotesize\textit{continued on next page}}\\
\endfoot
\bottomrule\endlastfoot

nominal & AS  & 1.001 & --- & 0 & $1.0\times10^{-3}$ & $9.9\times10^{-4}$ & 6.91 & 3.45 & D\\*
        & FT  & 0.969 & --- & 0 & 0 & 0 & 2.00 & 1.33 & D\\*
        & FxT & 0.377 & --- & 0 & 0 & 0 & 8.00 & 4.27 & D\\*
        & PdT & 0.480 & --- & 0 & 0 & 0 & 6.28 & 3.35 & D\\*
        & PT  & 0.901 & --- & 0 & 0 & 0 & 3.00 & 1.80 & D\\
\midrule
$u_{\max}=0.5$ & AS  & $>2$ & --- & 0.93 & 0.50 & 0.50 & 0.50 & 0.48 & N\\*
               & FT  & $>2$ & --- & 0.94 & 0.50 & 0.50 & 0.50 & 0.49 & N\\*
               & FxT & $>2$ & --- & 0.99 & 0.50 & 0.50 & 0.50 & 0.50 & N\\*
               & PdT & $>2$ & --- & 0.98 & 0.50 & 0.50 & 0.50 & 0.50 & N\\*
               & PT  & $>2$ & --- & 0.83 & 0.50 & 0.50 & 0.50 & 0.45 & N\\
\midrule
$u_{\max}=1$ & AS  & 1.576 & 0.575 & 0.43 & $5.3\times10^{-2}$ & $5.3\times10^{-2}$ & 1.00 & 0.93 & L\\*
             & FT  & 1.219 & 0.250 & 0.38 & $6.3\times10^{-2}$ & $6.3\times10^{-2}$ & 1.00 & 0.92 & L\\*
             & FxT & 1.045 & 0.668 & 0.47 & $1.5\times10^{-2}$ & $1.5\times10^{-2}$ & 1.00 & 0.98 & L\\*
             & PdT & 1.076 & 0.596 & 0.46 & $2.3\times10^{-2}$ & $2.3\times10^{-2}$ & 1.00 & 0.97 & L\\*
             & PT  & 0.999 & 0.098 & 0.50 & $1.0\times10^{-6}$ & $1.0\times10^{-6}$ & 1.00 & 1.00 & D\\
\midrule
$u_{\max}=1.1$ & AS  & 1.499 & 0.498 & 0.38 & $3.1\times10^{-2}$ & $3.1\times10^{-2}$ & 1.10 & 1.01 & L\\*
               & FT  & 1.153 & 0.184 & 0.32 & $3.4\times10^{-2}$ & $3.4\times10^{-2}$ & 1.10 & 0.99 & L\\*
               & FxT & 0.960 & 0.583 & 0.42 & 0 & 0 & 1.10 & 1.07 & D\\*
               & PdT & 0.994 & 0.514 & 0.41 & $4.3\times10^{-4}$ & $4.3\times10^{-4}$ & 1.10 & 1.06 & D\\*
               & PT  & 0.963 & 0.062 & 0.43 & 0 & 0 & 1.10 & 1.08 & D\\
\midrule
$u_{\max}=1.6$ & AS  & 1.269 & 0.268 & 0.24 & $6.4\times10^{-3}$ & $6.4\times10^{-3}$ & 1.60 & 1.42 & L\\*
               & FT  & 0.994 & 0.025 & 0.11 & $6.3\times10^{-4}$ & $6.3\times10^{-4}$ & 1.60 & 1.26 & D\\*
               & FxT & 0.701 & 0.324 & 0.27 & 0 & 0 & 1.60 & 1.53 & D\\*
               & PdT & 0.748 & 0.268 & 0.26 & 0 & 0 & 1.60 & 1.49 & D\\*
               & PT  & 0.916 & 0.015 & 0.22 & 0 & 0 & 1.60 & 1.41 & D\\
\midrule
$u_{\max}=2$ & AS  & 1.176 & 0.175 & 0.18 & $3.4\times10^{-3}$ & $3.4\times10^{-3}$ & 2.00 & 1.71 & L\\*
             & FT  & 0.969 & 0.000 & 0 & 0 & 0 & 2.00 & 1.33 & D\\*
             & FxT & 0.595 & 0.218 & 0.21 & 0 & 0 & 2.00 & 1.87 & D\\*
             & PdT & 0.651 & 0.171 & 0.19 & 0 & 0 & 2.00 & 1.80 & D\\*
             & PT  & 0.906 & 0.005 & 0.13 & 0 & 0 & 2.00 & 1.60 & D\\
\midrule
$h=7.3\times10^{-3}$ & AS  & 0.978 & --- & --- & --- & $7.9\times10^{-4}$ & 6.91 & 6.91 & D\\*
                     & FT  & 0.956 & --- & --- & --- & $5.3\times10^{-5}$ & 2.00 & 6.09 & D\\*
                     & FxT & 0.365 & --- & --- & --- & $2.1\times10^{-4}$ & 8.00 & 34.1 & D\\*
                     & PdT & 0.467 & --- & --- & --- & $1.3\times10^{-4}$ & 6.28 & 21.3 & D\\*
                     & PT  & 0.898 & --- & --- & --- & $1.1\times10^{-8}$ & 3.00 & 3.00 & D\\
\midrule
$h=6.7\times10^{-2}$ & AS  & 0.804 & --- & --- & --- & $4.8\times10^{-5}$ & 6.91 & 6.91 & D\\*
                     & FT  & $>2$ & --- & --- & --- & $4.5\times10^{-3}$ & 2.00 & 6.48 & P\\*
                     & FxT & $>2$ & --- & --- & --- & $1.9\times10^{-2}$ & 8.00 & 37.2 & P\\*
                     & PdT & $>2$ & --- & --- & --- & $1.1\times10^{-2}$ & 6.28 & 22.8 & P\\*
                     & PT  & 0.871 & --- & --- & --- & $6.2\times10^{-5}$ & 3.00 & 3.01 & D\\
\end{longtable}
}

\paragraph{Bounded disturbance.}
A continuous-time run with $d(t)=0.05\sin 5t$ illustrates the residual-set qualification of Section~\ref{subsec:specification_to_design}. For the three autonomous class laws, equating dissipation to $\bar d=0.05$ gives boundary radii $6.3\times10^{-4}$, $1.6\times10^{-4}$, and $2.5\times10^{-4}$ (FT, FxT, PdT). These are the sets the trajectories approach, whereas finite-time entry is certified only for the larger $\sigma$-margin sets; the observed amplitudes lie inside the boundary sets because the sinusoid does not realize its bound persistently, and since the radii are below $\varepsilon$ the observed times change by only about $3$-$5\%$: exact settling is lost, entry into a computable set in essentially the nominal time is retained. The asymptotic radius $\bar d/k_a=7.2\times10^{-3}$ exceeds $\varepsilon$, so that run is practical: without a finite-time mechanism the residual set is set by the gain alone. The prescribed-time law separates two events. Before $T_c$, $|x(t)|\le(1-t)^{3}+\tfrac{\bar d}{2}\bigl[(1-t)-(1-t)^{3}\bigr]$, so the divergent gain rejects the bounded disturbance at the terminal time (observed $E_T=2.4\times10^{-8}$ at the cutoff against the bound $2.5\times10^{-8}$); after $T_c$ the continuation $u=-3x$ sustains an oscillation of amplitude $\bar d/\sqrt{34}\approx9\times10^{-3}$, so retention is practical and set by the continuation gain, not by the prescribed-time mechanism. Arrival and retention are distinct properties, as in Table~\ref{tab:temporal_catalogue}.

\begin{figure*}[pos=!t]
\centering
\reviewfigure{0.92\textwidth}{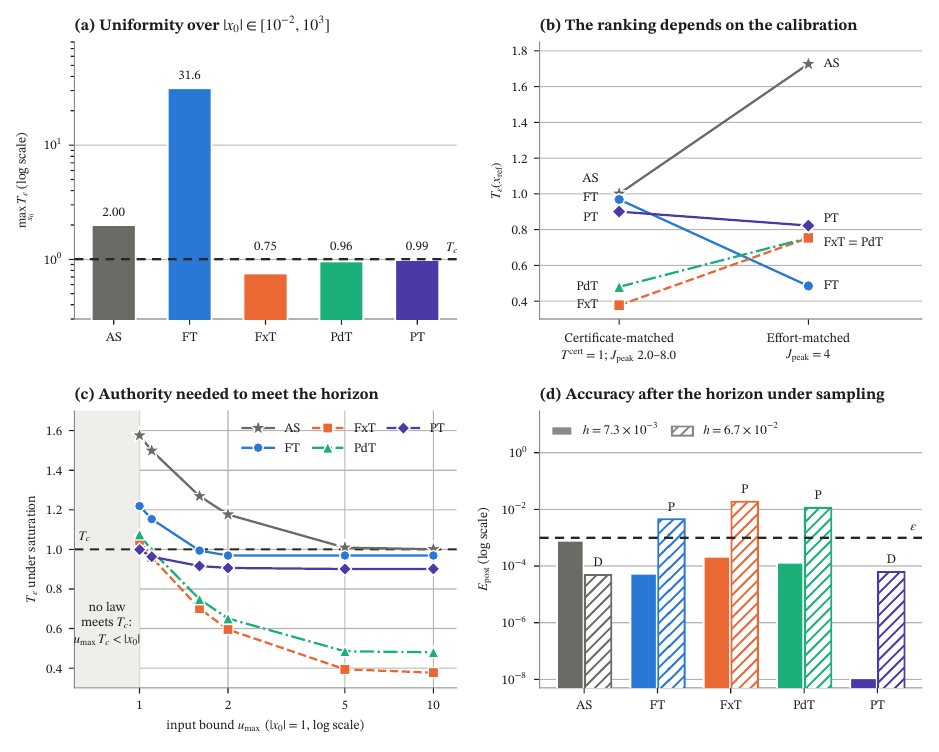}
\caption{Measured trade-offs of the five benchmark laws (values from Tables~\ref{tab:certificate_vs_observed}-\ref{tab:constrained_indices}).}
\label{fig:benchmark_tradeoffs}
\end{figure*}

\paragraph{Interpretation of the comparison.}
Under the choices stated in Section~\ref{subsec:benchmark_protocol}, the four class laws meet the same deadline wherever authority and sampling rate permit, whereas the asymptotic baseline meets only an $\varepsilon$-dependent entry time and needs its full nominal authority. The class laws differ in how the observed time depends on the initial condition (Fig.~\ref{fig:x0_sweeps}), what peak input a given time costs (Fig.~\ref{fig:S2_effort}), and how the guarantee degrades, late, practical, or lost, and through which term of the decay law when a constraint binds (Figs.~\ref{fig:S3_sampling} and~\ref{fig:S4_feasibility_map}). Superiority in observed settling time therefore does not imply superiority in effort, robustness, or implementability, and a temporal certificate should be read together with the region in which it can be realized. Figure~\ref{fig:benchmark_tradeoffs} collects these differences with values from Tables~\ref{tab:certificate_vs_observed}-\ref{tab:constrained_indices}; it describes this benchmark and these calibrations, not intrinsic properties of the classes.

\section{Open Problems and Future Research Directions}
\label{sec:open-problems}

The directions collected here are the unresolved structural questions exposed by the preceding sections, not a catalogue of application topics. They share one organizing question: under which initial set, accuracy, uncertainty, control-authority, reference, and implementation conditions can a selected convergence time actually be guaranteed? Answering it requires separating quantities that are routinely reported under the same name: a plant may reach the target within a horizon that a selected controller family cannot achieve, a controller may achieve it while the available certificate fails to prove it, and a continuous-time certificate may fail after sampling, saturation, filtering, estimation, or regularization.

Three questions condition most of the others. \emph{Feasibility}: a designer-assigned horizon has physical meaning only after the admissible initial set, actuator magnitude and rate, uncertainty, and safety or envelope requirements are shown to be compatible with it. \emph{Composition}: observers, filters, reaching phases, adaptation laws, learned models, and reference generators cannot remain outside the certificate of a loop described as having a single predefined or prescribed convergence time. \emph{Implementation}: sampling, computation, quantization, event triggering, regularization, and continuation at the terminal time should enter the temporal specification rather than be added after the proof. Sections~\ref{subsec:open_temporal}-\ref{subsec:open_digital} treat these questions in turn; Section~\ref{subsec:open_classes} collects extensions beyond finite-dimensional ordinary differential equations, and Section~\ref{subsec:open_evidence} the evidence needed to measure progress. Table~\ref{tab:open_problems}, at the end of the section, summarizes the problems in order of priority, together with the best available results, the missing element, and an indicator of progress.

\subsection{Feasible Horizons: Limits, Certificate Tightness, and Robust Accuracy}
\label{subsec:open_temporal}

\paragraph{Tightness of temporal certificates.}
Assignability and tightness are distinct properties (Sections~\ref{subsubsec:time_normalized_design} and~\ref{sec:numerical}): the benchmark two-power certificate \eqref{eq20} exceeds the actual supremum by $4/\pi$, whereas the normalized predefined-time bound is attained only as $|x_0|\to\infty$. Recursive designs add conservatism that belongs to the architecture, since norm and power-mean conversions accumulate with the system order (Section~\ref{subsubsec:recursive}). Existing results sharpen scalar comparison estimates \citep{aldana2019enhancing} and characterize tightness for autonomous normalized laws \citep{jimenez2020lyapunov}, but a broadly applicable theory remains lacking for composite Lyapunov functions of recursive designs and for augmented states with observer, filter, adaptation, or actuator dynamics. Reported results quantify the gap: in experiment, a fixed-time ISS formation bound of hundreds of seconds against much faster observed convergence \citep{Cheah2026FTISSFXISSFormation}, and, in simulation, microgrid restoration in about $3$~s against a certified $18.7$~s \citep{Yang2026FixedTimeMicrogridDoS}. The useful next step is not another sufficient settling-time inequality but a decomposition of
\[
T^{\mathrm{cert}}-\sup_{x_0\in\mathcal X_0}T^{\mathrm{exact}}(x_0)
\]
into contributions of the proof technique, the controller family, and the plant, together with Lyapunov or comparison functions that minimize the index \eqref{eq:settling_time_conservatism_index}, already a natural reporting variable, over a certified initial set; optimization-based, sum-of-squares, data-assisted, or trajectory-informed constructions are candidates when tightness, a common initial set, uncertainty, and implementation constraints must be treated jointly. Progress would appear as a ratio reproducibly close to one,
\[
\frac{T^{\mathrm{cert}}}
     {\sup_{x_0\in\mathcal X_0}T^{\mathrm{exact}}(x_0)}
\]
reported separately from numerical tolerance and implementation effects.

\paragraph{Converse results and characterizations.}
Sufficient Lyapunov conditions dominate the literature, whereas necessary conditions are scarce. For finite-time stability of continuous autonomous systems a converse Lyapunov theorem is available \citep{bhat2000finite}, and predefined-time stability of autonomous systems admits a Lyapunov-like characterization \citep{jimenez2020lyapunov}. No comparable characterization is known for the explicitly time-varying certificates of prescribed-time control, for practical designer-assigned-time properties, or for the augmented states of observer-based and recursive designs. Converse results would decide whether the failure to find a certificate reflects the property or the search, would identify which comparison functions can be used without loss of generality, and would supply the reference against which the conservatism index \eqref{eq:settling_time_conservatism_index} is measured. The open targets are converse theorems for prescribed-time convergence together with its continuation, including the regularity class of the resulting Lyapunov function, and robust converse results that retain the temporal class in the presence of the residuals of Section~\ref{subsec:specification_to_design}.

\paragraph{Physical lower bounds, controller-family limits, and admissible horizons.}
Section~\ref{subsec:feasible_decay} distinguished the minimum time achievable by the plant, the minimum horizon within a controller family, and the horizon certified by a particular construction. The benchmark made the distinction explicit through $u_{\max}T\ge |x_0|$, which the prescribed-time law with $\eta=1$ attains whereas other designs remain above it. Fixed-time control under input constraints has been treated by optimization \citep{garg2022fixedqp}, and prescribed-time safety formulations can yield explicit minimum recovery times under bounded authority \citep{constricting2026tubes}, but sharp lower bounds with nonlinear drift, actuator magnitude and rate, uncertainty, and a common initial set remain difficult to compute. Most temporal results constrain only the magnitude,
\[
\|u(t)\|\le U_{\max},
\]
whereas a physical actuator also imposes a rate constraint,
\[
\|\dot u(t)\|\le R_{\max}.
\]
An input admissible in magnitude may then be unrealizable, the minimum-time solution may contain a ramp-limited transient, and the minimum time depends on the initial input: the initial condition must then include the actuator state, $z_0=(x_0,u_0)$, and requiring $u(0)=0$ or leaving $u(0)$ free gives different limits. Below, $\mathcal X_0$ is a set of such initial conditions $z_0$, with $z_0=x_0$ when no actuator state is modeled. With the same quantifiers throughout, an initial set $\mathcal X_0$, an admissible input class $\mathcal U$ (magnitude, rate), and a family $\mathcal C$ of feedback laws whose applied inputs lie in $\mathcal U$, define
\[
T_{\mathrm{plant}}^{\min}(\mathcal X_0)=\sup_{z_0\in\mathcal X_0}\,\inf_{u\in\mathcal U}T(z_0,u),
\qquad
T_{\mathcal C}^{\min}(\mathcal X_0)=\inf_{\kappa\in\mathcal C}\,\sup_{z_0\in\mathcal X_0}T_\kappa(z_0),
\]
where $T(z_0,u)$ is the first time the open-loop input $u$ steers the plant from $z_0$ to the target and $T_\kappa(z_0)$ the settling time under $\kappa$. The first is a worst case of pointwise open-loop minima; the second requires one feedback law for the whole set, and $T_{\mathrm{plant}}^{\min}\le T_{\mathcal C}^{\min}$ because every closed-loop trajectory is generated by an admissible input. Since $T(z_0,u)$ is a first-arrival time and $T_\kappa(z_0)$ a settling time, the gap may also reflect the requirement of retaining the target; under uncertainty, the infimum is taken over non-anticipative strategies against the worst admissible disturbance. The design object is the admissible-horizon set
\[
\mathcal T_{\mathrm{adm}}(\mathcal X_0)
=\{T>0:\ \exists\,\kappa\in\mathcal C\text{ such that } T_\kappa(z_0)\le T\ \ \forall z_0\in\mathcal X_0\},
\]
whose lower boundary is $T_{\mathcal C}^{\min}$. Its dual is the deadline-dependent viable set
\[
\mathcal X_{\mathrm{adm}}(T_c,\varepsilon)
=
\left\{
z_0:
\text{the complete specification can be satisfied within }T_c
\right\},
\]
which, under actuation, safety, or rate constraints, is generally smaller than the domain of the unconstrained controller; a temporal statement without it can conceal a feasibility restriction, and characterizing how it shrinks with the horizon would permit fair comparison of controllers with the same nominal guarantee. The open problem is to make the lower boundary of $\mathcal T_{\mathrm{adm}}$ computable for broad nonlinear classes and to quantify the gaps
\[
T_{\mathrm{plant}}^{\min}(\mathcal X_0)
\le
T_{\mathcal C}^{\min}(\mathcal X_0)\le T_{\kappa,\sup}(\mathcal X_0)\le T^{\mathrm{cert}}_\kappa(\mathcal X_0),
\qquad \kappa\in\mathcal C,
\]
where $T_{\kappa,\sup}(\mathcal X_0)=\sup_{z_0\in\mathcal X_0}T_\kappa(z_0)$ and $T^{\mathrm{cert}}_\kappa$ is the certified bound, all four for the same initial set, accuracy, uncertainty, and admissible inputs. Such a theory would replace ``choose $T_c$ and increase the gains until the proof closes'' by ``compute the feasible temporal region and design within it,'' and would yield impossibility results in which no admissible controller, not merely none in the proposed family, meets the requested deadline.

\paragraph{Structural obstructions: internal dynamics, underactuation, and control direction.}
Feasibility is not only a question of authority. The mechanisms of Section~\ref{sec:mechanisms} act on a chain that the input can drive directly, and the prescribed-time designs reviewed here mostly assume normal-form or strict-feedback-like structures \citep{song2017time,krishnamurthy2020dynamic}. When the regulated output has internal (zero) dynamics, an assigned horizon can be imposed on the output only; after the output arrives, the internal state evolves under the zero dynamics, whose convergence the output horizon does not set, and if these dynamics are unstable, as in non-minimum-phase plants, exact output tracking within an arbitrarily short horizon may require unbounded internal states or inputs. Underactuated vehicles pose the same question for the unactuated degrees of freedom, and the corresponding designs typically certify a practical property of selected outputs \citep{liu2025auv}. Unknown control directions have been handled through a prescribed-time Nussbaum lemma \citep{Li2026EventTriggeredUnknownCoefficientsPsT}, in which the Nussbaum variable is shown only to remain bounded, so its transient is not part of the temporal claim. The open problem is to state, for a given structure, relative degree, stability and dimension of the internal dynamics, actuation rank, and sign uncertainty, which variables admit an assigned-time property, of which strength, and with which admissible input class. Such results would complement the authority-based limits above with structure-based possibility and impossibility statements.

\paragraph{Uniform margins and the meaning of an assigned time.}
The inequality $T(x_0)<T_c$ for every initial condition does not imply a common positive margin
\[
T_c-\sup_{x_0}T(x_0)>0;
\]
for the benchmark predefined-time law the supremum equals $T_c$ although every finite initial condition settles earlier, and the analogous issue arises for the initialization time $t_0$ when a task is restarted (Section~\ref{subsec:prescribed_time_control}). Few certificates provide a reserve: the frozen-gain criterion of \citet{Cui2026PrescribedTimeMorphingHMV} bounds every settling time by $T(1-e^{-k_2/T})<T$, whereas designs that settle exactly at the assigned instant, nominally in \citet{Shao2026TTRPrescribedPerformanceSMC} and by construction in \citet{Ding2026PDFSuperTwistingPsT}, have none. Since a zero nominal margin leaves no room for sampling, estimation, computation, or disturbance-induced delay, a certificate should state the guaranteed arrival, practical-entry, or restart margin over the certified initial set and initialization interval.

\paragraph{Horizon-dependent robustness and practical accuracy.}
A nonvanishing residual replaces exact settling by a residual set (Section~\ref{subsec:specification_to_design}), and shortening the horizon has competing effects: stronger dissipation may reduce the deterministic error, while larger gains amplify noise, discretization error, unmodeled dynamics, and actuator imperfections. The terminal accuracy is therefore a function
\[
r =
r\!\left(
T_c,
\delta,
h,
\nu,
U_{\max},
R_{\max},
\ldots
\right),
\]
of model or disturbance bounds $\delta$, sampling period $h$, and measurement-noise level $\nu$. The conflict is sharpest for prescribed-time gains, which multiply a nonvanishing measurement error directly. Inherent robustness limitations have been established for prescribed-time algorithms with an absolute deadline \citep{aldana2023inherent}, and bounded-gain redesigns have been proposed \citep{aldana2023redesign,su2025prescribed}, including exact settling with bounded gains through periodic delayed feedback, where noise again leaves a residual \citep{Ding2026PDFSuperTwistingPsT}: exactness does not require an unbounded gain, but it does not remove the noise problem. Missing is a comparison of fixed-, predefined-, and prescribed-time mechanisms under equal physical authority, reported as worst-case noise-to-terminal-error gains and horizon-accuracy curves that show where shorter horizons cease to improve practical accuracy.

\paragraph{Recovery limits after disturbances, faults, and model changes.}
Most formulations pose convergence from an initial time with a fixed disturbance model. When the disturbance, reference, actuator effectiveness, or model changes abruptly during operation, the relevant quantities are the peak deviation and the recovery time. If the corrective input, an adaptive gain, or an actuator state has a bounded rate, a nonzero transient is unavoidable, suggesting lower bounds
\[
E_{\mathrm{peak}}
\ge
\underline E
\left(
\Delta_d,
R_{\max},
\Gamma,
\mathcal X
\right),
\qquad
T_{\mathrm{rec}}
\ge
\underline T_{\mathrm{rec}}
\left(
\Delta_d,
R_{\max},
\Gamma,
\mathcal X
\right),
\]
where $\Delta_d$ characterizes the disturbance change and $\Gamma$ the adaptation or internal-controller rate. A recovery theory for designer-assigned-time control should state which changes can be rejected within a recovery horizon, what actuator reserve this requires, what transient error is unavoidable, and whether repeated events require a minimum dwell time.

\paragraph{Zero-width envelopes, safety, and actuator feasibility.}
Prescribed-performance and funnel architectures constrain the whole transient, but a nonzero terminal width yields a practical result, and saturation-tolerant or flexible-bound designs that widen the envelope or the nominal input interval online \citep{tan2026flexible} no longer meet the original specification. Likewise, when safety and a temporal CLF constraint conflict, the temporal condition is usually relaxed without computing the resulting deadline degradation. Fixed-time optimization relates initial set, authority, and horizon \citep{garg2022fixedqp}; prescribed-time safety designs assign the time at which a safe set is reached \citep{abel2024prescribedSafety}; constricting tubes expose a minimum recovery time under input constraints \citep{constricting2026tubes}; prescribed-time proximity control with collision-avoidance and visibility constraints has been obtained without input bounds \citep{Shao2026PsTSpacecraftSafety}; and rigid-formation control with collision avoidance and connectivity has been validated experimentally with a practical deadline \citep{Zhou2026PracticalPsTRigidFormation}. Open is a joint feasibility theory for
\[
\text{safety}
+
\text{transient envelope}
+
\text{terminal accuracy}
+
\text{deadline}
+
\text{input magnitude/rate}
\]
that distinguishes remaining safe, returning to a safe set, and reaching a target without leaving it, and that returns the induced change of deadline or accuracy as an output of the analysis rather than as an unspecified slack variable.

\paragraph{Fundamental deadline-effort and peaking-deadline trade-offs.}
Even the scalar benchmark exhibits nontrivial Pareto relations between peak input and settling time (Section~\ref{subsec:numerical_constraints}). Inverse-optimal prescribed-time designs \citep{li2021stochastic} and optimal predefined-time stabilization \citep{jimenez2017optimal} address optimality within selected classes, but the physical cost of different mechanisms has not been compared in general. Peak amplitude is one metric among energy,
\[
\int_0^\infty \|u(t)\|^2\,dt,
\]
total variation, slew rate, saturation duration, and intermediate state peaking, and reducing a feedback gain, the nominal command, or the applied input are different objectives that damping, feedforward, saturation, and actuator dynamics can move in different directions. The open question is the Pareto surface connecting deadline, accuracy, peak input, energy, input rate, and state excursion, and in particular lower bounds on the effort or peaking that is unavoidable as the horizon approaches the physical minimum, the counterpart of fundamental bandwidth and peaking limits in other areas of feedback theory.

\paragraph{Relation to optimal and predictive control.}
Finite-horizon optimal control and model predictive control with terminal constraints also generate inputs that reach a target at a specified time, and the minimum-time problem defines the plant limit $T_{\mathrm{plant}}^{\min}$ directly. Optimal predefined-time stabilization \citep{jimenez2017optimal}, inverse-optimal prescribed-time design \citep{li2021stochastic}, safe predefined-time optimal feedback \citep{kokolakis2024safe}, and fixed-time quadratic programs \citep{garg2022fixedqp} connect the two views only for particular classes. A systematic comparison is missing: under the same initial set, authority, and accuracy, how far Lyapunov-shaped temporal laws are from the optimal or predictive solution in effort and robustness, and whether receding-horizon schemes with a shrinking horizon can inherit a certified deadline under recursive feasibility. An answer would measure designer-assigned-time controllers against established optimal-control baselines rather than only against asymptotic designs.

\subsection{Composition: Auxiliary Dynamics Inside the Certificate}
\label{subsec:open_augmented}

\paragraph{A separation substitute for nonlinear temporal architectures.}
The linear separation principle is generally unavailable for nonsmooth or explicitly time-varying feedback (Section~\ref{subsubsec:observer_output}), and the same problem arises for reaching phases, dynamic-surface or command-filtered backstepping, adaptive compensation, disturbance observers, reference filters, and actuator dynamics. Common-horizon constructions \citep{tan2026decentralized,que2024disturbance} and sequential exact observer and tracking horizons \citep{ji2026prescribed} are strong but architecture-specific, while prescribed-time input-to-state stability for cascades \citep{krishnamurthy2024cascade} provides a compositional route. A temporal property established for each subsystem separately does not by itself imply the same property for the interconnection. The open problem is a general replacement for separation: conditions through temporal ISS gains, small-gain inequalities, residual bounds, or relative horizons under which an auxiliary subsystem with a weaker property preserves the guarantee of the regulated state, covering heterogeneous components, a finite-time observer with a predefined-time controller, or a practical filter with an exact nominal design, and ending in a certificate for the augmented state that names the property proved for each auxiliary error.

\paragraph{Hidden transients and allocation of temporal budgets.}
Observer peaking, reaching transients, command-filter errors, adaptation transients, and actuator lag can worsen as their assigned horizons shorten but are invisible in a nominal Lyapunov inequality, so a controller can satisfy its terminal-time requirement with an unacceptable intermediate excursion. Budget relations such as
\[
T_1+T_2+\cdots+T_m\le T_d
\]
are useful for sequential phases but conservative for interacting subsystems. The open problem is to distribute a global deadline among estimation, reaching, filtering, adaptation, and plant dynamics so as to minimize peak error, input demand, noise sensitivity, or computation while preserving the final guarantee; quantitative peaking-deadline and transient-deadline bounds would turn this tuning heuristic into an optimization problem.

\paragraph{Excitation, adaptation, and unavoidable identification time.}
Composite schemes such as \citet{jiang2025composite} show that excitation can strengthen a practical regulation result, but the time to acquire informative data and the subsequent estimator convergence time are distinct quantities that certificates rarely separate (Section~\ref{subsubsec:adaptive_learning}). Two problems are open: the minimal excitation level and duration for finite- or fixed-time identification, characterized independently of the regulation horizon, and lower bounds on the transient after an abrupt uncertainty or disturbance change when both the adaptive law and the actuator are rate-limited. An excitation-level-horizon relation would make the identification time an explicit part of the temporal claim instead of an implicit assumption about the data.

\paragraph{Learning that is certified rather than evaluated.}
An approximately solved Hamilton-Jacobi-Bellman equation or a neural approximation may improve nominal performance without establishing a temporal guarantee, and approximation errors that enter practical bounds are not always tied to the region visited by the closed loop. A stronger framework needs a temporal certificate in which the approximation error is a named parameter and an independently verifiable bound on that error over the certified reachable set; when the learned model is updated online, maintaining the certificate under distribution shift becomes part of the same problem. Progress would take the form
\[
T_\varepsilon
\le
\mathcal T
\left(
\bar e_{\mathrm{model}},
\mathcal X_0,
U_{\max},
R_{\max},
\ldots
\right)
\]
together with a procedure for validating $\bar e_{\mathrm{model}}$ on the relevant closed-loop region.

\paragraph{Feasibility of moving references.}
In tracking, the reference itself consumes control authority: a kinematically admissible reference may require an equivalent input beyond the magnitude or rate limits, so that no controller can enforce the error deadline. Necessary conditions should connect reference derivatives ($\dot r$, $\ddot r$, or higher for larger relative degree), plant dynamics, disturbance bounds, and actuator constraints before a convergence law is selected. Reference governors and trajectory filters restore feasibility but change the task, so the guarantee should distinguish convergence to the original reference from convergence to a modified one; the least modification that retains a deadline under bounded authority, and a reference-feasibility region reported separately from controller feasibility, are natural targets.

\subsection{Implementation: Terminal Behavior and Finite-Rate Realizations}
\label{subsec:open_digital}

\paragraph{Terminal regularity, continuation, and post-arrival invariance.}
Exact finite-time convergence of a continuous autonomous system requires loss of local Lipschitz regularity at the equilibrium, whereas time-varying prescribed-time feedback can remain Lipschitz in the state (Section~\ref{subsec:homogeneous_mechanisms}); the boundary between specifications achievable in each way is not fully characterized. For laws with terms that become undefined or unbounded at the terminal time, a result on $[0,T_c)$ is incomplete without the solution concept and the continuation at and after $T_c$ (Sections~\ref{subsec:prescribed_time_control} and~\ref{subsubsec:time_varying_pt}), and the benchmark showed that the continuation decides what happens after a nonzero terminal miss (Section~\ref{subsec:numerical_constraints}). Switching the scaling to a constant at $t_f$ \citep{Li2026EventTriggeredUnknownCoefficientsPsT}, gating it off after the observer and controller horizons \citep{ji2026prescribed}, and periodic delayed feedback with bounded gains \citep{Ding2026PDFSuperTwistingPsT} give constructions on $[0,\infty)$ for particular classes; generalized Filippov formulations offer another route \citep{seeber2023generalized}; practical prescribed-time formulations exist on an infinite interval \citep{cao2022practical}; and practical entry followed by asymptotic convergence \citep{Golestani2026PracticalPrescribedTimePPC,Zhou2026PracticalPsTRigidFormation}, or exact arrival with zero state and input afterwards \citep{Li2026EventTriggeredUnknownCoefficientsPsT}, cover both phases for particular designs without treating disturbances after arrival.

The open problem is one certificate on $[0,\infty)$ that states the solution concept and covers convergence, the terminal transition, and invariance after arrival, with the terminal miss as the initial condition of the continuation and later disturbances entering through a post-arrival robustness bound. It would separate three statements often conflated, reaching the target by the deadline, remaining there, and recovering after a later perturbation, and matters most for hybrid, sampled, and switched implementations, where the mathematical terminal event and the physical controller transition differ.

\paragraph{The joint price of regularization and digital implementation.}
Boundary layers, continuous approximations, gain caps, dead zones, and smoothing replace exact nonsmooth behavior by a practical one, and sampling adds a second regularization: the benchmark emulation of Section~\ref{sec:numerical} exhibits the characteristic two-cycle scale
\[
a_h=\left(\frac{hk}{2}\right)^{1/(1-p)}
\]
of the held sublinear update, a property of that emulation rather than a general accuracy bound; in the spacecraft design of \citet{Shao2026PsTSpacecraftSafety}, truncating the time transformation before $T$ restores continuity but turns exact arrival into entry into a neighborhood. These parameters are usually tuned independently of the horizon. Beyond the one-step conditions derived for the benchmark (two-cycle, large-state contraction boundary, loss of contraction near the terminal time), the open target is a computable map from
\[
(T_c,\varepsilon,h)
\]
to the minimum regularization and actuator requirements, and in particular a horizon-dependent sampling limit
\[
h_{\max}
=
h_{\max}
\left(
T_c,\varepsilon,
\mathcal X_0,
U_{\max},
R_{\max},
\delta
\right)
\]
for nonlinear multidimensional plants, recursive designs, and observer coupling, with computation delay, quantization, finite word length, and actuator update rate in the same bound. Such a result would show when an ``exact'' continuous-time certificate becomes only a practical $\varepsilon$-settling statement after implementation.

\paragraph{Class-preserving discretization.}
Emulation, direct sampled-data design, and discretization constructed to preserve the continuous-time mechanism are different problems. Consistent discretizations preserve finite- and fixed-time properties for homogeneous systems \citep{polyakov2019consistent,polyakov2023consistent}; exact nominal discretization exists for a class of predefined-time systems \citep{jimenez2020consistent}; implicit discretizations of the super-twisting algorithm, including saturation, remove numerical chattering \citep{brogliato2020implicit,seeber2025implicitSTA}; and implicit-Euler-based digital prescribed-time designs, with output-feedback extensions under noise, are a first step for time-varying control \citep{orlov2026digital}. What remains is a discrete-time theory that carries the temporal class without relying on small-$h$ emulation, in particular for recursive architectures, where the discretization of one virtual-control level enters the next and observer and adaptation states add further clocks; the target is a class-preserving certificate for the complete augmented loop.

\paragraph{Computational synthesis and verification of temporal certificates.}
Temporal certificates are almost always constructed by hand, and their validity after implementation is usually checked by simulation. Two computational problems are open. The first is synthesis: searching for Lyapunov or comparison functions that certify a stated class with minimal conservatism, for which sum-of-squares and other optimization-based constructions (Section~\ref{subsec:open_temporal}) are natural candidates but must accommodate fractional powers, time-varying scaling, and the nonsmooth or terminal behavior discussed above. The second is verification: establishing, by reachability analysis or other set-based methods, that a sampled, quantized, and saturated implementation enters the accuracy region within the deadline from every point of the certified initial set, rather than from simulated initial conditions only. Verified statements of this kind would turn observed feasibility classes such as those of Section~\ref{sec:numerical} into certified ones and are a prerequisite for using designer-assigned-time controllers where formal certification is required.

\paragraph{Event triggering, finite communication rates, and Zeno behavior.}
Exact finite-time convergence and Zeno-free event triggering can be in formal tension \citep{zeno2020finite}, and prescribed-time gains push communication frequency up near the terminal time. Prescribed-time event- and self-triggered schemes exclude Zeno behavior under particular conditions \citep{ning2023eventPrescribed,ning2023adaptiveEventPrescribed,Li2026EventTriggeredUnknownCoefficientsPsT,Zhang2026DynamicSelfTriggeredBipartitePsT}, and periodic event triggering gives a trivial lower bound on the inter-event interval. The broader problem is to make the minimum inter-event time
\[
\tau_{\min}
=
\tau_{\min}(T_c,\varepsilon,\delta,\ldots)
\]
a design variable, with the control or accuracy penalty required to enlarge it, and to fit the computation of the trigger or online optimization within that interval.

\paragraph{Scaling over networks and heterogeneous resources.}
Beyond retaining a convergence class as the number of agents grows (Section~\ref{subsubsec:distributed}), the required gain or effort must be related to connectivity, network size, delay, packet loss, clock mismatch, and quantization, and agents may differ in actuator capacity and local sampling rate. Delay- and DoS-dependent prescribed-time inequalities \citep{Zhu2026PrescribedTimeDelayDoS,Zhang2026PrescribedTimeConsensusDelayDoS} are a first step, but their gains are not related to network size or actuator capacity. A scalable theory should decide whether a common physical deadline can be retained under heterogeneous resources and, if not, how agent-specific horizons are assigned, through laws relating the gains to the algebraic connectivity $\lambda_2$ of the graph Laplacian, compositional communication-layer certificates, and conditions valid under switching or intermittent graphs.

\subsection{Temporal Guarantees Beyond the Finite-Dimensional ODE Setting}
\label{subsec:open_classes}

The framework of Sections~\ref{sec:notions}-\ref{sec:mechanisms} is built around finite-dimensional ordinary differential equations with classical, Carath\'eodory, or Filippov solutions. With memory, stochasticity, distributed parameters, or hybrid evolution, even the object called ``settling time'' can change, so extending the vocabulary requires more than carrying an ODE Lyapunov inequality into another class.

\paragraph{Memory, fractional dynamics, and delay.}
For systems with memory, arrival of the visible state at the target does not erase the influence of the past. For Caputo-type fractional-order systems with a continuous right-hand side and the classical equilibrium concept, exact finite-time stability is excluded under standard assumptions \citep{shen2014nonexistence}; consequently, finite-, predefined-, or prescribed-time claims in fractional-order systems \citep{munoz2021predefined,munoz2022second} must be interpreted with particular care in terms of derivative definition, target concept, solution concept, memory compensation, or practical accuracy.

This leads to a broader open question than the existence of finite-time convergence itself: the \emph{memory cost} of maintaining a target after nominal arrival. If the implementation retains only a finite history window, truncates the fractional memory, or resets an internal state at the deadline, the post-arrival error should be related explicitly to the memory horizon and the selected convergence horizon. A complete certificate should therefore cover both target entry and persistence. Time-delay systems pose the analogous problem for functional states: reaching the current-state target is not generally equivalent to settling the complete history segment. Delay-dependent prescribed-time inequalities keep the delayed term in the certificate \citep{Zhu2026PrescribedTimeDelayDoS}, but their conclusion concerns the current state, not the history segment. Appropriate temporal definitions for these systems, together with implementable memory-aware certificates, remain open.

\paragraph{Stochastic temporal guarantees.}
Prescribed-time and finite-time results for stochastic nonlinear systems exist \citep{li2022prescribed,li2021stochastic,yin2011finite}, but the temporal statements are fundamentally different. With $T_0$ the stochastic settling time, an expected settling-time bound,
\[
\mathbb E[T_0]\le T,
\]
a probabilistic deadline,
\[
\mathbb P(T_0\le T)\ge 1-\delta_p,
\]
and an almost-sure deadline are not interchangeable. Recent criteria bound $\mathbb E[T_0]$ and let the diffusion term itself supply the decay, so that $\mathcal LV$ need not be negative \citep{Peng2026NoiseControlFiniteTime}; a predefined bound in expectation is not a deadline for every sample path. Comparisons across designs are therefore meaningful only when the same probabilistic object is certified.

The open direction is a unified specification and comparison framework in which the probability level, noise intensity, certified initial set, and control authority are explicit. High-probability and almost-sure temporal guarantees under bounded actuation are particularly important because increasing the gain cannot arbitrarily suppress diffusion without physical cost.

\paragraph{Distributed-parameter, switched, and hybrid systems.}
For partial differential equations, convergence must be stated in a specified norm and the relation between a reduced-order controller and the distributed state must be quantified \citep{polyakov2018homogeneous,coron2013local}. Recent networked PDE-ODE designs still pair a finite-time performance envelope with only asymptotic tracking \citep{Yao2026FlexibleManipulatorsPrescribedPerformance}, and adaptive event-triggered control of hyperbolic PDE-ODE cascades under deception attacks proves only ultimate boundedness \citep{Zhang2026NussbaumHyperbolicPDE}; boundary actuation adds a constraint without analogue in an unconstrained scalar ODE. For switched and hybrid systems, physical time, flow time, jump count, and dwell-time conditions must be separated \citep{you2024finite,ding2026switching}. The common open problem is which temporal mechanisms survive model reduction, switching, jumps, or infinite-dimensional dynamics, judged by a class-appropriate temporal object and a certificate stated in the native solution concept of the class.

\subsection{Benchmarks, Failure Boundaries, and Experimental Validation}
\label{subsec:open_evidence}

\paragraph{What the audit reveals.}
Within the studies audited in Table~\ref{tab:audit_full}, hardware and hardware-in-the-loop validation concentrate on fast-sampled electromechanical platforms, while cooperative, constraint-aware, and implementation-aware architectures are rarely tested at that level. Reported quantities are also not standardized: analytical bounds, tolerance-based numerical times, nominal commands, applied inputs, and practical terminal errors are often compared as if equivalent. The criteria of Section~\ref{subsec:comparison_criteria} and the protocol of Section~\ref{subsec:benchmark_protocol}, certificate- and effort-matched calibration with a persistence-based settling measure, provide a reporting basis; observed-time matching alone can conceal a large gap between guarantee and transient.

\paragraph{Benchmarks beyond the scalar mechanism.}
The scalar integrator isolates the mechanism layer but cannot reveal architectural issues. A shared suite should include at least a relative-degree-two system under sliding-mode control (reaching versus sliding), a strict-feedback system under backstepping (recursive certificates and command filters), an output-feedback system (observer peaking and separation failure), and a constrained nonlinear plant with actuator amplitude and rate limits (feasible-horizon boundaries), all run over common initial sets, disturbance families, sampling grids, and accuracy definitions. Besides successful trajectories, the suite should report the boundary between success and failure: an infeasible requested deadline is informative, since it identifies a physical or architectural limit that a single successful simulation cannot reveal.

\paragraph{Evidence levels and reproducible implementation.}
Continuous-time simulation, sampled simulation, real-time or hardware-in-the-loop validation, and physical experiment are different evidence levels and should be identified explicitly, with sampling period, solver, actuator model, saturation and rate limits, regularization parameters, continuation law, filtering, initialization, and failed trials reported together with the gains. A reproducible study should let another researcher reconstruct the certified region, the experimentally successful region, and the failure boundary without re-deriving the paper; open implementations and standardized evaluation scripts would allow mechanisms and architectures to be compared on the same physical basis rather than on isolated trajectories.

\paragraph{Measuring progress.}
Taken together, progress on these problems can be judged by three criteria. The first is computable feasibility limits: plant, controller-family, and certified minimum times reported for a stated initial set, input class, and accuracy, so that an assigned horizon can be checked against what is attainable before a law is tuned. The second is temporal certificates that contain the auxiliary dynamics: observers, filters, reaching phases, adaptation, learned models, and reference generators enter the certificate with named properties instead of being assumed fast enough. The third is guarantees that are preserved under implementation: sampled, event-triggered, saturated, and regularized loops, including the continuation after the deadline, carry a certified temporal class rather than an emulated approximation of it. Results that meet all three would turn a designer-assigned convergence time from a nominal parameter into a verifiable specification. Table~\ref{tab:open_problems} collects the problems of this section with these criteria in view.

{\setlength{\tabcolsep}{3pt}\renewcommand{\arraystretch}{1.12}\footnotesize
\begin{longtable}{@{}>{\raggedright\arraybackslash}p{1.45cm} >{\raggedright\arraybackslash}p{2.7cm} >{\raggedright\arraybackslash}p{3.8cm} >{\raggedright\arraybackslash}p{3.8cm} >{\raggedright\arraybackslash}p{3.4cm}@{}}
\caption{Open problems at a glance. Priority: high if the problem limits the interpretation of most existing temporal claims; medium if it matters mainly for specific classes or settings.}\label{tab:open_problems}\\
\toprule
\textbf{Priority} & \textbf{Problem (Section)} & \textbf{Best available results} & \textbf{Missing element} & \textbf{Indicator of progress}\\
\midrule
\endfirsthead
\multicolumn{5}{@{}l}{\footnotesize\textit{Table~\ref{tab:open_problems} (continued)}}\\
\toprule
\textbf{Priority} & \textbf{Problem (Section)} & \textbf{Best available results} & \textbf{Missing element} & \textbf{Indicator of progress}\\
\midrule
\endhead
\midrule\multicolumn{5}{r@{}}{\footnotesize\textit{continued on next page}}\\
\endfoot
\bottomrule\endlastfoot
High & Computable feasible horizons and impossibility (\ref{subsec:open_temporal}) & QP-based fixed-time design under input constraints \citep{garg2022fixedqp}; minimum recovery time under bounded authority \citep{constricting2026tubes}; scalar plant limit $u_{\max}T\ge|x_0|$ (Section~\ref{sec:numerical}) & Computable $T_{\mathcal C}^{\min}$ and $\mathcal X_{\mathrm{adm}}(T_c,\varepsilon)$ with nonlinear drift and magnitude and rate limits & Chain $T_{\mathrm{plant}}^{\min}\le T_{\mathcal C}^{\min}$ $\le T_{\kappa,\sup}\le T_\kappa^{\mathrm{cert}}$ reported on one initial set\\
High & Structural obstructions (\ref{subsec:open_temporal}) & Normal-form and strict-feedback-like designs \citep{song2017time,krishnamurthy2020dynamic}; prescribed-time Nussbaum lemma \citep{Li2026EventTriggeredUnknownCoefficientsPsT}; practical results for underactuated vehicles \citep{liu2025auv} & Assigned-time results with internal dynamics, non-minimum phase, and underactuation & Possibility and impossibility statements per variable and input class\\
High & Separation substitute and temporal budgets (\ref{subsec:open_augmented}) & Common-horizon constructions \citep{tan2026decentralized,que2024disturbance}; sequential horizons \citep{ji2026prescribed}; prescribed-time ISS cascades \citep{krishnamurthy2024cascade} & General ISS or small-gain conditions for heterogeneous subsystems; peaking-deadline bounds & Certificate for the augmented state that names the property of each auxiliary error\\
High & Terminal continuation on $[0,\infty)$ (\ref{subsec:open_digital}) & Switched or gated scaling \citep{Li2026EventTriggeredUnknownCoefficientsPsT,ji2026prescribed}; periodic delayed feedback \citep{Ding2026PDFSuperTwistingPsT}; generalized Filippov solutions \citep{seeber2023generalized} & One certificate for arrival, terminal transition, retention, and post-arrival robustness & Post-arrival robustness bound with the terminal miss as initial condition\\
High & Implementation-aware certificates (\ref{subsec:open_digital}) & Consistent discretization \citep{polyakov2019consistent,polyakov2023consistent,jimenez2020consistent}; implicit digital prescribed-time design \citep{orlov2026digital} & $h_{\max}(T_c,\varepsilon,\mathcal X_0,$ $U_{\max},R_{\max},\delta)$; class-preserving discrete certificates for recursive loops; computational verification & Verified sampled-data statements from every point of the initial set\\
High & Benchmarks and evidence (\ref{subsec:open_evidence}) & Scalar benchmark (Section~\ref{sec:numerical}); hardware evidence concentrated on electromechanical platforms (Table~\ref{tab:audit_full}) & Shared multi-architecture suite with failure boundaries and open implementations & Reproducible success and failure regions across laboratories\\
Medium & Tightness and converse characterizations (\ref{subsec:open_temporal}) & Converse theorem for finite-time stability \citep{bhat2000finite}; predefined-time characterization \citep{jimenez2020lyapunov}; sharpened estimates \citep{aldana2019enhancing} & Converse results for prescribed-time and augmented or recursive states; tight composite certificates & $T^{\mathrm{cert}}/\sup T^{\mathrm{exact}}$ close to one\\
Medium & Robustness, recovery, and effort trade-offs (\ref{subsec:open_temporal}) & Inherent prescribed-time limitations \citep{aldana2023inherent}; bounded-gain redesigns \citep{aldana2023redesign,su2025prescribed}; optimal and inverse-optimal designs \citep{jimenez2017optimal,li2021stochastic} & Noise-to-terminal-error gains and effort lower bounds under equal authority; comparison with optimal and predictive baselines & Horizon-accuracy and deadline-effort curves\\
Medium & Certified learning and identification time (\ref{subsec:open_augmented}) & Composite learning with excitation \citep{jiang2025composite} & Verifiable model-error bound on the certified reachable set; excitation-horizon relation & $T_\varepsilon\le\mathcal T(\bar e_{\mathrm{model}},\ldots)$ with validated $\bar e_{\mathrm{model}}$\\
Medium & Networks, triggering, and non-ODE classes (\ref{subsec:open_digital}, \ref{subsec:open_classes}) & Zeno-free prescribed-time triggering \citep{ning2023eventPrescribed,Zhang2026DynamicSelfTriggeredBipartitePsT}; delay- and DoS-dependent inequalities \citep{Zhu2026PrescribedTimeDelayDoS}; nonexistence for Caputo systems \citep{shen2014nonexistence}; expected settling-time bounds \citep{Peng2026NoiseControlFiniteTime} & $\tau_{\min}$ and gains versus $\lambda_2$, network size, and resources; memory-aware and probabilistic deadlines & Deadlines stated together with network, memory, or probability parameters\\
\end{longtable}
}

\section{Conclusions}
\label{sec:conclusions}

The progression from asymptotic to designer-assigned-time control is a progression in the temporal information a guarantee supplies, not a ranking of controllers. Reading the literature in the three layers of this review leads to four conclusions.

First, a temporal claim must state its quantifiers and target: a finite settling time for each admissible initial condition, a uniform bound, a selectable bound whose tightness is a separate property, or a deadline or terminal limit whose exact first arrival must be stated separately. Practical convergence concerns entry into and retention within a specified set, not exact convergence to the origin.

Second, the decay law is shared across architectures, and the architectural realization determines whether the nominal guarantee is preserved. Reaching transients, composite recursive functions, observer separation, adaptation residuals, and graph dependence change the variable the certificate concerns and the auxiliary dynamics it must contain; Table~\ref{tab:audit} and Appendix~\ref{app:audit} record these qualifications for the examined set, not prevalence in the literature.

Third, plant reachability, the limit of a controller family, the performance of a fixed tuning, and its certified bound are different quantities. The scalar benchmark separates them at the mechanism layer only; it does not validate the architectural synthesis numerically, and none of its comparisons orders the stability classes.

Fourth, implementation is part of the theorem. Saturation, regularization, sampling, and event triggering each create a new closed loop, and a continuous-time certificate carries over only when the implemented loop is certified.

These conclusions define the three foundational problems of Section~\ref{sec:open-problems}, feasible horizons, compositional certificates, and class-preserving implementation, on which the conversion of practical and emulated evidence into certified temporal statements depends.

\printcredits

\section*{Declaration of competing interest}
The author declares that he has no known competing financial interests or personal relationships that could have appeared to influence the work reported in this paper.

\section*{Funding}
This research received no specific grant from any funding agency in the public, commercial, or not-for-profit sectors.

\section*{Data availability}
No new experimental data were generated. The MATLAB scripts that reproduce the numerical comparison of Section~\ref{sec:numerical} are available from the author upon reasonable request.

\section*{Declaration of generative AI and AI-assisted technologies in the writing process}
 During the preparation of this work the author used Claude by Anthropic in order to language editing. After using this tool, the author reviewed and edited the content as needed and takes full responsibility for the content of the published article.

\appendix
\setcounter{table}{0}
\renewcommand{\thetable}{A\arabic{table}}
\section{Architectural Audit of the Examined Studies}
\label{app:audit}

Table~\ref{tab:audit_full} records the study-level audit summarized in Section~\ref{subsec:design_guidance}; Table~\ref{tab:audit} reproduces the entries that mark the boundaries of the certificates discussed in Section~\ref{sec:architectures}.

\noindent Architectural audit of the design studies examined in Section~\ref{sec:architectures}, together with additional representative designs, coded by the criteria of Section~\ref{subsec:comparison_criteria}; studies cited in Section~\ref{sec:architectures} only for a single property are not coded. Each design is placed in the group in which its distinguishing element is discussed: the group of the architecture that generates its physical input, or a group of Sections~\ref{subsec:augmented_architectures}-\ref{subsec:implementation_architectures} when the distinguishing element is an observer, adaptation or learning, cooperation, or constraint handling; mechanisms and augmentations may overlap. Entries are coded from the strongest property its theorems prove for the certified variable, which is the regulated variable unless stated otherwise (Section~\ref{subsec:comparison_criteria}), not from its title or conclusions. \emph{Source class}: FT finite-, FxT fixed-, PdT predefined-, PT prescribed-time; AT arbitrary-time (Remark~\ref{rem:arbitrary_time}); ST short-time stability (Remark~\ref{rem:two_fts}); B bounded signals only. \emph{G} (guarantee): E exact for the certified variable; E\textsubscript{r} exact for the regulated variable, auxiliary errors only bounded or residual; P practical (entry into a residual set); P$^\ast$ practical with a qualified horizon (see text); U uniform ultimate boundedness or envelope result; E$^\dagger$ exact for a safety target; E$^\ddagger$ exact while the safety filter is inactive; - not a settling result; a slash separates the codes of two certified variables listed in that order. \emph{Aux.}: auxiliary dynamics in the certificate (Obs observer, DO disturbance observer, ESO extended-state observer, Est estimator, Ad adaptation, NN neural, Fz fuzzy, Cr actor-critic/identifier, Flt command filter, Dif differentiator, AW antisaturation/anti-windup state, Trg trigger, Sw switching, Reg regularization, PPF performance transformation, STA super-twisting states, Qnt quantizer, Enc encoder, CL composite learning, FC fault compensation, BLF barrier Lyapunov function, QP online quadratic program, TV scaling time-varying scaling, TS transf.\ time-scale transformation, S-Obs sampled-data observer, Dyn.\ gains dynamic observer/controller gains, graph gains graph-dependent gains, data stored-data regressor, Sat saturation compensation by bounded design or envelope relaxation). \emph{V}: N numerical, HIL hardware-in-the-loop, Hw physical experiment. Symbols: $e$ tracking error, $x$ plant state, $s$ sliding variable, $\tilde x$ estimation error; other symbols in the certified-variable column follow the cited source. Abbreviations in the realization column: ACC adaptive cruise control; ADP adaptive dynamic programming; ADRC active disturbance rejection control; ADT average dwell time; APF artificial potential field; AUV/UUV autonomous/unmanned underwater vehicle; CAV connected automated vehicle; CPS cyber-physical system; DDPG deep deterministic policy gradient; DOF degree of freedom; DPCC deadbeat predictive current control; DSC dynamic surface control; EL Euler-Lagrange; ET/ETC/DETC event-triggered/event-triggered control/dynamic ETC; ETM event-triggering mechanism; EV electric vehicle; FDI false-data injection; FNN fuzzy neural network; FTISS/FXISS finite-/fixed-time input-to-state stability; GRNN general regression neural network; LFC load frequency control; LOS line of sight; LSM linear synchronous motor; LTV linear time-varying; MAS multi-agent system; MIMO multi-input multi-output; NCS networked control system; NDO nonlinear disturbance observer; NMPC nonlinear model predictive control; NODE neural ordinary differential equation; OF output feedback; PD proportional-derivative; PDE partial differential equation; PMA pneumatic muscle actuator; PMSM/SPMSM (surface-mounted) permanent-magnet synchronous motor; PP/PPC prescribed performance/prescribed performance control; QUAV quadrotor UAV; RBFNN radial-basis-function NN; RISE robust integral of the sign of the error; RL reinforcement learning; SF state feedback; SM/SMC sliding mode/sliding-mode control; TD3 twin-delayed DDPG; TSM terminal sliding mode; TTR terminal-time regulator; TVHG time-varying high gain; UAV/UGV/USV unmanned aerial/ground/surface vehicle; VTOL vertical take-off and landing; WMR wheeled mobile robot; ZND/ZNN zeroing neural dynamics/network.

\setlength{\tabcolsep}{2.4pt}\renewcommand{\arraystretch}{1.10}\fontsize{9}{11}\selectfont
\begin{longtable}{@{}>{\raggedright\arraybackslash}p{2.25cm} >{\raggedright\arraybackslash}p{4.7cm} l >{\raggedright\arraybackslash}p{2.6cm} l >{\raggedright\arraybackslash}p{2.05cm} l@{}}
\caption{Architectural audit of the examined studies.}\label{tab:audit_full}\\
\toprule \textbf{Study} & \textbf{Realization (primary + augmentation; platform)} & \textbf{Source class} & \textbf{Certified variable} & \textbf{G} & \textbf{Aux.} & \textbf{V}\\ \midrule\endfirsthead
\multicolumn{7}{@{}l}{\textit{Table~\ref{tab:audit_full} (continued)}}\\ \toprule \textbf{Study} & \textbf{Realization (primary + augmentation; platform)} & \textbf{Source class} & \textbf{Certified variable} & \textbf{G} & \textbf{Aux.} & \textbf{V}\\ \midrule\endhead
\midrule\multicolumn{7}{r@{}}{\textit{continued on next page}}\\ \endfoot \bottomrule\endlastfoot
\multicolumn{7}{@{}l}{\textbf{Direct state feedback (Section \ref{subsubsec:direct_feedback})}}\\
\citealp{nezhad2026finite} & Robust voltage-based SF; flexible-joint manipulator & FT & $e$ & E & -- & HIL\\
\citealp{ning2025global} & SF with PT adjustment function; fully actuated sys. & PT & $x$ & E & -- & N\\
\citealp{zhang2026prescribed} & LTV SF from parametric Lyapunov eq. & PT & $x$ (bounded after $T_p$) & E & -- & N\\
\citealp{you2024finite} & Hybrid SF (intermittent/impulsive); network estimation & ST & state bound on window & -- & -- & N\\
\citealp{CruzZavala2026OperationalSpaceFiniteTime} & Transposed-Jacobian nonlinear PD, bounded torque; operational space & FT & Cartesian error & E & -- & N\\
\midrule
\multicolumn{7}{@{}l}{\textbf{Sliding-mode realizations (Section \ref{subsubsec:smc})}}\\
\citealp{kong2025predefined} & Fast TSM + predictive speed control + DO; PMSM & PdT & $e_1$ on $s_\omega=0$, DO error & E & DO & Hw\\
\citealp{kumar2025prescribed} & Adaptive SMC; DC-DC boost converter & PT & $s$, $\tilde\zeta$ / voltage error (asymptotic) & E/-- & Est, Ad, Sw & Hw\\
\citealp{lei2024fixed} & Dual SMC (integral-terminal + dynamic surface); LSM maglev & FxT & $w$, $s$, levitation-gap error & E & -- & Hw\\
\citealp{shi2024prescribed} & Nonsingular TSM, bounded TV gains; PMSM & PT & reaching + sliding phases & E & -- & Hw\\
\citealp{yin2025nonsingular} & Nonsingular fast SMC + DO; construction robot & PdT & $x$ ($\tilde x$ uncertified) & E\textsubscript{r} & DO & N\\
\citealp{hu2024finite} & Adaptive super-twisting SMC, faults; manipulator & FT & $e$ & E & Ad, STA & N\\
\citealp{li2025fastfinite} & Fast integral TSM + DO; SPMSM & FT & $s$, speed error & E\textsubscript{r} & DO & Hw\\
\citealp{ding2026switching} & Switching adaptive second-order SMC & PT & $s$, $x$ & P$^\ast$ & Ad, Sw & N\\
\citealp{wu2024adaptive} & Adaptive fuzzy integral SMC; boiler unit & FT & $s$ & -- & -- & N\\
\citealp{alsinai2026finite} & Adaptive super-twisting SMC; off-road EV & FT & $e$ & P & Ad & N\\
\citealp{liu2025practical} & Variable-exponent nonsingular SMC, regularized & FxT & $e$ & P & Reg & N\\
\citealp{jia2025prescribed} & Nonsingular SMC + NN & PT & $e$ & P & NN & N\\
\citealp{su2025second} & 2nd-order nonsingular SMC + PPF; nanopositioning & PdT & $e$ in envelope & P & PPF & Hw\\
\citealp{asadollahi2024fixed} & TSM, (free-will) arbitrary time; finance model & AT & $s$ & E & -- & N\\
\citealp{Omiloli2026RectifierFiniteTime} & Disturbance-adaptive FT sliding surface + current loop; rectifier & FT & voltage, current errors & E & Est & Hw\\
\citealp{Ren2027CompositeFiniteTimeQuadrotor} & FNITSMC + FT DO + adaptive barrier on $s$; quadrotor attitude & FT & attitude error (local) & E & DO, BLF & Hw\\
\citealp{Stihi2026FTPPFTVSMC} & Time-varying SMC with FT performance function; manipulator & FT & $e$ in envelope & U & PPF & Hw\\
\citealp{Shi2026AdaptiveFractionalFiniteTimeQuadrotor} & Fractional-order terminal SMC + STA observer; quadrotor & FT & $s$, $e$ (residual) & P & DO, STA & HIL\\
\citealp{Dong2026DynamicPowersFixedTimeSpacecraft} & Dynamic-power nonsingular SMC, boundary layer; spacecraft & FxT & $(\sigma_e,\omega_e)$ to nbhd & P & Reg & N\\
\citealp{Zhao2026FixedTimeVTOLVariablePPF} & Terminal SMC + FxT ESO + variable PPF, faults; VTOL UAV & FxT & $e$ (boundary layer) & P & ESO, Ad, PPF & HIL\\
\citealp{Zhao2026FixedTimeSmoothCanardUAV} & Smooth terminal SMC, relaxed PPF, allocation; rotor/wing UAV & FxT & $e$ (boundary layer) & P & Ad, PPF & HIL\\
\citealp{Xu2026AdaptiveFixedTimePMSM} & Integral TSM + FxT DO + improved DPCC; PMSM & FxT & speed error, $\tilde d$ & E & DO, Ad & Hw\\
\citealp{Pang2026NeuralFixedTimeSteerByWire} & Segmented nonsingular SMC + GRNN dead-zone comp.; steer-by-wire & FxT & steering error & P & NN & Hw\\
\citealp{Liang2026UUVPracticalFixedTimeFTC} & Nonsingular fast TSM + FxT DO + appointed-time PPC; UUV & FxT & $e$ in envelope & P & DO, Ad, PPF & N\\
\citealp{Shen2026VECFixedTimeDCDC} & Varying-exponent SMC + FxT DO; DC-DC buck converter & FxT & $s$ after observer transient & E & DO & Hw\\
\citealp{Yin2026ManipulatorPTESO} & PdT integral SMC + PdT ESO + FT PPF, faults; manipulator & PdT & $e$ & E & ESO, PPF, Ad & N\\
\citealp{Lai2026PredefinedCMNN} & Adaptive integral SMC; cyclic memristive NN synchronization & PdT & synchronization error & E & Ad & N\\
\citealp{Ding2026PTABFSMCPMSM} & PdT adaptive-barrier reaching law + FT DO; PMSM & PdT & speed error & P & DO, Ad & Hw\\
\citealp{Lv2026PredefinedLFCFOSMC} & Fractional-order SMC + specified-time PPF; single/multi-area LFC & PdT & frequency deviation & E & PPF & N\\
\citealp{Shao2026TTRPrescribedPerformanceSMC} & TTR-based performance tube + asymmetric barrier SMC; 2nd-order systems, platoon & PT & $x_1$ (nominal exact at $t_f$) & P & PPF & N\\
\citealp{Ding2026PDFSuperTwistingPsT} & Periodic-delayed-feedback STA + DO / state observer; high-order systems & PT & $y$ & E & STA, DO, Obs & N\\
\citealp{Liu2026ElectromagneticTakeover} & FxT reaching law + RBFNN SMC, PT sliding phase; cooperative spacecraft takeover & PT & $s$ (FxT); $e$ residual within $\sqrt2T$ & P & NN & N\\
\citealp{Zhang2026SecurityReachableSetSMC} & Adaptive neural SMC; semi-Markov jump CPS, delays, actuator attacks & FT & $s$ / $x$ (mean-square reachable set) & E/U & NN, Ad & N\\
\midrule
\multicolumn{7}{@{}l}{\textbf{Backstepping realizations (Section \ref{subsubsec:recursive})}}\\
\citealp{zeng2025backstepping} & Composite backstepping; two-time-scale systems & PT & slow + fast states, $u$ & E & TV scaling & N\\
\citealp{liu2025auv} & Backstepping + PdT DO + PdT filters; AUV & PdT & $e$, velocity, filter errors & P & DO, Flt & N\\
\citealp{luo2025glider} & Backstepping + cmd filter + ESO + PPF; glider & FxT & attitude error & P & Flt, ESO, PPF & N\\
\citealp{ding2024rectifier} & Backstepping + FxT SM-DO + differentiator; rectifier & FT & plant errors & P & DO, Dif & Hw\\
\citealp{li2024autonomous} & Decentralized backstepping + ext.\ observers; DC microgrid & FT & plant errors & P & Obs & Hw\\
\citealp{shi2024adaptive} & Tunable backstepping + fuzzy + boundary adaptation & PdT & $e$ & P & Fz, Ad & N\\
\citealp{hao2026predefined} & Backstepping + identifier-actor-critic + sat.\ comp. & PdT & composite errors & P$^\ast$ & Cr, AW & N\\
\citealp{wang2025optimal} & ADP optimal backstepping; reconfig.\ manipulator & FT & force/position errors & P & Cr & Hw\\
\citealp{zhang2025command} & Cmd-filtered adaptive fuzzy + BLF + ETC; QBall2 UAV & FT & $e$ (constraints kept) & P & Flt, Fz, Trg & Hw\\
\citealp{peng2024antisaturation} & Antisaturation backstepping + BLF + PPF; slung load & FxT & variables in envelope & U & AW, BLF, PPF & N\\
\citealp{Wang2026PracticalFiniteTimeTMR} & Backstepping with auxiliary variable, unknown reference velocity; mobile robot & FT & $e$ & P & Ad & Hw\\
\citealp{Ebrahimpour2026AdaptiveFiniteTimeFixedWing} & Backstepping + fast terminal 2nd-order SMC + differentiators; fixed-wing UAV & FT & $e$, differentiator errors & E/P & Dif, NN, AW & N\\
\citealp{Niu2026FixedTimeBacksteppingEHS} & Cmd-filtered backstepping + single RBFNN; electro-hydraulic & FxT & $e$ & P & Flt, NN & N\\
\citealp{Li2026FixedTimeAntiSwingQuadrotor} & Three-loop cascade, online mass/length estimation; slung load & FxT & position, swing errors & P & Est & N\\
\citealp{Liu2026AdaptivePTPPCQuadrotor} & Cmd-filtered backstepping + modified PdT law + PPF; quadrotor & PdT & $e$ in envelope & P & Fz, Flt, PPF & N\\
\citealp{Zhang2026USVOptimizedBackstepping} & Optimized backstepping + actor-critic + FxT DO + ETC; USV & FxT & $\tilde d$ / $e$ (SGUUB) & E/U & DO, NN, Cr, Trg & N\\
\citealp{Chu2026UnifiedNeuralEventTriggered} & Neural recursive design + FT differentiators + ETC; switched systems, state constraints & B & $e$ in time-varying set & -- & NN, Dif, Trg, Sw & N\\
\midrule
\multicolumn{7}{@{}l}{\textbf{Observer-based and output feedback (Section \ref{subsubsec:observer_output})}}\\
\citealp{you2025visual} & DO-based control, visual feedback; bolt tightening & FT & $e$ & E\textsubscript{r} & DO & Hw\\
\citealp{sui2025usv} & ESO-based tracking + PPF + saturation; USV & PT & $e$ in envelope & U & ESO & N\\
\citealp{cai2026predefined} & Fuzzy distributed optimal OF via RL; MAS & PdT & sync.\ error, Nash obj. & E\textsubscript{r} & Obs, Cr & N\\
\citealp{tan2026decentralized} & Decentralized OF, local outputs; interconnected sys. & PT & $x$, $\tilde x$, $u$ & E & Obs & N\\
\citealp{que2024disturbance} & DO-based tracking, nonvanishing uncertainty & PT & $e$, $\tilde x$ & E & DO & N\\
\citealp{farhat2024walking} & NDO-based TSM; NAO humanoid & FxT & walking position error & E & DO & Hw\\
\citealp{xu2025autotuned} & Autotuned nonlinear ESO; motor-drive servo & FxT & speed error & P & ESO & Hw\\
\citealp{qin2024enhanced} & Enhanced ESO + tracking; wheeled mobile robot & PT & $e$ & P & ESO & N\\
\citealp{ye2024fdi} & ESO-based tracking under FDI attacks; WMR & FT & $e$ & P & ESO & N\\
\citealp{liu2026hydraulic} & ESO-based composite anti-disturbance; hydraulic arm & PdT & $e$ & P & ESO & Hw\\
\citealp{lai2026output} & OF + hysteresis observer, predefined accuracy; piezo & FxT & $e$ within accuracy & P & Obs & Hw\\
\citealp{wang2026global} & LTV observer + LTV feedback, meas.\ uncertainty & PT & $x$, $\tilde x$ & E & Obs & N\\
\citealp{mao2025fixed} & Dual dynamic-gain OF; planar systems & FxT & $x$, $\tilde x$ & E & Obs, Dyn.\ gains, Sw & N\\
\citealp{Zhong2026AMVFiniteTime} & FT velocity observer + FT LOS guidance + adaptive FTSMC; marine vehicle & FT & $s$; heading error UUB & P & Obs, Ad & Hw\\
\citealp{Hou2026FiniteTimeFaultTolerantEVTOL} & FT fault-factor observer + FT velocity/attitude control; tilt-rotor eVTOL & FT & fault-factor, tracking errors & E & Obs, FC & Hw\\
\citealp{Wang2026FiniteTimeSynchronizedLFC} & FT ESO + FT-synchronized ADRC; LFC, heterogeneous units & FT & ACE / per-unit power consensus & P/E & ESO & N\\
\citealp{Meng2026FixedTimeAmphibiousVehicle} & Switched FxT ESO + switched FxT SMC; amphibious USV & FxT & position, velocity, $\tilde x$ & E & ESO, Sw & Hw\\
\citealp{Trinh2026ALRNNFixedTimePMSM} & Adaptive-learning-rate NN FxT observer + FxT SMC; PMSM & FxT & speed error & P & Obs, NN & Hw\\
\citealp{Cao2026CompositeObserverPT} & Fuzzy state observer + ESO + PdT filter, NSDF constraints; manipulator & PdT & $e$ & P & Obs, ESO, Fz, Flt & N\\
\citealp{Zhang2026FNNPTESOUSV} & FNN PT ESO (bounded gain) + BLF adaptive formation, faults; USVs & PT & formation error & P & ESO, NN, Fz, FC & N\\
\citealp{Cai2026MSVPracticalPsT} & Adaptive PT ESO + bounded TVHG backstepping; marine surface vessel & PT & $e$ & P & ESO, Ad & N\\
\citealp{Cui2026PrescribedTimeMorphingHMV} & cosh-type PT law + PT DO + sliding-tube PPF, TD3 morphing; hypersonic vehicle & PT & attitude error & E & DO, PPF, Cr & N\\
\midrule
\multicolumn{7}{@{}l}{\textbf{Adaptive and learning-based (Section \ref{subsubsec:adaptive_learning})}}\\
\citealp{wang2025predefined} & Backstepping + NN + cmd filters, faults, delay & PdT & $e$ & P & NN, Ad, Flt & N\\
\citealp{cai2016fuzzy} & Adaptive fuzzy backstepping; mechanical systems & FT & $x$, $e$ & P & Fz, Ad & N\\
\citealp{cuan2023cps} & Adaptive NN control; CPS under attacks & FT & $e$ & P & NN & N\\
\citealp{fu2025inverter} & Cascade adaptive control; stand-alone inverter & FT & voltage, current errors & P & Ad & HIL\\
\citealp{rodriguez2022usv} & Adaptive SMC; USV & FT & $e$ & P & Ad & Hw\\
\citealp{chen2023uam} & Adaptive SM-DO + PPC; aerial manipulator & FT & $e$ in envelope & P & DO, Ad, PPF & Hw\\
\citealp{cui2022quav} & Cmd-filtered adaptive PPC; QUAV & FxT & $e$ & P & Flt, Ad & N\\
\citealp{guo2025hydraulic} & SM adaptive control, deadzone; hydraulic manipulator & FxT & $e$ & P & Ad & Hw\\
\citealp{liu2025robot} & Adaptive NN control, saturation + PPC; robot & FxT & $e$ & P & NN & Hw\\
\citealp{wang2023maglev} & Adaptive position control; magnetic levitation & FxT & position ($\tilde d$ residual) & E\textsubscript{r} & DO & Hw\\
\citealp{hua2021adaptive} & Adaptive control; uncertain nonlinear systems & PT & $x$ ($\hat\theta$ bounded) & E\textsubscript{r} & Ad & N\\
\citealp{ni2021global} & NN control, time and accuracy, dead zone & PdT & $e$ & P & NN & N\\
\citealp{cui2025observer} & Observer-based adaptive fuzzy PPC; QUAV & PdT & $e$ & P & Obs, Fz & N\\
\citealp{liu2025global} & Fuzzy adaptive + global PP + FxT DO; manipulator & PdT & $e$ & P & DO, Fz & N\\
\citealp{cao2026parameter} & Parameter estimation + adaptive control; PMA & PT & $e$, $\tilde\theta$ & P & Est & Hw\\
\citealp{jiang2025composite} & Composite fuzzy, interval excitation; Kinova Gen3 & PdT & $e$ ($\tilde\theta$ to nbhd) & E\textsubscript{r} & Fz, data & Hw\\
\citealp{chen2023spacecraft} & Adaptive control; 6-DOF spacecraft formation & PT & regulated variables & E\textsubscript{r} & Ad & N\\
\citealp{ye2025armfinger} & Decentralized control; robotic arm-finger & PT & regulated variables & E\textsubscript{r} & Ad & N\\
\citealp{li2025platoon} & Fuzzy adaptive cooperative control; platoon & PT & spacing errors & E\textsubscript{r} & Fz & N\\
\citealp{sun2023appointed} & Appointed-time, disturbance negation; hypersonic & PdT & $e$ & P & DO & N\\
\citealp{song2026rl} & RL optimized + PdT DO + DSC + ETC; 2-DOF helicopter & PdT & composite errors & P & Cr, DO, Trg & N\\
\citealp{Phothongkum2026ConstrainedRLFiniteTime} & Constrained DDPG with FT weight conditions; benchmark systems & FT & $x$ & E & NN & N\\
\citealp{Chang2026FiniteTimeHinfServo} & Adaptive critic, FT weight update; servo $H_\infty$ & FT & weights; $x$ to bounded set & P & Cr & N\\
\citealp{Wang2026SpacecraftFiniteTimePE} & Critic learning, FT capture cost; spacecraft pursuit-evasion & FT & relative state, weights & E & Cr & N\\
\citealp{Lv2026MultiFiniteTimeHinfServo} & Mixed zero-sum critic learning; multi-drive servo & FT & states, weights to bounded set & P & Cr & N\\
\citealp{Wang2026FixedTimeRLFaultTolerant} & Actor-critic + fault compensation, saturation; quadrotor, launcher & FxT & $e$ & P & Cr, FC & HIL\\
\citealp{Xue2026NAGFixedTimeActorCritic} & Nesterov actor-critic concurrent learning; nonlinear systems & FxT & weights; $x$ UUB & U & Cr & N\\
\citealp{Niu2026SecurityPracticalFixedTimeCPS} & Nussbaum adaptive design, deception attacks; high-power CPS & FxT & closed-loop signals & P & Ad & N\\
\citealp{Yu2026ContinuumPTSS} & NODE model + PdT-synchronized ZND; continuum robot & PdT & task-space error & E & NN & Hw\\
\citealp{Mai2026PrescribedTimeRLAUV} & PT hybrid observer + APF + actor-critic backstepping, intermittent comm.; multi-AUV & PT & observer/\allowbreak parameter errors & P & Obs, NN, Cr, Flt & N\\
\citealp{Liu2026NNPracticalPrescribedTimeNCS} & NN attack compensator + adaptive backstepping + prescribed-performance boundary; NCS under deception & PT & $e$ in envelope & P & NN, Ad, Dif, PPF & N\\
\citealp{Li2026EventTriggeredUnknownCoefficientsPsT} & Adaptive PT control, PT Nussbaum lemma, event triggering; unknown control direction & PT & $x$, $u$ & E & Ad, Trg & N\\
\midrule
\multicolumn{7}{@{}l}{\textbf{Distributed and cooperative (Section \ref{subsubsec:distributed})}}\\
\citealp{wang2018prescribed} & Distributed TV scaling; consensus, containment & PT & consensus errors & E & graph gains & N\\
\citealp{liu2024prescribed} & Formation control, switching topology; USV & PT & mean-square formation errors & U & Sw, Ad & N\\
\citealp{chen2024prescribed} & Coop.\ output regulation + PT exosystem observer & PT & outputs, $\tilde x$ & E & Obs & N\\
\citealp{zhang2025dptuio} & Distributed unknown-input observer; digraph & PT & state at every node & E & TS transf. & N\\
\citealp{yue2025predefined} & Distributed resource allocation; balanced digraph & PdT & optimizer error & E & -- & N\\
\citealp{liu2023multiobjective} & Multiagent multiobjective optimization & PdT & nbhd of Pareto sol. & P & -- & N\\
\citealp{jin2025practical} & Practical output allocation; heterogeneous MAS & PdT & allocation error & P & -- & N\\
\citealp{an2025finite} & Fully distributed formation tracking + DO & FT & formation error & E & Ad, DO & N\\
\citealp{zhou2026formation} & FxT leader observer + regulator-eq.\ solver & FxT & formation error & E & Obs & N\\
\citealp{liu2023resource} & Optimal resource management; switching graphs & FxT & dispatch error & E & -- & N\\
\citealp{duan2022uav} & Cooperative UAV control, faults, input limits & FxT & formation error & P & NN, FC & N\\
\citealp{zhang2024quantized} & Adaptive formation tracking, quantized inputs & B & formation error & P & Qnt, Ad & N\\
\citealp{zhao2025collision} & Collision-free tracking, limited data rate & B & $e$, safety & -- & Obs, Qnt, Ad, Flt & N\\
\citealp{Li2026FiniteTimeBipartiteConsensus} & Smooth switched fuzzy backstepping, signed graph, input delay & FT & bipartite error & P & Fz & N\\
\citealp{Cheah2026FTISSFXISSFormation} & Bearing-only FTISS/FXISS formation + disturbance estimate & FT/FxT & formation error & P & Ad & Hw\\
\citealp{Chen2026FTBCMultiAgentAttacks} & Observer-based ET bounded consensus, DoS/FDI attacks & ST & consensus error on window & -- & Obs, Trg & N\\
\citealp{Xu2026FixedTimeOptimalBipartiteContainment} & ICA learning, sensor/actuator faults, saturation; bipartite containment & FxT & containment error & P & Cr, AW, DO & N\\
\citealp{Yang2026FixedTimeMicrogridDoS} & DoS-resilient intermittent secondary control + dynamic ETC; AC microgrid & FxT & frequency, power sharing & E/P & Trg & N\\
\citealp{Liang2026EventTriggeredPracticalFixedTimeConsensus} & Observer + NN + DSC, event-triggered consensus tracking & FxT & consensus error & P & Obs, NN, Trg & N\\
\citealp{Liu2026DynamicEdgeFixedTimeConsensus} & Dynamic edge/self-triggered consensus tracking & FxT & error in $R$-nbhd at $T$ & P & Trg & N\\
\citealp{Tang2026FixedTimeOptimalBipartiteContainment} & Optimized backstepping, identifier-actor-critic, saturation & FxT & containment error & P & Cr, AW & N\\
\citealp{Feng2026PredefinedAGHMAS} & PdT DO + PdT SMC + quantitative PPF + DETC; UAV-UGV formation & PdT & formation and UAV $z$-tracking errors & P & DO, PPF, Trg & N\\
\citealp{Tian2026PredefinedUSV} & Topology switching + PdT observer + RL, saturation, faults; USVs & PdT & formation errors & P & Obs, Cr, AW, Trg & N\\
\citealp{Sun2026PredefinedUAVSpatiotemporal} & Time-base-generator observer and SMC + APF avoidance; UAV swarm & PdT & formation error ($\delta$-residual) & P & Obs & N\\
\citealp{Ma2026PredefinedSynchronizedBipartite} & Hierarchical estimator + algebraic disturbance reconstruction; EL agents & PdT & formation error, synchronized & E & Est & N\\
\citealp{Zhu2026PredefinedZNN} & Event-triggered fuzzy-parameter ZNN; multi-robot formation & PdT & formation residual & E & Trg & N\\
\citealp{Guo2026RLPredefinedAUV} & Actor-critic RL + PdT terminal SMC + anti-windup; AUV formation & PdT & formation errors & P & Cr, AW & N\\
\citealp{Li2026FilterPredefinedConsensus} & Critic-actor RL + PdT filter, faults, DETC; consensus & PdT & consensus error & P & Cr, Flt, FC, Trg & N\\
\citealp{Chen2026PredefinedDeception} & Fuzzy adaptive PdT SMC, deception attacks; 2nd-order MAS & PdT & formation error & P & Fz, Ad & N\\
\citealp{Zhang2026DCMicrogridPT} & ACK-based time gain + FDI compensation; DC microgrids & PdT & voltage, current-sharing errors & P & Ad, TV scaling & HIL\\
\citealp{Cui2026AdaptiveFuzzyPTOptimalConsensus} & PdT gradient/Hessian estimator + fuzzy backstepping; optimal consensus & PdT & output to time-varying optimum & P & Est, Fz, Flt & N\\
\citealp{Ding2026GeneralizedPTUAV} & Arc-length consensus + PdT state estimator; UAV elliptical formation & PdT & arc-length error & P & Obs & N\\
\citealp{Luo2026HiTLUAVUSV} & Dual PdT PPF, human in the loop, ETC; UAV-USV formation-containment & PdT & errors in envelopes & U & PPF, Trg & N\\
\citealp{Zhao2026FunnelPTPlatoon} & Funnel + fuzzy observer + PdT filter + DETC; vehicle platoon & PdT & spacing error in funnel & P & Obs, Fz, Flt, Trg & N\\
\citealp{Cai2026PTDistributedOptimalRL} & Graphical game, critic learning with replay, saturation; MAS & PdT & synchronization error & E & Cr & N\\
\citealp{Xu2026AdaptivePrescribedPerformanceMAS} & Adding a power integrator + shifted tube PPF + DoS filter; constrained MAS & PT & $e$ in envelope & U & Ad, Flt, PPF & N\\
\citealp{Zhang2026PrescribedTimeConsensusDelayDoS} & Delay-compensating PT consensus + distributed observer; DoS, node/link delays & PT & consensus error & E & Obs, Sw & N\\
\citealp{Zhu2026PrescribedTimeDelayDoS} & Event-triggered PT consensus, delay-dependent PT lemma; DoS & PT & consensus error & E & Trg & N\\
\citealp{Liu2026DistributedPrescribedTimeOptimalFormation} & PT observer + flexible PPF + actor-critic + switching-threshold ETM; AUV-USV & PT & $e$ in envelope & P & Obs, NN, Cr, Flt, Trg, PPF & N\\
\citealp{Zhang2026DynamicSelfTriggeredBipartitePsT} & Adaptive NN backstepping + dynamic self-triggering, faults; signed-graph formation & PT & formation error & P & NN, Ad, Trg, FC, Dif & N\\
\citealp{Yao2026FlexibleManipulatorsPrescribedPerformance} & Switched adaptive observer + PDE boundary control + FT PPF; flexible manipulators, DoS & FT & $e$ in envelope (asympt. tracking) & U & Obs, NN, Sw, PPF & N\\
\citealp{Zhu2026OptimizedSlidingModePsTFormation} & Optimized SMC + identifier-critic-actor + PT performance function; heterogeneous MAS & PT & $e$ in envelope & U & NN, Cr, PPF & N\\
\citealp{Zhou2026PracticalPsTRigidFormation} & Frozen time-varying scaling + two-power law + distance PPF; rigid formation & PT & distance errors & P & PPF & Hw\\
\citealp{Li2026PsTClusterConsensusHMAS} & PT cluster estimator + NN PT synchronization, faults, saturation; USV-UAV, FDI & PT & cluster consensus error & P & Est, NN, FC, Sw & N\\
\citealp{An2026PsTPerformancePlatoon} & Variable-gain ESO + tracking differentiator + PT performance function; CAV platoon & PT & spacing error in envelope & U & ESO, Dif, PPF & N\\
\midrule
\multicolumn{7}{@{}l}{\textbf{Constraint- and implementation-aware (Section \ref{subsec:implementation_architectures})}}\\
\citealp{garg2022fixedqp} & CLF-CBF QP; ACC, multi-robot planning & FxT & safe set, target set & E & QP & N\\
\citealp{abel2024prescribedSafety} & Safety design; strict-feedback systems & PT & distance to safe set & E$^\dagger$ & -- & N\\
\citealp{zhang2023finiteRobot} & Composite learning + asymmetric output constr.; robot & FT & $e$ in envelope & E & CL & Hw\\
\citealp{sun2021finiteSaturation} & Saturated nonstrict-feedback, error constraint & FT & $e$ in envelope & P & AW & N\\
\citealp{song2026fixedConverter} & Current-constrained voltage regulation; DC-DC & FxT & voltage error & P & DO, Est, Ad & Hw\\
\citealp{chen2025predefinedMobile} & Deferred full-state constraints; mobile robot & PdT & $e$, state constr. & P & BLF & Hw\\
\citealp{zhang2025predefinedUSV} & Cluster formation, global output constraint; USVs & PdT & formation error & P & Est, DO & N\\
\citealp{liu2024commandFilter} & Cmd-filtered, state constraints, preassigned perf. & PdT & $e$ in envelope & P & Flt & N\\
\citealp{ning2024prescribedMobile} & Tracking with performance constraints; WMR & PT & $e$, envelope & E & -- & N\\
\citealp{nasab2025safeMobile} & PT controller + CBF safety filter; WMR & PT & $e$, safe set & E$^\ddagger$ & QP & N\\
\citealp{ma2025boundedFinite} & Bounded SM feedback; saturated linear systems & FT & $x$ (regional) & E & -- & N\\
\citealp{liu2023nonholonomic} & Saturated tracking, state constraints; nonholonomic & FT & $e$ & P & Sat & Hw\\
\citealp{gong2022predefinedSaturation} & Tracking + antisaturation system; high-order & PdT & $e$, $\xi$ & E\textsubscript{r} & AW, DO, Flt & N\\
\citealp{yang2024actorCriticPredefined} & Actor-critic fuzzy optimal, saturation & PdT & $e$ & P & Cr, AW & N\\
\citealp{zhou2025fixedAntiwindup} & Anti-windup SM tracking; piezoelectric stage & FxT & AW state $\lambda_1$ / $e$ (asymptotic) & P/-- & AW & Hw\\
\citealp{ji2022saturationTolerant} & Saturation-tolerant PPC; MIMO systems & B & $e$ in relaxed envelope & P & Sat & N\\
\citealp{lin2025practicalPrescribed} & Time-base generator, saturation; nonaffine & PT & output error & P & -- & N\\
\citealp{mao2023finiteSampled} & Sampled-data OF; feedforward nonholonomic & FT & $x$, $\tilde x$ & E & S-Obs & N\\
\citealp{mao2025fixedSampled} & Neural sampled-data OF & FxT & $x$ ($h$-dep.\ residual) & P & NN, S-Obs & N\\
\citealp{liu2020fixedEvent} & Event/self-triggered leader-follower consensus & FxT & consensus error & E & Trg & N\\
\citealp{wu2024predefinedEvent} & Event-triggered adaptive tracking, state constr. & PdT & $e$ & P & Trg, Ad & N\\
\citealp{ning2023eventPrescribed} & Event-triggered control; uncertain nonlinear & PT & $x$ & E & Trg & N\\
\citealp{ning2023adaptiveEventPrescribed} & Event-triggered adaptive, TV parameters & PT & $x$ & E\textsubscript{r} & Trg, Ad & N\\
\citealp{Jiang2026EventTriggeredFiniteTime} & Power-integrator adaptive backstepping + BLF + ETC; stochastic & FT & $x$ (in probability) & E & Ad, BLF, Trg & N\\
\citealp{Pan2026NCSMPCFiniteTimeAUV} & Lyapunov NMPC with FT terminal constraint + adaptive STA + ETM; AUV & FT & kinematic errors, $s$ & E & Trg, STA, Ad & Hw\\
\citealp{Wang2026MemristiveFiniteTimeQuasiSync} & Quantized intermittent ETC, output measurements; memristive NN & FT & sync.\ error (quasi) & P & Qnt, Trg & N\\
\citealp{Tian2026AdaptiveEventTriggeredHNS} & Adaptive ETC, unknown control coefficients, deferred constraint & FT & $x$ (global) & E & Ad, Trg & N\\
\citealp{Wang2026HFNCBFRoboticSafeTracking} & NN sliding-mode tracking + high-order FT neural CBF QP; manipulator & FT & safe set; $e$ to nbhd & E$^\dagger$/P & NN, QP & Hw\\
\citealp{Liu2026MultiLoopNCSFiniteTime} & Dual switching time-delay model, ADT scheduling; multi-loop NCS & ST & state bound on window & -- & Sw & N\\
\citealp{Xu2026NeuroadaptiveFixedTimePrescribedPerformance} & NN + DO + DSC + BLF with asymmetric FxT PPF; DC motor & FxT & $e$ in envelope & U & NN, DO, Flt, BLF & Hw\\
\citealp{Guo2026ADPFixedTimePrescribedPerformance} & FxT PPF + ESO + critic-only ADP + ETC; spacecraft attitude & FxT & $e$ in envelope & U & ESO, Cr, Trg & N\\
\citealp{Nguyen2026FixedTimeCableDrivenRobots} & FxT wrench control + $H_\infty$ observer + QP tension allocation; cable robot & FxT & position, velocity errors & E & Obs, QP & N\\
\citealp{Zhao2026FixedTimeSelfTriggered} & Limited-time output constraints, NN, FxT DSC filter, self-triggering & FxT & $e$ & P & NN, Flt, Trg & N\\
\citealp{Lin2026FixedTimePPCMMCPS} & FxT PPF + BLF + saturation compensation; macro-micro stage & FxT & $e$ in envelope & U & AW, Flt, PPF & N\\
\citealp{Hua2026SafetyCriticalFixedTimeUAV} & FxT PPF in nonsingular SMC, dynamic ETC; quadrotor & FxT & transformed error to set & P & PPF, Trg & N\\
\citealp{Liu2026DynamicMemoryPT} & NN backstepping + Nussbaum + memory ETC, hysteresis; stochastic delay & PdT & $e$ (in probability) & P & NN, Flt, Trg & N\\
\citealp{Li2026TeleoperationPT} & PdT torque estimator + sEMG admittance + ETC; teleoperation & PdT & $e$ ($\varepsilon$-residual) & P & Est, PPF, Trg & Hw\\
\citealp{Zhou2026PTQuasiSynchronization} & Intermittent self-triggered control, bounded gain; stochastic fuzzy networks & PdT & $\mathbb E\|e\|^2$ (quasi) & P & Trg & N\\
\citealp{Golestani2026PracticalPrescribedTimePPC} & Adaptive DSC + bounded performance-rate PPF + vanishing $\sigma$-modification & PT & $e$ (then asymptotic) & P & Ad, Flt, PPF & N\\
\citealp{Kim2026EventImpulsivePrescribedTimeUUV} & Dual-channel dynamic ETC + impulsive PT tracker + saturation compensator; UUV & PT & $e$ & P & NN, Flt, Trg, AW & N\\
\citealp{Shao2026PsTSpacecraftSafety} & RISE DO + APF + time-scale transformation; spacecraft proximity, collision/visibility & PT & relative position & P$^\ast$ & DO, Reg & N\\
\end{longtable}\normalsize\setlength{\tabcolsep}{6pt}\renewcommand{\arraystretch}{1.0}

\bibliographystyle{cas-model2-names}
\bibliography{references}

\end{document}